\documentclass[12pt]{article}

\usepackage{latexsym}
\usepackage{amsmath}
\usepackage{amssymb}
\usepackage{amsfonts}
\usepackage{amsthm}
\usepackage{amscd}
\usepackage{epsfig}
\usepackage{verbatim}
\usepackage{fancybox}
\usepackage{moreverb}
\usepackage{graphicx}
\usepackage{psfrag}
\usepackage[english]{babel}
\usepackage{hyperref}
\usepackage[subnum]{cases}
\usepackage{appendix}

\newcommand{\R}{{\mathbb R}}

\newcommand{\Z}{{\mathbb Z}}

\newcommand{\D}{{\partial}}

\newtheorem{theorem}{Theorem}

\title{\Large Asymptotic-Preserving scheme for a bi-fluid Euler-Lorentz model.}
\date{}
\author{St\'ephane Brull, Pierre Degond, Farice Deluzet, Alexandre Mouton.}

\begin{document}

\maketitle\begin{abstract}
The present work is devoted to the simulation of a strongly magnetized plasma considered as a mixture of an ion fluid and an electron fluid. For the sake of
simplicity, we assume that the model is isothermal and described by Euler equations coupled with a term representing the Lorentz force. Moreover we assume that both Euler systems are coupled through a quasi-neutrality constraint of the form $n_{i}=n_{e}$. The numerical method which is described in the present document is based on an Asymptotic-Preserving semi-discretization in time of a variant of this two-fluid Euler-Lorentz model with a small perturbation of the quasi-neutrality constraint. Firstly, we present the two-fluid model and the motivations for introducing a small perturbation into the quasi-neutrality equation, then we describe the time semi-discretization of the perturbed model and a fully-discrete finite volume scheme based on it. Finally, we present some numerical results which have been obtained with this method.
\end{abstract}

\maketitle



\noindent \textbf{Acknowledgments:}


\section{Introduction.}

This paper is devoted to the construction of a numerical scheme for the simulation of a two-fluid Euler-Lorentz model: such a model represents the evolution of a mixture of an ion gas and an electron gas which are submitted to the \textit{Lorentz force}. More precisely, we focus in this paper on a situation involving a strong Lorentz force and a low Mach number regime for both ion and electron fluids, \textit{i.e.} we assume that pressure and Lorentz forces are of the same order as $\tau^{-1}$ where $\tau > 0$ is the square of the ion Mach number and also represents the ratio between the ion gyro-period and the characteristic time scale of the experiment. When $\tau$ converges to 0, we reach an asymptotic regime which is referred to as the \textit{drift-fluid regime} or \textit{gyro-fluid regime}: in this limit regime, the pressure force for the ions and for the electrons is balanced by Lorentz force. In return, the momentum equations within the two-fluid Euler system degenerate into a pair of equations in which the parallel components of the ion and electron velocities can be viewed as Lagrange multipliers of the zero total force equations in the direction of the magnetic field (see \cite{BDD,DDSV}). \\
\indent Such a model describes plasma physics experiments involving strong external magnetic fields, such as Magnetic Confinement Fusion (MCF) experiments in tokamak reactors. In such a case, the rescaled gyro-period of the confined particles, which is denoted by $\tau$, can be close to 0. The assumption that $\tau$ is very small leads to a singularly perturbed Euler-Lorentz model. The limit $\tau \to 0$ is referred to as the \textit{gyro-fluid} limit (see \cite{Beer-Hammett, Dorland-Hammett, Hammett-Beer-Dorland-Cowley-Smith}). It is also possible to consider a \textit{gyro-kinetic} approach when a kinetic model is considered from the onset, instead of fluid equations (see \cite{Brizard_PhD, Hahm-Brizard, Dhaeseleer, Dubin, Frenod-Mouton, Grandgirard_JCP, Grandgirard, Heikkinen, Lee, Lee_2, Littlejohn}). For generalities on asymptotic regimes for fusion plasmas physics, we refer to \cite{Miyamoto}. \\
\indent In many experimental cases, the value of $\tau$ is not uniform and can vary a lot between subdomains of the tokamak. Additionally, it may depend on the time variable: in most of MCF experiments, $\tau$ is very small in the plasma core whereas it can be of order 1 far from the plasma core. From a numerical point of view, the usual approach for simulating both cases together consists in a domain decomposition according to the local value of $\tau$. More precisely, we choose to simulate the initial $\tau$-dependent model in the regions where $\tau = \mathcal{O}(1)$, and we choose the limit model in the regions where $\tau \ll 1$. Such an approach involves different numerical methods for solving either the Euler-Lorentz model or its drift-fluid limit according to the value of $\tau$. Generally, the coupling of these methods is not straightforward and presents several drawbacks such as the treatment of the interface position (or cross-talk region): indeed, it can depend on the time variable and, in most cases, costly algorithms are required to simulate the motion of the interface and to couple it with the space mesh. \\
\indent We choose a different method based on the resolution of the $\tau$-dependent Euler-Lorentz model and on the design of a scheme which is able to handle both the cases $\tau = \mathcal{O}(1)$ and $\tau \ll 1$. Then this so-called \textit{Asymptotic-Preserving (AP) method} provides consistent approximation of the Euler-Lorentz model when $\tau = \mathcal{O}(1)$ and of its limit regime when $\tau \to 0$, and does not require a $\tau$-dependent stability condition. As a consequence, such a method can be used on the whole simulation domain for both the $\tau = \mathcal{O}(1)$ and $\tau \ll 1$ regimes. AP schemes have been introduced by S. Jin \cite{Jin} and were applied on tranport models and their diffusive limits. Other applications can be found in plasma physics (see \cite{Belaouar-Crouseilles-Degond-Sonnendrucker, crisp1, crisp2, crisp3, Navoret, Degond-Deluzet-Negulescu, Savelief, Degond-Liu-Vignal} for quasi-neutrality regimes and \cite{BDD, DDSV} for strong magnetic fields regimes), low Mach number fluid dynamics (see \cite{Degond-Tang, McClarren-Lowrie}), or other types of transport problems (see \cite{Buet-Cordier-Lucquin-Mancini, Buet-Despres, Carrillo-Goudon-Lafitte, Crouseilles-Lemou, Filbet-Jin_stiff, Filbet-Jin_BGK, Klar, Lemou-Mieussens}) or diffusion problems (see \cite{Narski}). \\

\indent The present paper has two main goals: the first one is to propose a new equivalent formulation of the two-fluid Euler-Lorentz model when $\tau > 0$. In this new formulation, the parallel velocity equation for $\tau = 0$ explicitly appears as the limit of the parallel velocity equation for $\tau > 0$ by contrast to the original formulation. This reformulation is the building block for the Asymptotic-Preserving scheme for the two-fluid Euler-Lorentz model. For the scheme being simple, we choose an isothermal pressure law for both ion and electron fluids. We also assume that the fluids are coupled through a quasi-neutrality constraint which allows us to compute the self-consistent electric field within the Lorentz term, the magnetic field being external and given. This work is the last development of a program started in \cite{BDD, DDSV}: in \cite{BDD} and \cite{DDSV}, the one-fluid isentropic Euler-Lorentz model with given electric and magnetic fields has been investigated (in \cite{DDSV}, under a uniform magnetic field and in \cite{BDD}, with any magnetic field and arbitrary coordinate system). Here, the specificity of this work is to consider a two-fluid Euler-Lorentz model with self-consistent electric field, computed through the quasi-neutrality hypothesis. This leads to a system of two coupled anisotropic diffusion equations, which brings some specific difficulties. In particular, it involves some singularity which will be treated through a regularization procedure. This work also bears relations with \cite{Narski, Degond-Deluzet-Negulescu} which are concerned with more general anisotropic diffusion equations (but not in the context of the Euler-Lorentz model). Again, \cite{Narski} deals with a uniform anisotropy direction and \cite{Degond-Deluzet-Negulescu} with an arbitrary anisotropy direction and arbitrary coordinate systems compared to the direction of the anisotropy. \\

\indent The present paper is organized as follows. In section 2, we present the isothermal two-fluid Euler-Lorentz model and the drift-fluid limit regime. The section 3 is devoted to the reformulation of the $\tau$-dependent Euler-Lorentz model leading to a new equivalent formulation of these equations when $\tau > 0$ which is equivalent to the drift-fluid limit when $\tau = 0$. In section 4, we present a time semi-discretization of the Euler-Lorentz model which is also consistent with the new formulation of the model. Since this time semi-discrete scheme involves an ill-posed diffusion problem for the electric potential, we choose to recover the well-posedness of this problem by introducing a regularization of the quasi-neutrality constraint. This is the subject of the second part of section 4. In section 5, we present a fully-discrete finite volume scheme based on the AP scheme for the Euler-Lorentz model coupled with the perturbed quasi-neutrality constraint. Finally, in section 6, we present some numerical results which have been obtained with this scheme.

\section{The isothermal two-fluid Euler-Lorentz model.}
\setcounter{equation}{0}
\subsection{Scaling.}
In this paragraph, we present the scaling of Euler-Lorentz equations which leads to the dimensionless following model:

\begin{subnumcases}{\label{EL_rescaled}}
\displaystyle \D_{t}n^{\tau} + \nabla_{\mathbf{x}} \cdot \mathbf{q}_{\alpha}^{\tau} = 0 \, , & \label{EL_rescaled_n} \\
\begin{split}
\displaystyle \epsilon_{\alpha}\,\tau\,\Big[\D_{t}\mathbf{q}_{\alpha}^{\tau} + \nabla_{\mathbf{x}} \cdot \Big( \cfrac{\mathbf{q}_{\alpha}^{\tau} \otimes \mathbf{q}_{\alpha}^{\tau}}{n^{\tau}} \Big) \Big] + & T_{\alpha}\,\nabla_{\mathbf{x}}n^{\tau} \\
& = \mathfrak{q}_{\alpha}\,\big[-n^{\tau} \, \nabla_{\mathbf{x}}\phi^{\tau} + \mathbf{q}_{\alpha}^{\tau} \times \mathbf{B}\big] \, , 
\end{split}
& \label{EL_rescaled_q} \\
\alpha \in \{i,e\}\, ,
\end{subnumcases}
where $\tau$ is the ratio between the ion gyro-period and the characteristic time scale but also the square value of the ion Mach number, $T_{i} = 1$ and $T_{e}$ are the dimensionleass ion and electron temperatures, and $\epsilon_{\alpha}$ and $\mathfrak{q}_{\alpha}$ are defined by
\begin{equation}
\epsilon_{\alpha} = \left\{
\begin{array}{ll}
1 \, , & \textnormal{if $\alpha = i$,} \\
\epsilon \, , & \textnormal{if $\alpha = e$,}
\end{array}
\right. \qquad \mathfrak{q}_{\alpha} = \left\{
\begin{array}{ll}
1 \, , & \textnormal{if $\alpha = i$,} \\
-1 \, , & \textnormal{if $\alpha = e$,}
\end{array}
\right.
\end{equation}
where $\epsilon$ is the ratio between the unit electron mass and the unit ion mass. Finally, $n^{\tau}$, $\mathbf{q}_{i}^{\tau}$, $\mathbf{q}_{e}^{\tau}$, $\phi^{\tau}$ and $\mathbf{B}$ correspond to the dimensionless ion and electron density, the dimensionless ion momentum, the dimensionless electron momentum, the dimensionless electric potential and the external magnetic field and are functions of the position $\mathbf{x} \in \R^{3}$ and of the time $t \geq 0$. \\


\indent Our starting point is the two-fluid Euler-Lorentz model describing a mixture of an ion gas and an electron gas. This model writes
\begin{equation} \label{full_EL_origin}
\left\{
\begin{array}{l}
\displaystyle \D_{t}n_{\alpha} + \nabla_{\mathbf{x}} \cdot \mathbf{q}_{\alpha} = 0 \, , \\
\displaystyle \D_{t}\mathbf{q}_{\alpha} + \nabla_{\mathbf{x}} \cdot \Big( \cfrac{\mathbf{q}_{\alpha} \otimes \mathbf{q}_{\alpha}}{n_{\alpha}} \Big) + \cfrac{1}{m_{\alpha}} \, \nabla_{\mathbf{x}}p_{\alpha} = \mathfrak{q}_{\alpha}\,\cfrac{e}{m_{\alpha}} \, ( n_{\alpha} \, \mathbf{E} + \mathbf{q}_{\alpha} \times \mathbf{B}) \, , \\
\displaystyle \D_{t}e_{\alpha} + \nabla_{\mathbf{x}} \cdot \Big( \cfrac{e_{\alpha}+p_{\alpha}}{n_{\alpha}} \, \mathbf{q}_{\alpha}\Big) = \mathfrak{q}_{\alpha}\,e\,\mathbf{E}\cdot\mathbf{q}_{\alpha} \, , \\
\alpha \in \{i,e\}\, ,
\end{array}
\right.
\end{equation}
and the ion-electron coupling is insured by the following quasi-neutrality constraint
\begin{equation} \label{QN_origin}
n_{i} = n_{e} = n \, .
\end{equation}
In this two-fluid model, $n_{i}$, $\mathbf{q}_{i}$, $e_{i}$ and $p_{i}$ (resp. $n_{e}$, $\mathbf{q}_{e}$, $e_{e}$ and $p_{e}$) are respectively the density, the momentum, the total energy per mass unit and the pressure for the ion (resp. electron) gas. The physical constants $m_{i}$, $m_{e}$, and $e$ stand for the unit ion mass, the unit electron mass and the absolute value of unit electron charge. The electric and magnetic fields are denoted with $\mathbf{E}$ and $\mathbf{B}$ and we assume that $\mathbf{B}$ is given whereas $\mathbf{E}$ is generated by the ions and the electrons through the constraint (\ref{QN_origin}). \\

\indent In the present context, we assume that both ion and electron gases are isothermal, \textit{i.e.} we assume that $p_{i}$ and $p_{e}$ are given by
\begin{equation}
p_{i} = k_{B}\,T_{i}\,n_{i} \, , \qquad p_{e} = k_{B} \, T_{e}\, n_{e} \, ,
\end{equation}
with constant temperatures $T_{i}$ and $T_{e}$, and we also assume that the electric field $\mathbf{E}$ derives from a potential $\phi$, \textit{i.e.} $\mathbf{E} = -\nabla_{\mathbf{x}} \phi$. \\
\indent Consequently, the model (\ref{full_EL_origin})-(\ref{QN_origin}) is reduced to
\begin{equation} \label{Euler-bifluid_isothermal}
\left\{
\begin{array}{l}
\displaystyle \D_{t}n + \nabla_{\mathbf{x}} \cdot \mathbf{q}_{\alpha} = 0 \, , \\
\begin{split}
\displaystyle \D_{t}\mathbf{q}_{\alpha} + \nabla_{\mathbf{x}} \cdot \Big( \cfrac{\mathbf{q}_{\alpha} \otimes \mathbf{q}_{\alpha}}{n} \Big) + &\cfrac{k_{B}\,T_{\alpha}}{m_{\alpha}} \, \nabla_{\mathbf{x}}n \\
& = \mathfrak{q}_{\alpha}\,\cfrac{e}{m_{\alpha}} \, ( -n \, \nabla_{\mathbf{x}} \phi + \mathbf{q}_{\alpha} \times \mathbf{B}) \, ,
\end{split}
\\
\alpha \in \{i,e\}\, .
\end{array}
\right.
\end{equation}
We introduce characteristic length $\overline{x}$, time $\overline{t}$, momentum $\overline{q}$, ion temperature $\overline{T}$, electric potential $\overline{\phi}$, and magnetic field $\overline{B}$ such that
\begin{equation}
\mathbf{x} = \overline{x} \, \mathbf{x}' \, , \qquad t = \overline{t}\,t' \, , \qquad T_{i} = \overline{T} \, , \qquad  T_{e} = \overline{T} \, T_{e}' \, ,
\end{equation}

\begin{equation}
\begin{array}{rclrcl}
n(\overline{x}\,\mathbf{x}',\overline{t}\,t') &=& \overline{n}\,n'(\mathbf{x}',t') \, , & \phi(\overline{x}\,\mathbf{x}',\overline{t}\,t') &=& \overline{\phi}\,\phi'(\mathbf{x}',t') \, , \\ \\
\mathbf{q}_{\alpha}(\overline{x}\,\mathbf{x}',\overline{t}\,t') &=& \overline{q}\,\mathbf{q}_{\alpha}'(\mathbf{x}',t') \, , & \mathbf{B}(\overline{x}\,\mathbf{x}',\overline{t}\,t') &=& \overline{B}\,\mathbf{B}'(\mathbf{x}',t') \, .
\end{array}
\end{equation}

We consider the natural ratio 
\begin{equation}
\overline{q} = \cfrac{\overline{x}\,\overline{n}}{\overline{t}} \, ,
\end{equation}
and we also assume that the electric and magnetic forces are of the same order, which means in terms of characteristic scales that
\begin{equation}
\overline{q}\overline{B} = \cfrac{\overline{n}\,\overline{\phi}}{\overline{x}} \, .
\end{equation}
We define the characteristic sound speed $\overline{c}$ for the ions, the characteristic Mach number for the ions $\overline{M}$ and the characteristic cyclotron frequency $\overline{\omega}$ for the ions by
\begin{equation}
\overline{c} = \sqrt{\cfrac{k_{B}\overline{T}}{m_{i}}} \, , \qquad \overline{M}
= \cfrac{\overline{q}}{\overline{n}\,\overline{c}} \, , \qquad \overline{\omega} = \cfrac{e\,\overline{B}}{m_{i}} \, .
\end{equation}
Considering a low Mach number regime induces 
\begin{equation}
\overline{M} = \sqrt{\tau} \, ,
\end{equation}
with $\tau \geq 0$ small and assuming that the applied magnetic field is strong allows us to take
\begin{equation}
\overline{t}\,\overline{\omega} = \cfrac{1}{\tau} \, .
\end{equation}
Finally, we denote the ratio $m_{e}/m_{i}$ with $\epsilon$ and we assume that it is a dimensionless fixed constant. Then, removing the primed notations and adding $\tau$ in exponent, we finally obtain the rescaled isothermal two-fluid Euler-Lorentz model writing (\ref{EL_rescaled}).

\subsection{The limit model}
If $\tau$ converges to 0 in (\ref{EL_rescaled}), we formally get the model
\begin{subnumcases}{\label{EL_limit}}
\displaystyle \D_{t}n^{0} + \nabla_{\mathbf{x}} \cdot \mathbf{q}_{\alpha}^{0} = 0 \, , & \label{EL_limit_n} \\
T_{\alpha}\,\nabla_{\mathbf{x}}n^{0} = \mathfrak{q}_{\alpha} \, \big[- n^{0} \, \nabla_{\mathbf{x}}\phi^{0} + \mathbf{q}_{\alpha}^{0} \times \mathbf{B}\big] \, , & \label{EL_limit_q} \\
\alpha \in \{i,e\} \, ,
\end{subnumcases}
in which the parallel part of $\mathbf{q}_{i}^{0}$ and $\mathbf{q}_{e}^{0}$ are implicit. Indeed, if we separate the parallel and perpendicular parts of (\ref{EL_limit_q}) for any $\alpha$, we get
\begin{subnumcases}{\label{EL_limit_decompo}}
\displaystyle \D_{t}n^{0} + \nabla_{\mathbf{x}} \cdot \mathbf{q}_{\alpha}^{0} = 0 \, , & \label{EL_limit_decompo_n} \\
T_{\alpha}\,\mathbf{b} \cdot \nabla_{\mathbf{x}}n^{0} = - \mathfrak{q}_{\alpha}\, n^{0} \, \mathbf{b} \cdot \nabla_{\mathbf{x}}\phi^{0} \, , & \label{EL_limit_decompo_zero_force} \\
(\mathbf{q}_{\alpha}^{0})_{\perp} = \cfrac{1}{\|\mathbf{B}\|} \, \mathbf{b} \times (\mathfrak{q}_{\alpha}\, T_{\alpha}\,\nabla_{\mathbf{x}} n^{0} + n^{0}\,\nabla_{\mathbf{x}}\phi^{0}) \, , & \label{EL_limit_decompo_qperp} \\
\alpha \in \{i,e\} \, , &
\end{subnumcases}
where $\mathbf{q}_{\perp} = \mathbf{b} \times (\mathbf{q} \times \mathbf{b})$, $\mathbf{b} = \cfrac{B}{\|\mathbf{B}\|}$ and $\|\mathbf{B}\|^{2} = B_{x}^{2} + B_{y}^{2} + B_{z}^{2}$. We observe in this reformulated limit model that $(\mathbf{q}_{i}^{0})_{\perp}$ and $(\mathbf{q}_{e}^{0})_{\perp}$ can be algebraically computed from $n^{0}$ and $\phi^{0}$, but we do not get any explicit constraint for $(\mathbf{q}_{i}^{0})_{||}$ and $(\mathbf{q}_{e}^{0})_{||}$. One way to answer to this difficulty is to couple the limit model (\ref{EL_limit_decompo}) with the following equations:
\begin{equation} \label{q_para_0}
\left\{
\begin{array}{l}
\begin{split}
\D_{t}\big((\mathbf{q}_{\alpha}^{0})_{||}\big) &- \big(\D_{t}(\mathbf{b}\otimes\mathbf{b})\big) \, \mathbf{q}_{\alpha}^{0} + (\mathbf{b}\otimes\mathbf{b}) \, \nabla_{\mathbf{x}} \cdot \Big( \cfrac{\mathbf{q}_{\alpha}^{0} \otimes \mathbf{q}_{\alpha}^{0}}{n^{0}} \Big) \\
&+ \lim_{\tau\,\to\,0} \Big[\cfrac{1}{\epsilon_{\alpha}\tau}\, (\mathbf{b}\otimes\mathbf{b})\, \big(T_{\alpha}\,\nabla_{\mathbf{x}}n^{\tau} + \mathfrak{q}_{\alpha}\,n^{\tau} \, \nabla_{\mathbf{x}}\phi^{\tau} \big) \Big] = 0 \, ,
\end{split}
\\
\alpha \in \{i,e\}\, ,
\end{array}
\right.
\end{equation}
These equations are not more than the parallel part of (\ref{EL_rescaled_q}) for any $\alpha \in \{i,e\}$, with $\tau \to 0$. However, such a coupling is permitted if we are insured that
\begin{equation} \label{eq_phin_tau}
\mathbf{b} \cdot \big(T_{\alpha}\,\nabla_{\mathbf{x}}n^{\tau} + \mathfrak{q}_{\alpha}\,n^{\tau} \, \nabla_{\mathbf{x}}\phi^{\tau} \big) = \mathcal{O}(\epsilon_{\alpha}\tau) \, , \qquad \forall\,\alpha \in\{i,e\} \, .
\end{equation}
These hypotheses are coherent with the parallel part of (\ref{EL_rescaled_q}) if we are insured that
\begin{equation}
\D_{t}\mathbf{q}_{\alpha}^{\tau} + \nabla_{\mathbf{x}} \cdot \big( \cfrac{\mathbf{q}_{\alpha}^{\tau} \otimes \mathbf{q}_{\alpha}^{\tau}}{n^{\tau}} \big) = \mathcal{O}(1) \, , \qquad \forall\,\alpha \in\{i,e\} \, .
\end{equation}
Furthermore, if we consider the limit $\tau \to 0$ of (\ref{eq_phin_tau}), we obtain
\begin{equation}
\mathbf{b} \cdot \big(T_{\alpha}\,\nabla_{\mathbf{x}}n^{0} + \mathfrak{q}_{\alpha}\, n^{0} \, \nabla_{\mathbf{x}}\phi^{0} \big) = 0 \, , \qquad \forall\,\alpha \in\{i,e\} \, ,
\end{equation}
which are exactly the equations (\ref{EL_limit_decompo_zero_force}).

\section{Reformulation of the $\tau$-dependent model and of the limit model.} \label{Reformulation}
\setcounter{equation}{0}

In order to validate the coupling between (\ref{EL_limit_decompo}) and (\ref{q_para_0}), we have to insure that
\begin{equation} \label{AP_property}
\mathbf{b} \cdot \big(T_{\alpha}\,\nabla_{\mathbf{x}}n^{\tau} + \mathfrak{q}_{\alpha}\,n^{\tau} \, \nabla_{\mathbf{x}}\phi^{\tau} \big) = \mathcal{O}(\epsilon_{\alpha}\tau) \, , \qquad \forall\,\alpha \in\{i,e\} \, .
\end{equation}
These properties are validated since the Euler-Lorentz equations (\ref{EL_rescaled}) are equivalent to
\begin{subnumcases}{\label{EL_reformulated_final}}
\begin{split}
&\D_{t}^{2}n^{\tau} - \cfrac{1}{\epsilon_{\alpha}\tau}\, \nabla_{\mathbf{x}} \cdot \big( (\mathbf{b}\otimes\mathbf{b})\, \big(T_{\alpha}\,\nabla_{\mathbf{x}}n^{\tau} + \mathfrak{q}_{\alpha}\,n^{\tau} \, \nabla_{\mathbf{x}}\phi^{\tau} \big) \big) \\
&\quad = \nabla_{\mathbf{x}}\cdot \Big(\big(\D_{t}(\mathbf{b}\otimes\mathbf{b})\big) \, \mathbf{q}_{\alpha}^{\tau} - (\mathbf{b}\otimes\mathbf{b}) \, \nabla_{\mathbf{x}} \cdot \Big( \cfrac{\mathbf{q}_{\alpha}^{\tau} \otimes \mathbf{q}_{\alpha}^{\tau}}{n^{\tau}} \Big) -\D_{t}((\mathbf{q}_{\alpha}^{\tau})_{\perp}) \Big) \, ,
\end{split}
& \label{EL_reformulated_final_zero_force} \\
\begin{split}
\D_{t}\big((\mathbf{q}_{\alpha}^{\tau})_{||}\big) - \big(\D_{t}(&\mathbf{b}\otimes\mathbf{b})\big) \, \mathbf{q}_{\alpha}^{\tau} + (\mathbf{b}\otimes\mathbf{b}) \, \nabla_{\mathbf{x}} \cdot \Big( \cfrac{\mathbf{q}_{\alpha}^{\tau} \otimes \mathbf{q}_{\alpha}^{\tau}}{n^{\tau}} \Big) \\
&+ \cfrac{1}{\epsilon_{\alpha}\tau}\, (\mathbf{b}\otimes\mathbf{b})\, \big(T_{\alpha}\,\nabla_{\mathbf{x}}n^{\tau} + \mathfrak{q}_{\alpha}\,n^{\tau} \, \nabla_{\mathbf{x}}\phi^{\tau} \big) = 0 \, ,
\end{split}
& \label{EL_reformulated_final_qpara} \\
\begin{split}
(\mathbf{q}_{\alpha}^{\tau})_{\perp} = \cfrac{1}{\|\mathbf{B}\|} \, \mathbf{b} \times \big(\mathfrak{q}_{\alpha}\,T_{\alpha}\,& \nabla_{\mathbf{x}}n^{\tau} + n^{\tau}\,\nabla_{\mathbf{x}}\phi^{\tau}\big) \\
&+ \cfrac{\mathfrak{q}_{\alpha}\,\epsilon_{\alpha}\,\tau}{\|\mathbf{B}\|} \,\mathbf{b} \times \Big[\D_{t}\mathbf{q}_{\alpha}^{\tau} + \nabla_{\mathbf{x}} \cdot \Big(\cfrac{\mathbf{q}_{\alpha}^{\tau} \otimes \mathbf{q}_{\alpha}^{\tau}}{n^{\tau}} \Big) \Big] \, ,
\end{split}
& \label{EL_reformulated_final_qiperp} \\
\alpha \in \{i,e\}\, ,
\end{subnumcases}
and more precisely thanks to the diffusion equations (\ref{EL_reformulated_final_zero_force}) for both $\alpha = i$ and $\alpha = e$. These equations are obtained for (\ref{EL_rescaled}) by a differentiation in time and position procedure and projections in the direction of $\mathbf{b}$ and perpendicularly to $\mathbf{b}$. More details about these computations can be found in Appendix \ref{reformulation_continuous}. \\
\indent When $\tau \to 0$, we obtain
\begin{equation} \label{EL_limit_reformulated_final}
\left\{
\begin{array}{l}
\displaystyle \nabla_{\mathbf{x}} \cdot \big( (\mathbf{b}\otimes\mathbf{b})\, \big(T_{\alpha}\,\nabla_{\mathbf{x}}n^{0} + \mathfrak{q}_{\alpha}\,n^{0} \, \nabla_{\mathbf{x}}\phi^{0} \big) \big) = 0 \, , \\
\displaystyle \D_{t}\big((\mathbf{q}_{\alpha}^{0})_{||}\big) - \big(\D_{t}(\mathbf{b}\otimes\mathbf{b})\big) \, \mathbf{q}_{\alpha}^{0} + (\mathbf{b}\otimes\mathbf{b}) \, \nabla_{\mathbf{x}} \cdot \Big( \cfrac{\mathbf{q}_{\alpha}^{0} \otimes \mathbf{q}_{\alpha}^{0}}{n^{0}} \Big) \\
\displaystyle \qquad \qquad \quad + \lim_{\tau\, \to \,0} \Big[\cfrac{1}{\epsilon_{\alpha}\,\tau}\, (\mathbf{b}\otimes\mathbf{b})\, \big(T_{\alpha}\,\nabla_{\mathbf{x}}n^{\tau} + \mathfrak{q}_{\alpha}\,n^{\tau} \, \nabla_{\mathbf{x}}\phi^{\tau} \big) \Big] = 0 \, , \\ \\
\displaystyle (\mathbf{q}_{\alpha}^{0})_{\perp} = \cfrac{1}{\|\mathbf{B}\|} \, \mathbf{b} \times \big(\mathfrak{q}_{\alpha}\,T_{\alpha}\,\nabla_{\mathbf{x}}n^{0} + n^{0}\,\nabla_{\mathbf{x}}\phi^{0}\big) \, , \\
\alpha \in \{i,e\} \, , 
\end{array}
\right.
\end{equation}
which is equivalent to (\ref{EL_limit_decompo})-(\ref{q_para_0}).

\section{Semi-discrete AP schemes.}

\setcounter{equation}{0}

In this section, we propose a semi-discretization of (\ref{EL_rescaled}) in time which is also consistent with the reformulated model (\ref{EL_reformulated_final}): proceeding in such a way insures us that the approximation which will be computed ought to this numerical method will be also consistent with (\ref{EL_limit_reformulated_final}) and, equivalently, with (\ref{EL_limit}) when $\tau$ converges to 0. \\
\indent More precisely, the semi-discretization we describe in the the next lines is based on semi-implicit mass fluxes and fully implicit pressure and Lorentz forces. This strategy is motivated by the fact that we want to preserve the balance between the pressure gradient and the Lorentz term, \textit{i.e.} to insure that, at every time step $t^{m}$,
\begin{equation} \label{AP_property_SD}
T_{\alpha}\,\nabla_{\mathbf{x}}n^{\tau,m} + \mathfrak{q}_{\alpha} \, \big[ n^{\tau,m} \, \nabla_{\mathbf{x}}\phi^{\tau,m} - \mathbf{q}_{\alpha}^{\tau,m} \times \mathbf{B}^{m} \big] = \mathcal{O}(\epsilon_{\alpha}\,\tau) \, , \quad \forall\,\alpha\in\{i,e\} \, , 
\end{equation}
where the notation $\theta^{m}$ stands for an approximation of the function $\theta=\theta(\mathbf{x},t)$ at the time step $t = t^{m}$. This methodology differs from the AP semi-discretizations which were described in \cite{BDD} and \cite{DDSV}: indeed, in these papers, the authors considered a fully explicit mass flux, a fully implicit Lorentz term and a semi-implicit pressure gradient, and this leads to a non-conservative discretization of the velocity equation. \\

\indent Firstly, we describe the semi-discretization and we reformulate the semi-discrete model which is obtained by following the same approach as in Section \ref{Reformulation}. Then we discuss the difficulties which are brought by this reformulation and we introduce a regularization of the mass conservation equations (\ref{EL_rescaled_n}) which allows us to bypass these difficulties.

\subsection{Time semi-discretization.} \label{usdicr}

\subsubsection{Asymptotic-Preserving property}

\indent The considered time semi-discretization is the following:
\begin{subnumcases}{\label{semi-discrete_origin}}
\begin{split}
\cfrac{n^{\tau,m+1}-n^{\tau,m}}{\Delta t} + \nabla_{\mathbf{x}} &\cdot \big( (\mathbf{b}^{m+1} \otimes \mathbf{b}^{m+1})\, \mathbf{q}_{\alpha}^{\tau,m+1}\big) \\
&+ \nabla_{\mathbf{x}} \cdot \big((\mathbb{I}-\mathbf{b}^{m+1}\otimes\mathbf{b}^{m+1})\,\mathbf{q}_{\alpha}^{\tau,m}\big) = 0 \, ,
\end{split}
& \label{semi-discrete_origin_n} \\
\begin{split}
\cfrac{\mathbf{q}_{\alpha}^{\tau,m+1}-\mathbf{q}_{\alpha}^{\tau,m}}{\Delta t} &+ \nabla_{\mathbf{x}} \cdot \Big(\cfrac{\mathbf{q}_{\alpha}^{\tau,m} \otimes \mathbf{q}_{\alpha}^{\tau,m}}{n^{\tau,m}} \Big) + \cfrac{T_{\alpha}}{\epsilon_{\alpha}\,\tau}\, \nabla_{\mathbf{x}}n^{\tau,m+1} \\
&= \cfrac{\mathfrak{q}_{\alpha}}{\epsilon_{\alpha}\,\tau}\, \big[ - n^{\tau,m+1}\,\nabla_{\mathbf{x}}\phi^{\tau,m+1} + \mathbf{q}_{\alpha}^{\tau,m+1} \times \mathbf{B}^{m+1} \big] \, ,
\end{split}
& \label{semi-discrete_origin_q} \\
\alpha \in \{i,e\} \, .
\end{subnumcases}
As it has been announced above, we chose to implicit the whole pressure and Lorentz terms in order to have (\ref{AP_property_SD}) at every time step. The choice of the implicitation of the parallel part of the mass fluxes is motivated by the fact that we have to reformulate the model (\ref{semi-discrete_origin}) by injecting the parallel part of (\ref{semi-discrete_origin_q}) with $\alpha = i$ (resp. $\alpha = e$) in (\ref{semi-discrete_origin_n}) with $\alpha = i$ (resp. $\alpha = e$). Such a procedure leads to the separate computation of $(\mathbf{q}_{i}^{\tau,m+1})_{||}^{m+1}$, $(\mathbf{q}_{i}^{\tau,m+1})_{\perp}^{m+1}$, $(\mathbf{q}_{e}^{\tau,m+1})_{||}^{m+1}$ and $(\mathbf{q}_{e}^{\tau,m+1})_{\perp}^{m+1}$ by applying projection operators in the $\mathbf{b}^{m+1}$ and orthogonal to $\mathbf{b}^{m+1}$ directions. This leads to
\begin{subnumcases}{\label{SD_qparaperp}}
\begin{split}
&(\mathbf{q}_{\alpha}^{\tau,m+1})_{\perp}^{m+1} - \cfrac{\mathfrak{q}_{\alpha}\,\epsilon_{\alpha}\,\tau}{\Delta t\, \|\mathbf{B}^{m+1}\|} \, \mathbf{b}^{m+1} \times (\mathbf{q}_{\alpha}^{\tau,m+1})_{\perp}^{m+1} \\
&\qquad \qquad = \cfrac{1}{\|\mathbf{B}^{m+1}\|} \, \mathbf{b}^{m+1} \times \big[ \mathfrak{q}_{\alpha}\,T_{\alpha}\,\nabla_{\mathbf{x}}n^{\tau,m+1} + n^{\tau,m+1}\,\nabla_{\mathbf{x}} \phi^{\tau,m+1}\big] \\
&\qquad \qquad \qquad + \cfrac{\mathfrak{q}_{\alpha}\,\epsilon_{\alpha}\,\tau}{\|\mathbf{B}^{m+1}\|} \, \mathbf{b}^{m+1} \times \Big[ -\cfrac{\mathbf{q}_{\alpha}^{\tau,m}}{\Delta t} + \nabla_{\mathbf{x}} \cdot \Big( \cfrac{\mathbf{q}_{\alpha}^{\tau,m} \otimes \mathbf{q}_{\alpha}^{\tau,m}}{n^{\tau,m}} \Big) \Big] \, , 
\end{split}
& \label{SD_qperp} \\
\begin{split}
&(\mathbf{q}_{\alpha}^{\tau,m+1})_{||}^{m+1} \\
&\quad = (\mathbf{b}^{m+1} \otimes \mathbf{b}^{m+1}) \, \mathbf{q}_{\alpha}^{\tau,m} - \Delta t \, (\mathbf{b}^{m+1} \otimes \mathbf{b}^{m+1}) \, \nabla_{\mathbf{x}} \cdot \Big( \cfrac{\mathbf{q}_{\alpha}^{\tau,m} \otimes \mathbf{q}_{\alpha}^{\tau,m}}{n^{\tau,m}} \Big) \\
&\qquad - \cfrac{\Delta t}{\epsilon_{\alpha}\,\tau} \, (\mathbf{b}^{m+1} \otimes \mathbf{b}^{m+1}) \, \big[ T_{\alpha}\,\nabla_{\mathbf{x}}n^{\tau,m+1} + \mathfrak{q}_{\alpha}\,n^{\tau,m+1}\,\nabla_{\mathbf{x}} \phi^{\tau,m+1} \big] \, ,
\end{split}
& \label{SD_qpara} \\
\alpha \in \{i,e\} \, ,
\end{subnumcases}
on one hand, and to a couple of anisotropic diffusion equations for $n^{\tau,m+1}$ and $\phi^{\tau,m+1}$ with an anisotropy carried by $\mathbf{b}^{m+1}$ (see the computations of Appendix \ref{reformulation_SD_appendix} with $C_{i} = C_{e} = 0$) on the other hand. These diffusion equations are of the form
\begin{equation} \label{eq_n}
-\nabla_{\mathbf{x}} \cdot \big((\mathbf{b}^{m+1} \otimes \mathbf{b}^{m+1}) \, \nabla_{\mathbf{x}}n^{\tau,m+1}\big) + \lambda \, \tau \, n^{\tau,m+1} = \tau\, R^{\tau,m+1} \, ,
\end{equation}
\begin{equation} \label{eq_phi}
-\nabla_{\mathbf{x}} \cdot \big(n^{\tau,m+1}\,(\mathbf{b}^{m+1} \otimes \mathbf{b}^{m+1}) \, \nabla_{\mathbf{x}}\phi^{\tau,m+1}\big) = \tau\, S^{\tau,m+1} \, ,
\end{equation}
where $\lambda$ only depends on $\epsilon$, $\Delta t$ and $T_{e}$, and where
\begin{equation}
R^{\tau,m+1} = R\big(\Delta t,T_{e},\epsilon,n^{\tau,m},\mathbf{q}_{i}^{\tau,m},\mathbf{q}_{e}^{\tau,m},\mathbf{b}^{m+1}\big) \, ,
\end{equation}
\begin{equation}
S^{\tau,m+1} = S\big(\Delta t,T_{e},\epsilon,n^{\tau,m+1},n^{\tau,m},\mathbf{q}_{i}^{\tau,m},\mathbf{q}_{e}^{\tau,m},\mathbf{b}^{m+1}\big) \, .
\end{equation}

Remark that the algebraic equations (\ref{SD_qperp}) can be solved for any value of $\tau$. Then the scheme (\ref{semi-discrete_origin}) is Asymptotic-Preserving if and only if
\begin{equation} \label{AP_property_para_SD}
\mathbf{b}^{m+1} \cdot (T_{\alpha}\,\nabla_{\mathbf{x}}n^{\tau,m+1}+\mathfrak{q}_{\alpha}\,n^{\tau,m+1}\,\nabla_{\mathbf{x}}\phi^{\tau,m+1}) = \mathcal{O}(\epsilon_{\alpha}\,\tau) \, , \quad \forall\,\alpha\in\{i,e\}\, .
\end{equation}
These hypotheses are validated if we take into account the following boundary conditions
\begin{subnumcases}{\label{BC_nphi}}
(\mathbf{b}^{m+1} \cdot \nabla_{\mathbf{x}}n^{\tau,m+1})\,(\mathbf{b}^{m+1} \cdot \mathbf{\nu}) = 0 \, , & \textnormal{on $\D\Omega$,} \label{BC_nphi_n} \\
(\mathbf{b}^{m+1} \cdot \nabla_{\mathbf{x}}\phi^{\tau,m+1})\,(\mathbf{b}^{m+1} \cdot \mathbf{\nu}) = 0 \, , & \textnormal{on $\D\Omega$,} \label{BC_nphi_phi}
\end{subnumcases}
alongwith the diffusion equations (\ref{eq_n}) and (\ref{eq_phi}). Indeed, the solutions of the problems (\ref{eq_n})-(\ref{BC_nphi_n}) and (\ref{eq_phi})-(\ref{BC_nphi_phi}) satisfy
\begin{equation}
\mathbf{b}^{m+1} \cdot \nabla_{\mathbf{x}}n^{\tau,m+1} = \mathcal{O}(\tau) \, , \qquad \mathbf{b}^{m+1} \cdot \nabla_{\mathbf{x}} \phi^{\tau,m+1} = \mathcal{O}(\tau) \, ,
\end{equation}
which is (\ref{AP_property_para_SD}) up to some linear combinations. \\

\subsubsection{Anisotropic diffusion problems}

Now we are insured that the semi-discrete scheme (\ref{semi-discrete_origin}) is Asymptotic-Preserving, we focus on the anisotropic diffusion problems which are satisfied by $n^{\tau,m+1}$ and $\phi^{\tau,m+1}$. \\

\indent We remark the diffusion equation (\ref{eq_n}) coupled with the Neumann boundary condition (\ref{BC_nphi_n}) is well posed for any $\tau > 0$ but becomes ill-posed when $\tau = 0$. To be more precise, the limit of (\ref{eq_n})-(\ref{BC_nphi_n}) writes
\begin{equation} \label{eq_diff_n_limit}
\left\{
\begin{array}{ll}
-\nabla_{\mathbf{x}} \cdot \big((\mathbf{b}^{m+1} \otimes \mathbf{b}^{m+1}) \, \nabla_{\mathbf{x}}\tilde{n}^{0,m+1}\big) = 0 \, , &\textnormal{on $\Omega$,} \\
(\mathbf{b}^{m+1} \cdot \nabla_{\mathbf{x}}\tilde{n}^{0,m+1})\,(\mathbf{b}^{m+1} \cdot \mathbf{\nu}) = 0 \, , & \textnormal{on $\D\Omega$,}
\end{array}
\right.
\end{equation}
and a solution $\tilde{n}^{0,m+1}$ of (\ref{eq_diff_n_limit}) is defined up to a function $c : \Omega \to \R$ such that $\mathbf{b}^{m+1} \cdot \nabla_{\mathbf{x}} c = 0$. Since we want to compute the particular solution $n^{0,m+1}$ of (\ref{eq_diff_n_limit}) which is exactly the limit of $(n^{\tau,m+1})_{\tau \, > \, 0}$ when $\tau \to 0$, we follow the same approach as in \cite{BDD} and we use the following theorem: \label{non-unique}

\begin{theorem}[Brull, Degond, Deluzet \cite{BDD}] \label{theo}
Let us consider the subspace $K \subset L^{2}(\Omega)$ defined by
\begin{equation}
K = \big\{ u \in L^{2}(\Omega) \, : \, \mathbf{b}^{m+1} \cdot \nabla_{\mathbf{x}} u = 0 \, \big\} \, ,
\end{equation}
and the functional space $\mathcal{W}_{0}$ defined by
\begin{equation}
\mathcal{W}_{0} = \big\{ u \in L^{2}(\Omega) \, : \, \nabla_{\mathbf{x}} \cdot (\mathbf{b}^{m+1}\,u) \in L^{2}(\Omega) \, , \, (\mathbf{b}^{m+1} \cdot \nu)\,u = 0 \,\, \textit{on $\D\Omega$}\,\big\} \, ,
\end{equation}
provided with the norm $\|u\|_{\mathcal{W}_{0}} = \big\|\nabla_{\mathbf{x}} \cdot (\mathbf{b}^{m+1}\,u)\big\|_{L^{2}(\Omega)}$. Then we have the following properties:
\begin{enumerate}
\item $K$ is a closed subset in $L^{2}(\Omega)$,
\item $\mathcal{W}_{0}$ is a Hilbert space and $\nabla_{\mathbf{x}} \cdot (\mathbf{b}^{m+1}\,\mathcal{W}_{0})$ is a closed subset of $L^{2}(\Omega)$,
\item $L^{2}(\Omega) = K \oplus K^{\perp}$ with $K^{\perp} = \nabla_{\mathbf{x}} \cdot (\mathbf{b}^{m+1}\,\mathcal{W}_{0})$.
\end{enumerate}
\end{theorem}

Having these results in hand and assuming that $n^{\tau,m+1}$ is in $L^{2}(\Omega)$, we write
\begin{equation}
n^{\tau,m+1} = \pi^{\tau,m+1} + q^{\tau,m+1} \, , 
\end{equation}
with $\pi^{\tau,m+1} \in K$ and $q^{\tau,m+1} \in K^{\perp}$. Since the solution $n^{\tau,m+1}$ of (\ref{eq_n})-(\ref{BC_nphi_n}) is unique when $\tau > 0$, the functions $\pi^{\tau,m+1}$ and $q^{\tau,m+1}$ are also unique as the projection of $n^{\tau,m+1}$ on $K$ and $K^{\perp}$ respectively. Then the diffusion problem (\ref{eq_n})-(\ref{BC_nphi_n}) writes
\begin{equation} \label{eq_n_decomposed}
\left\{
\begin{array}{ll}
\begin{split}
-\nabla_{\mathbf{x}} \cdot \big((\mathbf{b}^{m+1} \otimes &\mathbf{b}^{m+1})\,\nabla_{\mathbf{x}}q^{\tau,m+1}\big) \\
&+ \lambda\,\tau\,(\pi^{\tau,m+1}+q^{\tau,m+1}) = \tau\,R^{\tau,m+1} \, ,
\end{split}
& \textnormal{on $\Omega$,} \\
(\mathbf{b}^{m+1} \cdot \nabla_{\mathbf{x}}q^{\tau,m+1})\,(\mathbf{b}^{m+1} \cdot \nu) = 0 \, , & \textnormal{on $\D\Omega$.}
\end{array}
\right.
\end{equation}
If we consider the variational formulation of (\ref{eq_n_decomposed}) over $K$, we find that $(\lambda\,\pi^{\tau,m+1} - R^{\tau,m+1}) \in K^{\perp}$, \textit{i.e.} there exists $h^{\tau,m+1}$ such that
\begin{equation} \label{def_pi_n}
\left\{
\begin{array}{ll}
\lambda \, \pi^{\tau,m+1} - R^{\tau,m+1} = \nabla_{\mathbf{x}} \cdot (\mathbf{b}^{m+1}\,h^{\tau,m+1}) \, , & \textnormal{on $\Omega$,} \\
(\mathbf{b}^{m+1} \cdot \nu)\,h^{\tau,m+1} = 0 \, , & \textnormal{on $\D\Omega$.}
\end{array}
\right.
\end{equation}
By applying the operator $\mathbf{b}^{m+1} \cdot \nabla_{\mathbf{x}}$ on this equation, we find that $h^{\tau,m+1}$ is the unique solution of
\begin{equation} \label{eq_h_n}
\hspace{-0.2cm}\left\{
\begin{array}{ll}
-\mathbf{b}^{m+1} \cdot \nabla_{\mathbf{x}} \big(\nabla_{\mathbf{x}} \cdot (\mathbf{b}^{m+1} \, h^{\tau,m+1})\big) = \cfrac{1}{\lambda}\, \mathbf{b}^{m+1} \cdot \nabla_{\mathbf{x}} R^{\tau,m+1} \, , & \textnormal{on $\Omega$,} \\
(\mathbf{b}^{m+1} \cdot \mathbf{\nu})\,h^{\tau,m+1} = 0 \, , & \textnormal{on $\D\Omega$,}
\end{array}
\right.
\end{equation}
which is a well-posed problem for any $\tau \geq 0$. \\
\indent Since $q^{\tau,m+1} \in K^{\perp}$, we claim that there exists $l^{\tau,m+1} \in L^{2}(\Omega)$ such that
\begin{equation} \label{def_q_n}
\left\{
\begin{array}{ll}
q^{\tau,m+1} = \nabla_{\mathbf{x}} \cdot (\mathbf{b}^{m+1} \,l^{\tau,m+1}) \, , &\textnormal{on $\Omega$,} \\
(\mathbf{b}^{m+1} \cdot \nu)\,l^{\tau,m+1} = 0 \, , &\textnormal{on $\D\Omega$.}
\end{array}
\right.
\end{equation}
If we consider now the variational formulation of (\ref{eq_n_decomposed}) over $K^{\perp}$, we find that $l^{\tau,m+1}$ is the solution of a fourth-order problem which can be written as two successive second-order problems of the form
\begin{equation} \label{eq_l1_n}
\left\{
\begin{array}{ll}
\begin{split}
-\mathbf{b}^{m+1} \cdot \nabla_{\mathbf{x}} \big(\nabla_{\mathbf{x}} \cdot (\mathbf{b}^{m+1}\,&L^{\tau,m+1})\big) + \tau\,\lambda\,L^{\tau,m+1} \\
&\qquad = - \tau\,\mathbf{b}^{m+1} \cdot \nabla_{\mathbf{x}} R^{\tau,m+1} \, ,
\end{split}
& \textnormal{on $\Omega$,} \\
(\mathbf{b}^{m+1} \cdot \mathbf{\nu})\, L^{\tau,m+1} = 0 \, , & \textnormal{on $\D\Omega$,} \\
\end{array}
\right.
\end{equation}
\begin{equation} \label{eq_l2_n}
\left\{
\begin{array}{ll}
-\mathbf{b}^{m+1} \cdot \nabla_{\mathbf{x}} \big(\nabla_{\mathbf{x}} \cdot (\mathbf{b}^{m+1}\,l^{\tau,m+1})\big) = L^{\tau,m+1} \, , & \textnormal{on $\Omega$,} \\
(\mathbf{b}^{m+1} \cdot \mathbf{\nu})\, l^{\tau,m+1} = 0 \, , & \textnormal{on $\D\Omega$.} \\
\end{array}
\right.
\end{equation}
Remark that these problems remain well-posed for any value of $\tau \geq 0$. Then, instead of solving the problem (\ref{eq_n})-(\ref{BC_nphi_n}) which becomes ill-posed when $\tau = 0$, we solve the problems (\ref{eq_h_n}), (\ref{eq_l1_n}) and (\ref{eq_l2_n}) for computing $h^{\tau,m+1}$ and $l^{\tau,m+1}$, then we compute $\pi^{\tau,m+1}$ and $q^{\tau,m+1}$ by using (\ref{def_pi_n}) and (\ref{def_q_n}) respectively, and we finally get $n^{\tau,m+1}$ as the sum of $\pi^{\tau,m+1}$ and $q^{\tau,m+1}$. \\

\indent Concerning the problem (\ref{eq_phi})-(\ref{BC_nphi_phi}) for the electric potential $\phi^{\tau,m+1}$, we remark that it remains ill-posed for any value of $\tau \geq 0$. Indeed, assuming that this diffusion problem admits at least one solution $\tilde{\phi}^{\tau,m+1}$, we can prove that this solution is not unique: for this purpose, we consider a function $c : \overline{\Omega} \to \R$ satisfying
\begin{equation}
\mathbf{b}^{m+1} \cdot \nabla_{\mathbf{x}} c = 0 \, , \qquad \textnormal{on $\overline{\Omega}$.}
\end{equation}
Then, it is straightforward that $\tilde{\phi}^{\tau,m+1}+c$ is also a solution of the problem (\ref{eq_phi})-(\ref{BC_nphi_phi}). Since this proof works for the problem (\ref{eq_phi})-(\ref{BC_nphi_phi}) and for any value of $\tau \geq 0$, the decomposition of its solution does not provide unique projections on $K$ and $K^{\perp}$ just as it is done for $n^{\tau,m+1}$. Then, we have to find a way to restore the uniqueness of the solution of the diffusion problem for the electric potential. This is what we do in the next paragraph.

\subsection{Regularized two-fluid Euler-Lorentz model.}

As it is explained in the previous paragraph, a classical Asymptotic-Preserving scheme based on the model (\ref{EL_rescaled}), \textit{i.e.} based on making implicit the parallel mass fluxes and Lorentz and pressure terms, leads to a non-unique solution problem for computing the electric potential at time step $t^{m+1}$. In order to bypass the difficulty and to restore the well-posedness of the problem in $\phi^{\tau,m+1}$, we choose to include a small regularization in the mass conservation equations (\ref{EL_rescaled_n}). That is why we introduce the terms 
$C_{i}\,\D_{t}\phi$ and $C_{e}\,\D_{t}\phi$ in such a way that this new model writes
\begin{equation} \label{ELPP_rescaled}
\left\{
\begin{array}{l}
\displaystyle \D_{t}n^{\tau} + C_{\alpha}\,\D_{t}\phi^{\tau} + \nabla_{\mathbf{x}} \cdot \mathbf{q}_{\alpha}^{\tau} = 0 \, , \\
\begin{split}
\epsilon_{\alpha}\,\tau\,\Big[\D_{t}\mathbf{q}_{\alpha}^{\tau} + \nabla_{\mathbf{x}} \cdot \Big( \cfrac{\mathbf{q}_{\alpha}^{\tau} \otimes \mathbf{q}_{\alpha}^{\tau}}{n^{\tau}} \Big) \Big] &+ T_{\alpha}\,\nabla_{\mathbf{x}}n^{\tau} \\
& = \mathfrak{q}_{\alpha} \, \big[- n^{\tau} \, \nabla_{\mathbf{x}}\phi^{\tau} + \mathbf{q}_{\alpha}^{\tau} \times \mathbf{B} \big] \, ,
\end{split}
\\
\alpha \in \{i,e\} \, .
\end{array}
\right.
\end{equation}
Here $C_{i},C_{e} > 0$ are two fixed small parameters which will be chosen later. Then we consider the same semi-discretization method as previously, \textit{i.e.}
\begin{subnumcases}{\label{semi-discrete_ELPP}}
\begin{split}
\cfrac{n^{\tau,m+1}-n^{\tau,m}}{\Delta t} + C_{\alpha}\, &\cfrac{\phi^{\tau,m+1}-\phi^{\tau,m}}{\Delta t} \\
& + \nabla_{\mathbf{x}} \cdot \big( (\mathbf{b}^{m+1} \otimes \mathbf{b}^{m+1}) \, \mathbf{q}_{\alpha}^{\tau,m+1} \big) \\
&+ \nabla_{\mathbf{x}} \cdot \big((\mathbb{I}-\mathbf{b}^{m+1} \otimes \mathbf{b}^{m+1})\,\mathbf{q}_{\alpha}^{\tau,m}\big) = 0 \, ,
\end{split}
& \label{semi-discrete_ELPP_n} \\
\begin{split}
\cfrac{\mathbf{q}_{\alpha}^{\tau,m+1}-\mathbf{q}_{\alpha}^{\tau,m}}{\Delta t} &+ \nabla_{\mathbf{x}} \cdot \Big(\cfrac{\mathbf{q}_{\alpha}^{\tau,m} \otimes \mathbf{q}_{\alpha}^{\tau,m}}{n^{\tau,m}} \Big) + \cfrac{T_{\alpha}}{\epsilon_{\alpha}\,\tau}\, \nabla_{\mathbf{x}}n^{\tau,m+1} \\
&= \cfrac{\mathfrak{q}_{\alpha}}{\epsilon_{\alpha}\,\tau}\, \big[ - n^{\tau,m+1}\,\nabla_{\mathbf{x}}\phi^{\tau,m+1} + \mathbf{q}_{\alpha}^{\tau,m+1} \times \mathbf{B}^{m+1} \big] \, ,
\end{split}
& \label{semi-discrete_ELPP_q} \\
\alpha \in \{i,e\} \, .
\end{subnumcases}
By splitting parallel and perpendicular parts of (\ref{semi-discrete_ELPP_q}) for $\alpha = i$ and $\alpha = e$, we get that $(\mathbf{q}_{\alpha}^{\tau,m+1})_{\perp}^{m+1}$ and $(\mathbf{q}_{\alpha}^{\tau,m+1})_{||}^{m+1}$ satisfy (\ref{SD_qparaperp}). Following the same procedure as in the previous paragraph (see Appendix \ref{reformulation_SD_appendix}), we inject (\ref{SD_qpara}) with $\alpha = i$ (resp. $\alpha = e$) in (\ref{semi-discrete_ELPP_n}) with $\alpha = i$ (resp. $\alpha = e$). Under the hypotheses
\begin{equation} \label{constraints_CiCe}
C_{i} + \epsilon\,C_{e} = 0 \qquad \textnormal{and} \qquad C_{i}-\cfrac{\epsilon}{T_{e}}\,C_{e} = C \, ,
\end{equation}
with $C > 0$ being given, we find that $n^{\tau,m+1}$ and $\phi^{\tau,m+1}$ satisfy the following diffusion equations:
\begin{equation} \label{ELPP_diffusion_n}
-\nabla_{\mathbf{x}} \cdot \big((\mathbf{b}^{m+1} \otimes \mathbf{b}^{m+1}) \, \nabla_{\mathbf{x}}n^{\tau,m+1}\big) + \tau \, \lambda_{1} \, n^{\tau,m+1} = \tau \, R^{\tau,m+1} \, ,
\end{equation}
\begin{equation} \label{ELPP_diffusion_phi}
-\nabla_{\mathbf{x}} \cdot \big(n^{\tau,m+1}\,(\mathbf{b}^{m+1} \otimes \mathbf{b}^{m+1}) \, \nabla_{\mathbf{x}}\phi^{\tau,m+1}\big) + \tau \, \lambda_{2} \, \phi^{\tau,m+1} = \tau \, S^{\tau,m+1} \, ,
\end{equation}
where $\lambda_{1}$, $\lambda_{2}$ only depend on $\epsilon$, $\Delta t$, $T_{e}$ and $C$, and where
\begin{equation}
R^{\tau,m+1} = R\big(\Delta t,T_{e},\epsilon,n^{\tau,m},\mathbf{q}_{i}^{\tau,m},\mathbf{q}_{e}^{\tau,m},\mathbf{b}^{m+1}\big) \, ,
\end{equation}
\begin{equation}
S^{\tau,m+1} = S\big(\Delta t,T_{e},\epsilon,C,n^{\tau,m+1},n^{\tau,m},\phi^{\tau,m},\mathbf{q}_{i}^{\tau,m},\mathbf{q}_{e}^{\tau,m},\mathbf{b}^{m+1}\big) \, .
\end{equation}

Remark that the constraints (\ref{constraints_CiCe}) are equivalent to
\begin{equation} \label{property_C1C2}
C_{i} = \cfrac{T_{e}\,C}{1+T_{e}} \, , \qquad C_{e} = -\cfrac{T_{e}\,C}{\epsilon\,(1+T_{e})} \, .
\end{equation}
As a consequence, it is necessary to take $C > 0$ small enough to insure that $C_{i}$ and $C_{e}$ are close to 0. For that, we can take $C = \mathcal{O}(\epsilon)$ provided that the ratio $\epsilon = m_{e}/m_{i}$ is small. \\

\indent We couple (\ref{ELPP_diffusion_n}) with the boundary condition given in (\ref{BC_nphi_n}). This diffusion problem is well-posed for any $\tau > 0$ and becomes ill-posed if $\tau = 0$ because of a lack of uniqueness of the solution (see page \pageref{non-unique}). As a consequence, we can apply Theorem \ref{theo} and write $n^{\tau,m+1}$ under the following form:
\begin{equation} \label{decompo_diffusion_ELPP_1}
n^{\tau,m+1} = \pi^{\tau,m+1} + q^{\tau,m+1} \, ,
\end{equation}
with $\pi^{\tau,m+1} \in K$, $q^{\tau,m+1} \in K^{\perp}$ defined by
\begin{equation}
\begin{split}
\pi^{\tau,m+1} &= \cfrac{1}{\lambda_{1}} \, \big[ R^{\tau,m+1} + \lambda_{1}\,\nabla_{\mathbf{x}} \cdot (\mathbf{b}^{m+1}\,h^{\tau,m+1})\big] \, , \\
q^{\tau,m+1} &= \nabla_{\mathbf{x}} \cdot (\mathbf{b}^{m+1}\,l^{\tau,m+1}) \, ,
\end{split}
\end{equation}
where $h^{\tau,m+1}$ and $l^{\tau,m+1}$ are the solutions of
\begin{equation}
\hspace{-0.4cm}\left\{
\begin{array}{ll}
-\mathbf{b}^{m+1} \cdot \nabla_{\mathbf{x}} \big(\nabla_{\mathbf{x}} \cdot (\mathbf{b}^{m+1} \, h^{\tau,m+1})\big) = \cfrac{1}{\lambda_{1}}\, \mathbf{b}^{m+1} \cdot \nabla_{\mathbf{x}} R^{\tau,m+1} \, , & \textnormal{on $\Omega$,} \\
(\mathbf{b}^{m+1} \cdot \mathbf{\nu})\, h^{\tau,m+1} = 0 \, , & \textnormal{on $\D\Omega$,}
\end{array}
\right.
\end{equation}
and
\begin{equation}
\left\{
\begin{array}{ll}
\begin{split}
-\mathbf{b}^{m+1} \cdot \nabla_{\mathbf{x}} \big(\nabla_{\mathbf{x}} \cdot (\mathbf{b}^{m+1}\,&L^{\tau,m+1})\big) + \tau\,\lambda_{1}\,L^{\tau,m+1} \\
&\qquad = - \tau\,\mathbf{b}^{m+1} \cdot \nabla_{\mathbf{x}} R^{\tau,m+1} \, ,
\end{split}
& \textnormal{on $\Omega$,} \\
(\mathbf{b}^{m+1} \cdot \mathbf{\nu})\, L^{\tau,m+1} = 0 \, , & \textnormal{on $\D\Omega$,} \\
\end{array}
\right.
\end{equation}
\begin{equation}
\left\{
\begin{array}{ll}
-\mathbf{b}^{m+1} \cdot \nabla_{\mathbf{x}} \big(\nabla_{\mathbf{x}} \cdot (\mathbf{b}^{m+1}\,l^{\tau,m+1})\big) = L^{\tau,m+1} \, , & \textnormal{on $\Omega$,} \\
(\mathbf{b}^{m+1} \cdot \mathbf{\nu})\, l^{\tau,m+1} = 0 \, , & \textnormal{on $\D\Omega$.} \\
\end{array}
\right.
\end{equation}

\indent Now, we couple (\ref{ELPP_diffusion_phi}) with the boundary condition given in (\ref{BC_nphi_phi}). This diffusion problem has the same properties of (\ref{ELPP_diffusion_n})-(\ref{BC_nphi_n}) which has been discussed above. Then, we can write $\phi^{\tau,m+1}$ under the following form:
\begin{equation}
\phi^{\tau,m+1} = \tilde{\pi}^{\tau,m+1} + \tilde{q}^{\tau,m+1} \, ,
\end{equation}
with $\tilde{\pi}^{\tau,m+1} \in K$ and $\tilde{q}^{\tau,m+1}$ defined by
\begin{equation}
\begin{split}
\tilde{\pi}^{\tau,m+1} &= \cfrac{1}{\lambda_{2}} \, \big[ S^{\tau,m+1} + \lambda_{2}\,\nabla_{\mathbf{x}} \cdot (\mathbf{b}^{m+1}\,\tilde{h}^{\tau,m+1})\big] \, , \\
\tilde{q}^{\tau,m+1} &= \nabla_{\mathbf{x}} \cdot (\mathbf{b}^{m+1}\,\tilde{l}^{\tau,m+1}) \, ,
\end{split}
\end{equation}
where $\tilde{h}^{\tau,m+1}$ and $\tilde{l}^{\tau,m+1}$ are the solutions of
\begin{equation}
\hspace{-0.4cm}\left\{
\begin{array}{ll}
-\mathbf{b}^{m+1} \cdot \nabla_{\mathbf{x}} \big(\nabla_{\mathbf{x}} \cdot (\mathbf{b}^{m+1} \, \tilde{h}^{\tau,m+1})\big) = \cfrac{1}{\lambda_{2}}\, \mathbf{b}^{m+1} \cdot \nabla_{\mathbf{x}} S^{\tau,m+1} \, , & \textnormal{on $\Omega$,} \\
(\mathbf{b}^{m+1} \cdot \mathbf{\nu})\, \tilde{h}^{\tau,m+1} = 0 \, , & \textnormal{on $\D\Omega$,}
\end{array}
\right.
\end{equation}
and
\begin{equation}
\left\{
\begin{array}{ll}
\begin{split}
-\mathbf{b}^{m+1} \cdot \nabla_{\mathbf{x}} \big(\nabla_{\mathbf{x}} \cdot (\mathbf{b}^{m+1}\,&\tilde{L}^{\tau,m+1})\big) + \tau\,\lambda_{2}\,\tilde{L}^{\tau,m+1} \\
& \qquad = - \tau\,\mathbf{b}^{m+1} \cdot \nabla_{\mathbf{x}} S^{\tau,m+1} \, ,
\end{split}
& \textnormal{on $\Omega$,} \\
(\mathbf{b}^{m+1} \cdot \mathbf{\nu})\, \tilde{L}^{\tau,m+1} = 0 \, , & \textnormal{on $\D\Omega$,} \\
\end{array}
\right.
\end{equation}
\begin{equation} \label{decompo_diffusion_ELPP_2}
\left\{
\begin{array}{ll}
-\mathbf{b}^{m+1} \cdot \nabla_{\mathbf{x}} \big(\nabla_{\mathbf{x}} \cdot (\mathbf{b}^{m+1}\,\tilde{l}^{\tau,m+1})\big) = \tilde{L}^{\tau,m+1} \, , & \textnormal{on $\Omega$,} \\
(\mathbf{b}^{m+1} \cdot \mathbf{\nu})\, \tilde{l}^{\tau,m+1} = 0 \, , & \textnormal{on $\D\Omega$.} \\
\end{array}
\right.
\end{equation}

\section{Fully-discrete scheme}

\setcounter{equation}{0}

In this section, we present the fully-discrete version of the Asymptotic-Preserving method for (\ref{ELPP_rescaled}) we have presented in the previous paragraph. Before going further, we introduce some notations which will be used throughout this section. \\
First, we consider a uniform mesh $(x_{i},y_{j},z_{k}) = (i\,\Delta x, j\,\Delta y, k\,\Delta z)$ on $\Omega$ and we define the following subsets of $\Z^{3}$:
\begin{equation}
\begin{split}
I &= \big\{ (i,j,k) \in \Z^{3} \, : \, (x_{i},y_{j},z_{k}) \in \Omega \big\} \, , \\
\overline{I} &= \{ (i+\alpha,j+\beta,k+\gamma) \, : \, (i,j,k) \in I, \, (\alpha,\beta,\gamma) \in \{-1,0,1\}^{3} \big\} \, .
\end{split}
\end{equation}
Then, we define the meshed domain $\Omega_{h}$ by
\begin{equation}
\Omega_{h} = \bigcup_{(i,j,k) \, \in \, I} [x_{i-1/2},x_{i+1/2}] \times [y_{j-1/2},y_{j+1/2}] \times [z_{k-1/2},z_{k+1/2}] \, .
\end{equation}
We also define the subsets $I_{*}$ and $\overline{I_{*}}$ of $\Z^{3}$ as follows:
\begin{equation}
\begin{split}
\overline{I_{*}} &= \big\{ (i,j,k) \in \Z^{3} \, : \, (x_{i+1/2},y_{j+1/2},z_{k+1/2}) \in \Omega_{h} \big\} \, , \\
I_{*} &= \big\{ (i,j,k) \in \overline{I_{*}} \, : \, (x_{i+1/2},y_{j+1/2},z_{k+1/2}) \notin \D\Omega_{h} \big\} \, .
\end{split}
\end{equation}
Finally, we assume that, for any $K = (K_{x},K_{y},K_{z}) \in \Z^{3}$, the notation "$|_{K}$" stands for an approximation at the point $(x_{K_{x}},y_{K_{y}},z_{K_{z}})$ and that the notation "$|_{K_{*}}$" stands for an approximation at the point $(x_{K_{x}+1/2},y_{K_{y}+1/2},z_{K_{z}+1/2})$. From now, we also denote the point $(x_{K_{x}},y_{K_{y}},z_{K_{z}})$ by a \textit{cell center} and the point $(x_{K_{x}+1/2},y_{K_{y}+1/2},z_{K_{z}+1/2})$ by a \textit{node}. \\

The next lines are structured as follows: firstly, we present the finite volume scheme based on the semi-discretization (\ref{semi-discrete_ELPP}). Then, we reformulate the obtained fully-discrete scheme by following the same approach as in section 4 and we suggest a numerical method for solving the fully-discrete diffusion equations for $n^{\tau,m+1}$ and $\phi^{\tau,m+1}$.

\subsection{Finite volume scheme}

First, we introduce some notations for the explicit and implicit fluxes for the hydrodynamic part of (\ref{semi-discrete_ELPP}):
\begin{equation}
\mathbf{f}_{\alpha,a}^{exp,\tau,m} = \left(
\begin{array}{c}
\mathbf{e}_{a} \cdot \big((\mathbb{I}-\mathbf{b}^{m+1} \otimes \mathbf{b}^{m+1})\,\mathbf{q}_{\alpha}^{\tau,m}\big) \\ \\ \cfrac{q_{\alpha,a}^{\tau,m} \, \mathbf{q}_{\alpha}^{\tau,m}}{n^{\tau,m}}
\end{array}
\right) \, ,
\end{equation}
\begin{equation}
\mathbf{f}_{\alpha,a}^{imp,\tau,m+1} = \left(
\begin{array}{c}
\mathbf{e}_{a} \cdot \big((\mathbf{b}^{m+1}\otimes\mathbf{b}^{m+1})\,\mathbf{q}_{\alpha}^{\tau,m+1}\big) \\ \\ \mathbf{e}_{a} \,\cfrac{T_{\alpha}\,n^{\tau,m+1}}{\epsilon_{\alpha}\,\tau}
\end{array}
\right) \, ,
\end{equation}
where $\alpha \in \{i,e\}$, $a \in \{x,y,z\}$, $\mathbf{e}_{a}$ is a vector of the canonical basis of $\R^{3}$, and where $\mathbb{I}$ is the $3 \times 3$ identity matrix. Then, we consider different notations for divergence and gradient operator depending on whether they are applied on some component of the implicit or the explicit fluxes. More precisely, we define the operators $\nabla_{h}^{FV}\cdot$, $\nabla_{h}\cdot$ and $\nabla_{h}$ by linking them to the fluxes by the following relations:
\begin{equation}
\left(
\begin{array}{c}
\nabla_{h}^{VF} \cdot \big((\mathbb{I}-\mathbf{b}^{m+1} \otimes \mathbf{b}^{m+1})\,\mathbf{q}_{\alpha}^{\tau,m}\big) \\
\nabla_{h}^{VF} \cdot \big( \cfrac{\mathbf{q}_{\alpha}^{\tau,m} \otimes \mathbf{q}_{\alpha}^{\tau,m}}{n^{\tau,m}} \big)
\end{array}
\right) = \sum_{a\,\in \, \{x,y,z\}} \D_{a}\mathbf{f}_{\alpha,a}^{exp,\tau,m} \, ,
\end{equation}
\begin{equation}
\left(
\begin{array}{c}
\nabla_{h} \cdot \big((\mathbf{b}^{m+1}\otimes\mathbf{b}^{m+1})\,\mathbf{q}_{\alpha}^{\tau,m+1}\big) \\
\cfrac{T_{\alpha}}{\epsilon_{\alpha}\,\tau}\,\nabla_{h}n^{\tau,m+1}
\end{array}
\right) = \sum_{a\,\in \, \{x,y,z\}} \D_{a}\mathbf{f}_{\alpha,a}^{imp,\tau,m+1} \, ,
\end{equation}
where $\alpha \in \{i,e\}$. As a consequence, the fully-discrete model obtained from (\ref{semi-discrete_ELPP}) is written as follows:
\begin{subnumcases}{\label{FD_ELPP}}
\begin{split}
\cfrac{{n^{\tau,m+1}}_{|_{K}}-{n^{\tau,m}}_{|_{K}}}{\Delta t} &+ \Big(\nabla_{h} \cdot \big((\mathbf{b}^{m+1}\otimes\mathbf{b}^{m+1})\,\mathbf{q}_{\alpha}^{\tau,m+1}\big)\big)_{|_{K}} \\
&\quad + \Big(\nabla_{h}^{FV} \cdot \big((\mathbb{I}-\mathbf{b}^{m+1} \otimes \mathbf{b}^{m+1})\,\mathbf{q}_{\alpha}^{\tau,m}\big)\Big)_{|_{K}} \\
&\qquad \qquad \qquad \qquad = -C_{\alpha}\,\cfrac{{\phi^{\tau,m+1}}_{|_{K}}-{\phi^{\tau,m}}_{|_{K}}}{\Delta t} \, ,
\end{split}
& \label{FD_ELPP_n} \\
\begin{split}
&\cfrac{{\mathbf{q}_{\alpha}^{\tau,m+1}}_{|_{K}}-{\mathbf{q}_{\alpha}^{\tau,m}}_{|_{K}}}{\Delta t} + \Big(\nabla_{h}^{FV} \cdot \big(\cfrac{\mathbf{q}_{\alpha}^{\tau,m} \otimes \mathbf{q}_{\alpha}^{\tau,m}}{n^{\tau,m}} \big) \Big)_{|_{K}} \\
& = -\cfrac{1}{\epsilon_{\alpha}\,\tau} \, \big[ T_{\alpha}\,\nabla_{h} n^{\tau,m+1} + \mathfrak{q}_{\alpha}\,( n^{\tau,m+1} \,\nabla_{h}\phi^{\tau,m+1} - \mathbf{q}_{\alpha}^{\tau,m+1} \times \mathbf{B}^{m+1})\big]_{|_{K}} \, , 
\end{split}
& \label{FD_ELPP_q} \\
\alpha \in \{i,e\} \, .
\end{subnumcases}

In finite volume terms, we have
\begin{equation} \label{def_derive}
\begin{split}
(\D_{a}\mathbf{f}_{\alpha,a}^{exp,\tau,m})_{|_{K}} = \cfrac{1}{\Delta a} \, \big({\mathbf{\mathcal{F}}_{\alpha,a}^{\tau,m}}_{|_{K+\mathbf{e}_{a}/2}} - {\mathbf{\mathcal{F}}_{\alpha,a}^{\tau,m}}_{|_{K-\mathbf{e}_{a}/2}}\big) \, , 
\end{split}
\end{equation}
with
\begin{equation} \label{def_flux_shock-capturing}
\begin{split}
{\mathbf{\mathcal{F}}_{\alpha,a}^{\tau,m}}_{|_{K+\mathbf{e}_{a}/2}} &= \cfrac{1}{2}\, \big({\mathbf{f}_{\alpha,a}^{exp,\tau,m}}_{|_{K}} + {\mathbf{f}_{\alpha,a}^{exp,\tau,m}}_{|_{K+\mathbf{e}_{a}}}\big) \\
&\qquad - \cfrac{1}{2}\,
{\mathbb{D}_{\alpha,a}^{\tau,m}}_{|_{K+\mathbf{e}_{a}/2}}\,
\big({\mathbf{W}_{\alpha}^{\tau,m}}_{|_{K+\mathbf{e}_{a}}} -
    {\mathbf{W}_{\alpha}^{\tau,m}}_{|_{K}} \big) \, .
\end{split}
\end{equation}
where $\alpha \in \{i,e\}$ and $a\in \{x,y,z\}$. In these formulae, ${\mathbf{W}_{\alpha}^{\tau,m}}_{|_{K}}$ is
\begin{equation}
{\mathbf{W}_{\alpha}^{\tau,m}}_{|_{K}} = \left(
\begin{array}{c}
{n^{\tau,m}}_{|_{K}} \\ {\mathbf{q}_{\alpha}^{\tau,m}}_{|_{K}}
\end{array}
\right) \, ,
\end{equation}
and $\mathbb{D}_{\alpha,a}^{\tau,m}$ is the numerical viscosity matrix linked with the flux $\mathbf{f}_{\alpha,a}^{exp,\tau,m}$ for any $\alpha \in \{i,e\}$. Since the main goal of the present paper is to validate the time semi-discretization presented in the paragraph 4.2, we choose to compute the viscosity matrices with Rusanov's method (see \cite{Rusanov} and \cite{LeVeque}). For this purpose, we denote the eigenvalues of the jacobian matrices $\textnormal{Jac}_{\mathbf{\mathbf{W}}_{\alpha}}(\mathbf{f}_{\alpha,a}^{exp,\tau,m})$ by $\lambda_{\alpha,k,a}^{\tau,m}$ ($k=1,2,3,4$). Then, $\mathbb{D}_{\alpha,a}^{\tau,m}$ is defined by
\begin{equation}
{\mathbb{D}_{\alpha,a}^{\tau,m}}_{|_{K+\mathbf{e}_{a}/2}} = \mathbb{I} \, \max_{k=1,2,3,4} \max\Big( |{\lambda_{\alpha,k,a}^{\tau,m}}_{|_{K+\mathbf{e}_{a}}}| \, , \, |{\lambda_{\alpha,k,a}^{\tau,m}}_{|_{K}}| \Big) \, .
\end{equation}

\subsection{Reformulation of the fully-discrete scheme}

Following the same approach as in sections 3 and 4, we reformulate the discretized model (\ref{FD_ELPP}) by computing separately the perpendicular part and the parallel part of ${\mathbf{q}_{i}^{\tau,m+1}}_{|_{K}}$ and ${\mathbf{q}_{e}^{\tau,m+1}}_{|_{K}}$ and by solving two diffusion equations to find ${n^{\tau,m+1}}_{|_{K}}$ and ${\phi^{\tau,m+1}}_{|_{K}}$. More precisely, under the hypotheses (\ref{constraints_CiCe}) for $C_{i}$ and $C_{e}$, we solve
\begin{equation} \label{eq_nmp1_FD}
-\Big(\nabla_{h}\cdot \big( (\nabla_{h}n^{\tau,m+1})_{||}^{m+1}\big)\Big)_{|_{K}} + \lambda_{1}\,\tau\,{n^{\tau,m+1}}_{|_{K}} = \tau\,{R^{\tau,m+1}}_{|_{K}} \, ,
\end{equation}
for finding ${n^{\tau,m+1}}_{|_{K}}$ and
\begin{equation} \label{eq_phimp1_FD}
-\Big(\nabla_{h}\cdot \big( n^{\tau,m+1}\,(\nabla_{h}\phi^{\tau,m+1})_{||}^{m+1}\big)\Big)_{|_{K}} + \lambda_{2}\,\tau\,{\phi^{\tau,m+1}}_{|_{K}} = \tau\,{S^{\tau,m+1}}_{|_{K}} \, ,
\end{equation}
for finding ${\phi^{\tau,m+1}}_{|_{K}}$. These discrete diffusion equations are obtained by injecting the parallel part of (\ref{FD_ELPP_q}) with $\alpha = i$ (resp. $\alpha = e$) according to ${\mathbf{b}^{m+1}}_{|_{K}}$ into (\ref{FD_ELPP_n}) with $\alpha = i$ (resp. $\alpha = e$), then performing some linear combinations of the obtained equations (see Appendix \ref{reformulation_FD}). \\
\indent Having ${n^{\tau,m+1}}_{|_{K}}$ and ${\phi^{\tau,m+1}}_{|_{K}}$ in hand, we can compute separately the parallel part and the perpendicular part of ${\mathbf{q}_{i}^{\tau,m+1}}_{|_{K}}$ and ${\mathbf{q}_{e}^{\tau,m+1}}_{|_{K}}$ by using the following formulae:
\begin{subnumcases}{\label{q_paraperp_FD}}
\begin{split}
\big((&\mathbf{q}_{\alpha}^{\tau,m+1})_{||}^{m+1} \big)_{|_{K}} \\
&= \big((\mathbf{q}_{\alpha}^{\tau,m})_{||}^{m+1} \big)_{|_{K}} - \Delta t \, \Big[ \Big(\nabla_{h}^{FV} \cdot \big(\cfrac{\mathbf{q}_{\alpha}^{\tau,m} \otimes \mathbf{q}_{\alpha}^{\tau,m}}{n^{\tau,m}} \big) \Big)_{||}^{m+1} \Big]_{|_{K}} \\
&\qquad - \cfrac{\Delta t}{\epsilon_{\alpha}\,\tau}\, \Big( \big( T_{\alpha}\,\nabla_{h}n^{\tau,m+1} + \mathfrak{q}_{\alpha}\, n^{\tau,m+1} \, \nabla_{h}\phi^{\tau,m+1} \big)_{||}^{m+1}\Big)_{|_{K}} \, ,
\end{split}
& \label{qpara_FD} \\
\begin{split}
&\big((\mathbf{q}_{i}^{\tau,m+1})_{\perp}^{m+1}\big)_{|_{K}} - \cfrac{\mathfrak{q}_{\alpha}\,\epsilon_{\alpha}\,\tau}{\Delta t \, \|{\mathbf{B}^{m+1}}_{|_{K}}\|} \, {\mathbf{b}^{m+1}}_{|_{K}} \times \big((\mathbf{q}_{i}^{\tau,m+1})_{\perp}^{m+1}\big)_{|_{K}} \\
& = \Big[ \cfrac{1}{\|\mathbf{B}^{m+1}\|} \, \mathbf{b}^{m+1} \times ( n^{\tau,m+1} \, \nabla_{h}\phi^{\tau,m+1} + \mathfrak{q}_{\alpha}\,T_{\alpha}\,\nabla_{h}n^{\tau,m+1} ) \Big]_{|_{K}} \\
&\quad + \cfrac{\mathfrak{q}_{\alpha}\,\epsilon_{\alpha}\,\tau}{\|{\mathbf{B}^{m+1}}_{|_{K}}\|} \, {\mathbf{b}^{m+1}}_{|_{K}} \times \Big[ -\cfrac{{\mathbf{q}_{\alpha}^{\tau,m}}_{|_{K}}}{\Delta t} + \Big(\nabla_{h}^{FV} \cdot \big(\cfrac{\mathbf{q}_{\alpha}^{\tau,m} \otimes \mathbf{q}_{\alpha}^{\tau,m}}{n^{\tau,m}} \big) \Big)_{|_{K}} \Big] \, ,
\end{split}
& \label{qperp_FD} \\
\alpha \in \{i,e\} \, .
\end{subnumcases}

\subsection{Three-point scheme}

In this paragraph, we focus on the resolution of (\ref{eq_nmp1_FD}) and (\ref{eq_phimp1_FD}) provided with a discretization of the Neumann-like boundary conditions (\ref{BC_nphi}). For solving these diffusion equations, we follow the approach of Degond \& Tang in \cite{Degond-Tang}: we choose a three-point scheme by replacing the equations (\ref{eq_nmp1_FD}) and (\ref{eq_phimp1_FD}) by
\begin{equation}
\left\{
\begin{array}{ll}
\begin{split}
-(\D_{h,*}^{m+1} \D_{h}^{m+1}n^{\tau,m+1})_{|_{K}} + \tau\, \lambda_{1}\,&{n^{\tau,m+1}}_{|_{K}} \\
&= \tau\, {R^{\tau,m+1}}_{|_{K}} \, ,
\end{split}
& \forall\, K \in I \, , \\
(\D_{h}^{m+1}n^{\tau,m+1})_{|_{K_{*}}} = 0 \, , & \forall\, K \in \overline{I_{*}} \backslash I_{*} \, ,
\end{array}
\right.
\end{equation}
and 
\begin{equation}
\left\{
\begin{array}{ll}
\begin{split}
-\big(\D_{h,*}^{m+1} (n_{*}^{\tau,m+1}\,\D_{h}^{m+1}\phi^{\tau,m+1})\big)_{|_{K}} + &\tau \, \lambda_{2}\,{\phi^{\tau,m+1}}_{|_{K}} \\
&= \tau\, {S^{\tau,m+1}}_{|_{K}} \, ,
\end{split}
& \forall\, K \in I \, , \\
(\D_{h}^{m+1}\phi^{\tau,m+1})_{|_{K_{*}}} = 0 \, , & \forall\, K \in \overline{I_{*}} \backslash I_{*} \, ,
\end{array}
\right.
\end{equation}
respectively. In these equations, ${n_{*}^{\tau,m+1}}_{|_{K_{*}}}$ stands for the average of $n^{\tau,m+1}$ on the node $(x_{K_{x}+1/2},y_{K_{y}+1/2},z_{K_{z}+1/2})$ defined by
\begin{equation}
{n_{*}^{\tau,m+1}}_{|_{K_{*}}} = \cfrac{1}{8} \, \sum_{\alpha,\beta,\gamma \, \in \, \{0,1\} } {n^{\tau,m+1}}_{|_{K+\alpha\mathbf{e}_{x}+\beta\mathbf{e}_{y}+\gamma\mathbf{e}_{z}}} \, ,
\end{equation}
and the operators $\D_{h}^{m+1}$ and $\D_{h,*}^{m+1}$ correspond to some approximations of $\mathbf{b}^{m+1} \cdot \nabla_{\mathbf{x}}$ and $\nabla_{\mathbf{x}} \cdot (\mathbf{b}^{m+1} \cdot )$ respectively. These operators are defined by
\begin{equation} \label{def_bdotgrad_FD}
\begin{split}
(\D_{h}^{m+1}p)_{|_{K_{*}}} &= {\mathbf{b}^{m+1}}_{|_{K_{*}}} \cdot \left(
\begin{array}{c}
\displaystyle \sum_{\beta,\gamma \, \in \, \{0,1\}} \cfrac{p_{|_{K+\mathbf{e}_{x}+\beta\mathbf{e}_{y}+\gamma\mathbf{e}_{z}}} - p_{|_{K+\beta\mathbf{e}_{y}+\gamma\mathbf{e}_{z}}}}{4\Delta x} \\
\displaystyle \sum_{\alpha,\gamma \, \in \, \{0,1\}} \cfrac{p_{|_{K+\alpha\mathbf{e}_{x}+\mathbf{e}_{y}+\gamma\mathbf{e}_{z}}} - p_{|_{K+\alpha\mathbf{e}_{x}+\gamma\mathbf{e}_{z}}}}{4\Delta y} \\
\displaystyle \sum_{\alpha,\beta \, \in \, \{0,1\}} \cfrac{p_{|_{K+\alpha\mathbf{e}_{x}+\beta\mathbf{e}_{y}+\mathbf{e}_{z}}} - p_{|_{K+\alpha\mathbf{e}_{x}+\beta\mathbf{e}_{y}}}}{4\Delta z}
\end{array}
\right) \, ,
\end{split}
\end{equation}
\begin{equation} \label{def_divbdot_FD}
\begin{split}
(\D_{h,*}^{m+1}p)_{|_{K}} &= \sum_{\beta,\gamma \, \in \, \{0,1\}} \cfrac{(b_{x}^{m+1}\,p)_{|_{K_{*}-\beta\mathbf{e}_{y}-\gamma\mathbf{e}_{z}}} - (b_{x}^{m+1}\,p)_{|_{K_{*}-\mathbf{e}_{x}-\beta\mathbf{e}_{y}-\gamma\mathbf{e}_{z}}}}{4\Delta x} \\
&\quad + \hspace{-0.3cm} \sum_{\alpha,\gamma \, \in \, \{0,1\}} \cfrac{(b_{y}^{m+1}\,p)_{|_{K_{*}-\alpha\mathbf{e}_{x}-\gamma\mathbf{e}_{z}}} - (b_{y}^{m+1}\,p)_{|_{K_{*}-\alpha\mathbf{e}_{x}-\mathbf{e}_{y}-\gamma\mathbf{e}_{z}}}}{4\Delta y} \\
&\quad + \hspace{-0.3cm} \sum_{\alpha,\beta \, \in \, \{0,1\}} \cfrac{(b_{z}^{m+1}\,p)_{|_{K_{*}-\alpha\mathbf{e}_{x}-\beta\mathbf{e}_{y}}} - (b_{z}^{m+1}\,p)_{|_{K_{*}-\alpha\mathbf{e}_{x}-\beta\mathbf{e}_{y}-\mathbf{e}_{z}}}}{4\Delta z} \, .
\end{split}
\end{equation}
Then, replacing $\mathbf{b}^{m+1} \cdot \nabla_{\mathbf{x}}$ and $\nabla_{\mathbf{x}} \cdot (\mathbf{b}^{m+1} \cdot)$ by $\D_{h}^{m+1}$ and $\D_{h,*}^{m+1}$ in the decomposition procedure (\ref{decompo_diffusion_ELPP_1})-(\ref{decompo_diffusion_ELPP_2}), we compute ${n^{\tau,m+1}}_{|_{K}}$ and ${\phi^{\tau,m+1}}_{|_{K}}$ that satisfy
\begin{equation} \label{AP_property_FD}
(T_{\alpha}\,\D_{h}^{m+1}n^{\tau,m+1} + \mathfrak{q}_{\alpha}\,n_{*}^{\tau,m+1} \, \D_{h}^{m+1}\phi^{\tau,m+1})_{|_{K_{*}}} = \mathcal{O}(\epsilon_{\alpha}\,\tau) \, , \quad \forall\,\alpha \in \{i,e\} \, .
\end{equation}
We remark that we have the good Asymptotic-Preserving property on the nodes. However, we need it on the cell centers, \textit{i.e.} 
\begin{equation}
\big(\mathbf{b}^{m+1} \cdot (T_{\alpha}\,\nabla_{h}n^{\tau,m+1} + \mathfrak{q}_{\alpha}\,n^{\tau,m+1} \, \nabla_{h}\phi^{\tau,m+1}) \big)_{|_{K}} = \mathcal{O}(\epsilon_{\alpha}\,\tau) \, ,
\end{equation}
for any $\alpha \in \{i,e\}$. To reach such a result, we introduce the discrete gradient $\nabla_{h,*}$ defined by
\begin{equation}
(\nabla_{h,*}p)_{|_{K_{*}}} = \left(
\begin{array}{c}
\displaystyle \sum_{\beta,\gamma \, \in \, \{0,1\}} \cfrac{p_{|_{K+\mathbf{e}_{x}+\beta\mathbf{e}_{y}+\gamma\mathbf{e}_{z}}} - p_{|_{K+\beta\mathbf{e}_{y}+\gamma\mathbf{e}_{z}}}}{4\Delta x} \\
\displaystyle \sum_{\alpha,\gamma \, \in \, \{0,1\}} \cfrac{p_{|_{K+\alpha\mathbf{e}_{x}+\mathbf{e}_{y}+\gamma\mathbf{e}_{z}}} - p_{|_{K+\alpha\mathbf{e}_{x}+\gamma\mathbf{e}_{z}}}}{4\Delta y} \\
\displaystyle \sum_{\alpha,\beta \, \in \, \{0,1\}} \cfrac{p_{|_{K+\alpha\mathbf{e}_{x}+\beta\mathbf{e}_{y}+\mathbf{e}_{z}}} - p_{|_{K+\alpha\mathbf{e}_{x}+\beta\mathbf{e}_{y}}}}{4\Delta z}
\end{array}
\right) \, ,
\end{equation}
and we use it to couple the three-point scheme we have presented with the formulae for $\big((\mathbf{q}_{i}^{\tau,m+1})_{\perp}^{m+1}\big)_{|_{K}}$, $\big((\mathbf{q}_{e}^{\tau,m+1})_{\perp}^{m+1}\big)_{|_{K}}$, $\big((\mathbf{q}_{i}^{\tau,m+1})_{||}^{m+1}\big)_{|_{K}}$ and $\big((\mathbf{q}_{e}^{\tau,m+1})_{||}^{m+1}\big)_{|_{K}}$: more precisely, we use the operator $\nabla_{h,*}$ for computing the following terms in (\ref{q_paraperp_FD}):
\begin{subnumcases}{}
\begin{split}
\Big( &\big( T_{\alpha}\,\nabla_{h}n^{\tau,m+1} + \mathfrak{q}_{\alpha}\,n^{\tau,m+1} \, \nabla_{h}\phi^{\tau,m+1} \big)_{||}^{m+1}\Big)_{|_{K}} \\
&= \cfrac{1}{8} \, \sum_{\alpha,\beta,\gamma\,\in\,\{0,1\}} \Big((T_{\alpha}\,\nabla_{h,*}n^{\tau,m+1} \\
& \qquad \qquad \qquad \qquad + \mathfrak{q}_{\alpha}\, n_{*}^{\tau,m+1} \, \nabla_{h,*}\phi^{\tau,m+1})_{||}^{m+1} \Big)_{|_{K_{*}-\alpha\mathbf{e}_{x}-\beta\mathbf{e}_{y}-\gamma\mathbf{e}_{z}}} \, ,
\end{split}
& \\
\begin{split}
&\Big( \cfrac{1}{\|\mathbf{B}^{m+1}\|} \, \mathbf{b}^{m+1} \times ( \mathfrak{q}_{\alpha}\,n^{\tau,m+1} \, \nabla_{h}\phi^{\tau,m+1} + T_{\alpha}\, \nabla_{h}n^{\tau,m+1} ) \Big)_{|_{K}} \\
&= \cfrac{1}{8} \sum_{\alpha,\beta,\gamma\,\in\,\{0,1\}} \Big(\cfrac{1}{\|{\mathbf{B}^{m+1}}\|} \, {\mathbf{b}^{m+1}} \times \big[ n_{*}^{\tau,m+1} \, \nabla_{h,*}\phi^{\tau,m+1} \\
&\qquad \qquad \qquad \qquad \qquad \qquad \qquad \qquad + \nabla_{h,*}n^{\tau,m+1} \big]\Big)_{|_{K_{*}-\alpha\mathbf{e}_{x}-\beta\mathbf{e}_{y}-\gamma\mathbf{e}_{z}}} \, ,
\end{split}
& \\
\alpha \in \{i,e\}\,.
\end{subnumcases}

As a consequence, we obtain the properties (\ref{eq_nmp1_FD}) and (\ref{eq_phimp1_FD}) on cell centers and we can compute the parallel part and the perpendicular part of $\mathbf{q}_{i}^{\tau,m+1}$ and $\mathbf{q}_{e}^{\tau,m+1}$ by using separately the formulae (\ref{q_paraperp_FD}).

\section{Numerical results}
\setcounter{equation}{0}

In this last section, we present some 2D numerical results which have been obtained with the AP scheme we have presented in sections 4.2 and 5 for the perturbed two-fluid euler-Lorentz model (\ref{ELPP_rescaled}).

\subsection{Validation of the three-point scheme for the diffusion problems}

\indent Since the AP scheme relies on the truthfulness of the properties (\ref{AP_property_para_SD}), we first present some numerical results from the three-point scheme used for the resolution the diffusion problems for $n^{\tau,m+1}$ and $\phi^{\tau,m+1}$ (see paragraph 5.3). More precisely, the main goal of the first test sequence is to solve the diffusion problems (\ref{ELPP_diffusion_n})-(\ref{BC_nphi_n}) and (\ref{ELPP_diffusion_phi})-(\ref{BC_nphi_phi}) for any value of $\tau \geq 0$ and to insure that
\begin{equation}
\forall \, K \, , \qquad 
{n^{\tau,m+1}}_{|_{K}} \to {n^{0,m+1}}_{|_{K}} \, , \quad {\phi^{\tau,m+1}}_{|_{K}} \to {\phi^{0,m+1}}_{|_{K}} \, ,
\end{equation}
as $\tau \to 0$. To perform this validation, we apply our method to the following diffusion problem:
\begin{equation} \label{diffusion_generic}
\left\{
\begin{array}{ll}
-\nabla_{\mathbf{x}} \cdot \big(H^{\tau}\,(\mathbf{b} \otimes \mathbf{b})\,\nabla_{\mathbf{x}} p^{\tau}) + \tau\,\lambda\,p^{\tau} = \tau\,f^{\tau} \, , & \textnormal{on $\Omega$,} \\
\big(H^{\tau}\,(\mathbf{b} \otimes \mathbf{b}) \cdot \nabla_{\mathbf{x}}p^{\tau}\big)\cdot \nu = 0 \, , & \textnormal{on $\D\Omega$,}
\end{array}
\right.
\end{equation}
where $\lambda > 0$, $f^{\tau} : \Omega \to \R$, $H^{\tau} : \overline{\Omega} \to \R_{+}^{*}$ and $\mathbf{b} : \overline{\Omega} \to \R^{3}$ are given. \\

Let us consider a function sequence $(p^{\tau})_{\tau \, \geq \, 0}$ defined by
\begin{equation} \label{def_ptau_diffusion}
p^{\tau} = p_{0} + \tau \,p_{1}^{\tau} \, ,
\end{equation}
with $p_{0}$ and $p_{1}^{\tau}$ satisfying
\begin{equation}
\mathbf{b} \cdot \nabla_{\mathbf{x}}p_{0} = 0 \, , \quad \textnormal{on $\overline{\Omega}$,}
\end{equation}
and
\begin{equation}
\big(H^{\tau}\,(\mathbf{b} \otimes \mathbf{b})\,\nabla_{\mathbf{x}}p_{1}^{\tau}\big) \cdot \nu = 0 \, , \quad \textnormal{on $\D\Omega$.}
\end{equation}
We assume from now that 
\begin{equation}
f^{\tau} = \lambda\,p^{\tau} - \nabla_{\mathbf{x}} \cdot \big( H^{\tau}\,(\mathbf{b} \otimes \mathbf{b})\,\nabla_{\mathbf{x}}p_{1}^{\tau}\big) \, ,
\end{equation}
which implies that $p^{\tau}$ is the analytic solution of the problem (\ref{diffusion_generic}). \\

\begin{figure}[ht]
\parbox{0.5\linewidth}{\includegraphics[scale=0.33]{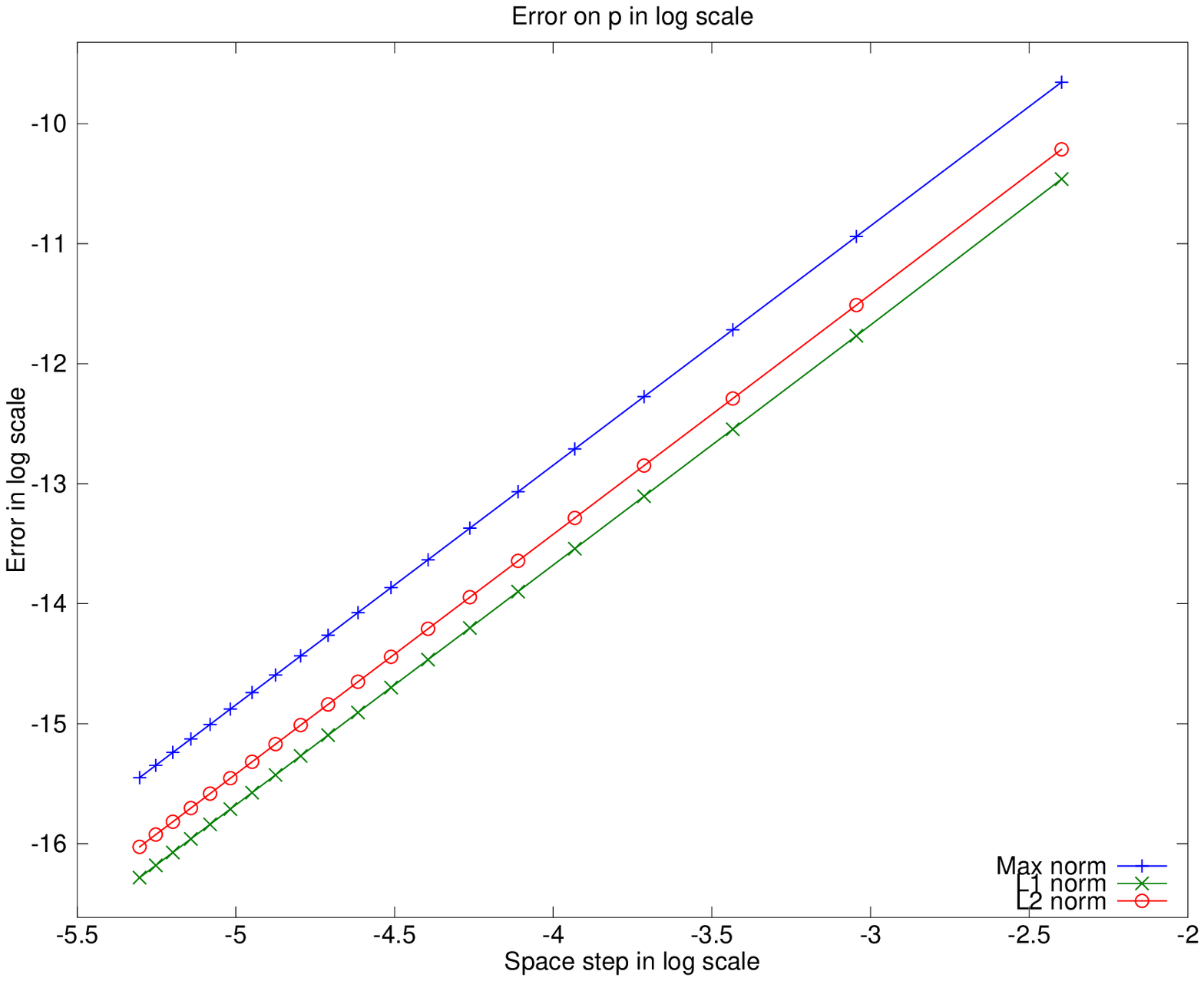}} \parbox{0.5\linewidth}{\includegraphics[scale=0.33]{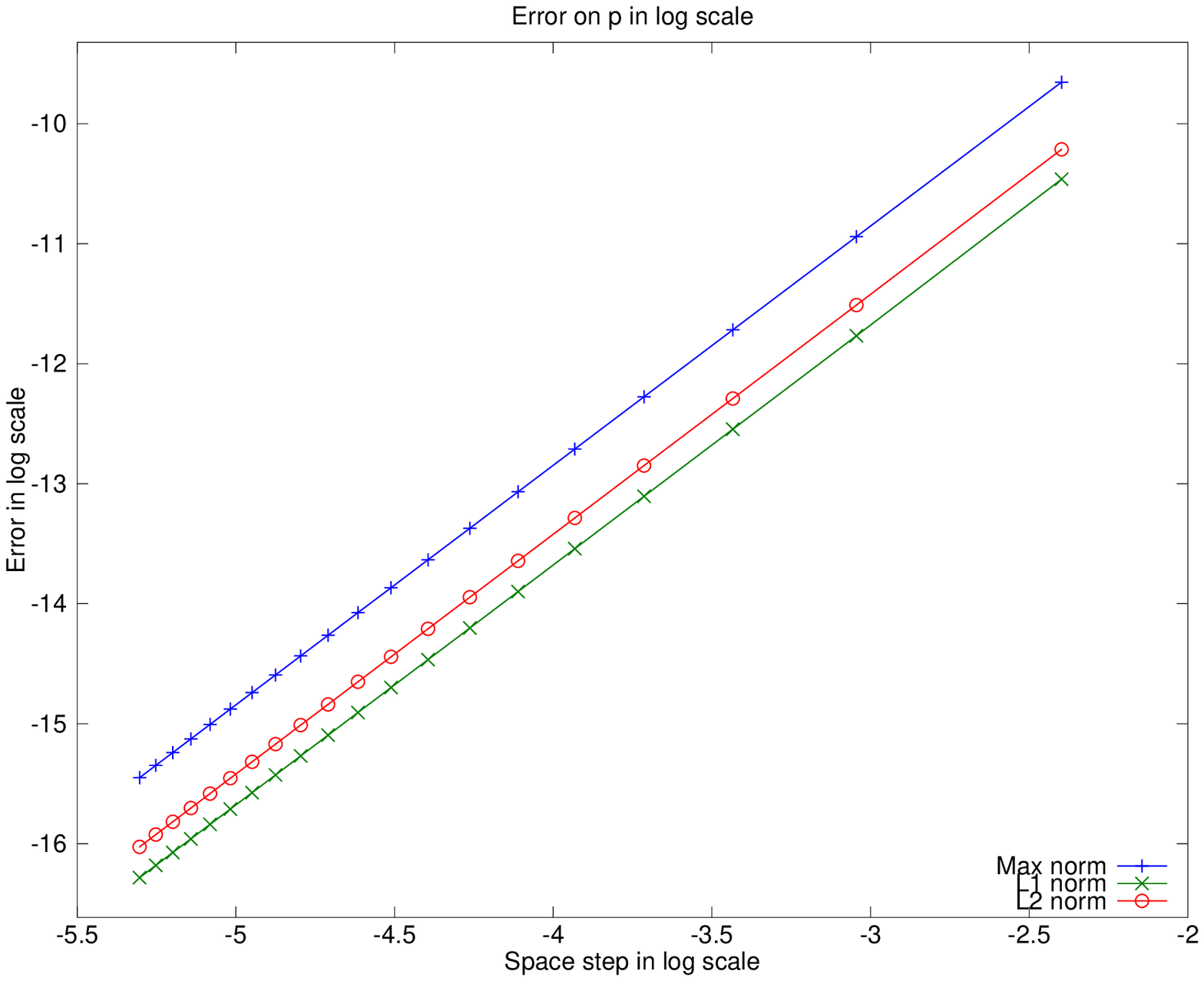}}
\caption{$L^{1}$, $L^{2}$ and $L^{\infty}$ norms of the error between $p^{\tau}$ and its approximation $p_{app}^{\tau}$ as functions of $h$: case with $\tau = 10^{-2}$ (left) and $\tau = 10^{-9}$ (right).} \label{error_diffusion}
\end{figure}

\begin{figure}[ht]
\begin{center}
\includegraphics[scale=0.33]{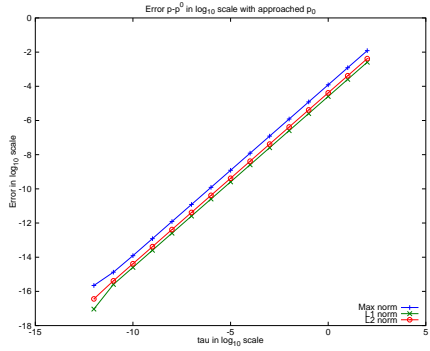}
\end{center}
\caption{$L^{1}$, $L^{2}$ and $L^{\infty}$ norms of the error between $p^{\tau}$ and $p^{0} = p_{0}$ as functions of $\tau$: case with a $100 \times 100$ uniform mesh.} \label{cv_tau_diffusion}
\end{figure}

\indent In Figure \ref{error_diffusion}, we plot the evolution of the error between $p^{\tau}$ and its approximation (denoted with $p_{app}^{\tau}$) as a function of the space step $h$. In these results which are presented in decimal logarithmic scale, we have chosen $\tau = 10^{-2}$ and $\tau = 10^{-9}$. $\Omega$ is set to $[1,2] \times [1,2] \subset \R^{2}$, $p_{0}$ and $p_{1}^{\tau}$ to
\begin{equation}
p_{0}(x,y) = 2 \, , \qquad p_{1}(x,y) = \big((x-1)(2-x)(y-1)(2-y)\big)^{3} \, ,
\end{equation}
and $\lambda$, $H^{\tau}$, and $\mathbf{b}$ are chosen as
\begin{equation}
\lambda = 1 \, ,
\end{equation}
\begin{equation}
H^{\tau}(x,y) = 1 + \sin^{2}(x)\,\sin^{2}(y) \, ,
\end{equation}
\begin{equation}
\mathbf{b} = (\sin\theta,-\cos\theta)\, , \quad \textnormal{with $\theta(x,y) = \arctan(y/x)$.}
\end{equation}
As we can remark in this figure, the solver for the diffusion problem (\ref{diffusion_generic}) based on the micro-macro decomposition presented in Section 4.1 and on the discrete differential operators $\D_{h}$ and $\D_{h,*}$ is second order accurate in $h$ since we observe that the error $\|p^{\tau}-p_{app}^{\tau}\|_{L^{p}}$ ($p = 1,2,\infty$) is linearly decreasing in $\log_{10}$ scale when $h \to 0$, with a slope which is equal to 2. This is due to the fact that, according to the definitions (\ref{def_bdotgrad_FD}) and (\ref{def_divbdot_FD}), the operators $\D_{h}$ and $\D_{h,*}$ are themselves second order accurate. Furthermore, we have this property for $\tau = 10^{-2}$ and $\tau = 10^{-9}$, so we can conclude that the second order accurate of the solver is not penalized by the smallness of $\tau$. \\
\indent Together with this convergence results in $h$, we plot in Figure \ref{cv_tau_diffusion} the error between $p_{app}^{\tau}$ provided by the solver and $p^{0} = p_{0}$ as a function of $\tau$ with a $100 \times 100$ uniform mesh, and we take the same values of $p_{0}$, $p_{1}^{\tau}$, $\lambda$, $H^{\tau}$, and $\mathbf{b}$ as above. By definition of the analytic solution of $p^{\tau}$ and $p_{0}$ (see (\ref{def_ptau_diffusion})), we except this error to be of the same order of $\tau$. This is confirmed by Figure \ref{cv_tau_diffusion}: indeed, the error $\|p_{app}^{\tau}-p_{0}\|_{L^{p}}$ ($p=1,2,\infty$) is linearly decreasing in $\log_{10}$ scale when $\tau$ converges to 0 with a slope which is equal to 1. \\
\indent From these two results, we can claim that
\begin{equation}
\lim_{h\,\to\,0} \lim_{\tau\,\to\,0} p_{app}^{\tau} = p_{0} \, ,
\end{equation}
which is exactly to say that the numerical solver for the diffusion problem (\ref{diffusion_generic}) is Asymptotic-Preserving when $\tau \to 0$. Then, we can use it for solving the diffusion problems (\ref{ELPP_diffusion_n})-(\ref{BC_nphi_n}) and (\ref{ELPP_diffusion_phi})-(\ref{BC_nphi_phi}) for $n^{\tau,m+1}$ and $\phi^{\tau,m+1}$.

\subsection{Numerical results for the two-fluid Euler-Lorentz model near the drift-fluid limit}

In order to validate the AP scheme we have developed for the perturbed Euler-Lorentz (\ref{ELPP_rescaled}), we compare the results which are computed by the AP scheme to those which can be produced with a fully explicit finite volume method. From now, we denote with \textit{classical method} a finite volume scheme which is based on the following time semi-discretization:
\begin{equation} \label{ELPP_FE}
\left\{
\begin{array}{l}
\cfrac{n^{\tau,m+1}-n^{\tau,m}}{\Delta t} + C_{\alpha}\,\cfrac{\phi^{\tau,m+1}-\phi^{\tau,m}}{\Delta t} + \nabla_{\mathbf{x}} \cdot \mathbf{q}_{\alpha}^{\tau,m} = 0 \, , \\ \\
\cfrac{\mathbf{q}_{\alpha}^{\tau,m+1}-\mathbf{q}_{\alpha}^{\tau,m}}{\Delta t} + \nabla_{\mathbf{x}} \cdot \Big( \cfrac{\mathbf{q}_{\alpha}^{\tau,m} \otimes \mathbf{q}_{\alpha}^{\tau,m}}{n^{\tau,m}} \Big) + \cfrac{T_{\alpha}}{\epsilon_{\alpha}\,\tau} \, \nabla_{\mathbf{x}}n^{\tau,m} \\
\qquad \qquad \qquad \qquad = \mathfrak{q}_{\alpha}\, \big[-\cfrac{1}{\epsilon_{\alpha}\,\tau} \, n^{\tau,m}\,\nabla_{\mathbf{x}} \phi^{\tau,m} + \mathbf{q}_{\alpha}^{\tau,m+1} \times \mathbf{B}^{m+1} \big] \, , \\ \\
\alpha \in \{i,e\} \, .
\end{array}
\right.
\end{equation}
We easily remark that the stability condition of such a method strongly depends on $\tau$: as in a low Mach number numerical experiment, the smaller $\tau$ is, the smaller the time step $\Delta t$ must be in order to insure that a method based on (\ref{ELPP_FE}) is stable. As a example, if we solve the hydrodynamic part of (\ref{ELPP_FE}) with Rusanov' scheme, we must have $\Delta t = \mathcal{O}(h\,\tau^{1/2})$ at least, where $h=\min(\Delta x,\Delta y,\Delta z)$. \\

\indent In the next lines, we distinguish two opposite situations:
\begin{itemize}
\item \textit{The resolved case:} The time step is small enough in order to insure that both classical and AP methods capture the fast time variations within the solution,
\item \textit{The under-resolved case:} The time step does not allow the capture of fast time variations but insures at least the stability of the AP scheme.
\end{itemize}


\indent The test case we present here is based on the perturbation of the following stationary case:
\begin{itemize}
\item The magnetic field is uniform and reads $\mathbf{B} = (\sin\alpha,-\cos\alpha,0)$ with $\alpha \in \R$ fixed,
\item $n^{\tau,0}(x,y) = n_{0}$, $\phi^{\tau,0}(x,y) = \phi_{0}$, $\mathbf{q}_{i}^{\tau,0} = \mathbf{q}_{e}^{\tau,0} = \mathbf{B}$ with some constants $n_{0}$ and $\phi_{0}$,
\item The whole system (\ref{ELPP_rescaled}) does not depend on the variable $z$.
\end{itemize}
Remarking that both classical and AP schemes compute the exact solution provided with these initial datas, we choose to introduce a small perturbation at the initial time step. More precisely, we choose to replace $n^{\tau,0}(x,y) = n_{0}$ by
\begin{equation}
n^{\tau,0}(x,y) = n_{0} + \tau\,\max \big( 0, 1-\eta\,(x-x_{0})^{2}-\eta\,(y-y_{0})^{2}\big) \, ,
\end{equation}
with $\eta \geq 0$ and $(x_{0},y_{0}) \in \Omega$. \\

\begin{center}
\begin{tabular}{cc}
\includegraphics[scale=0.25]{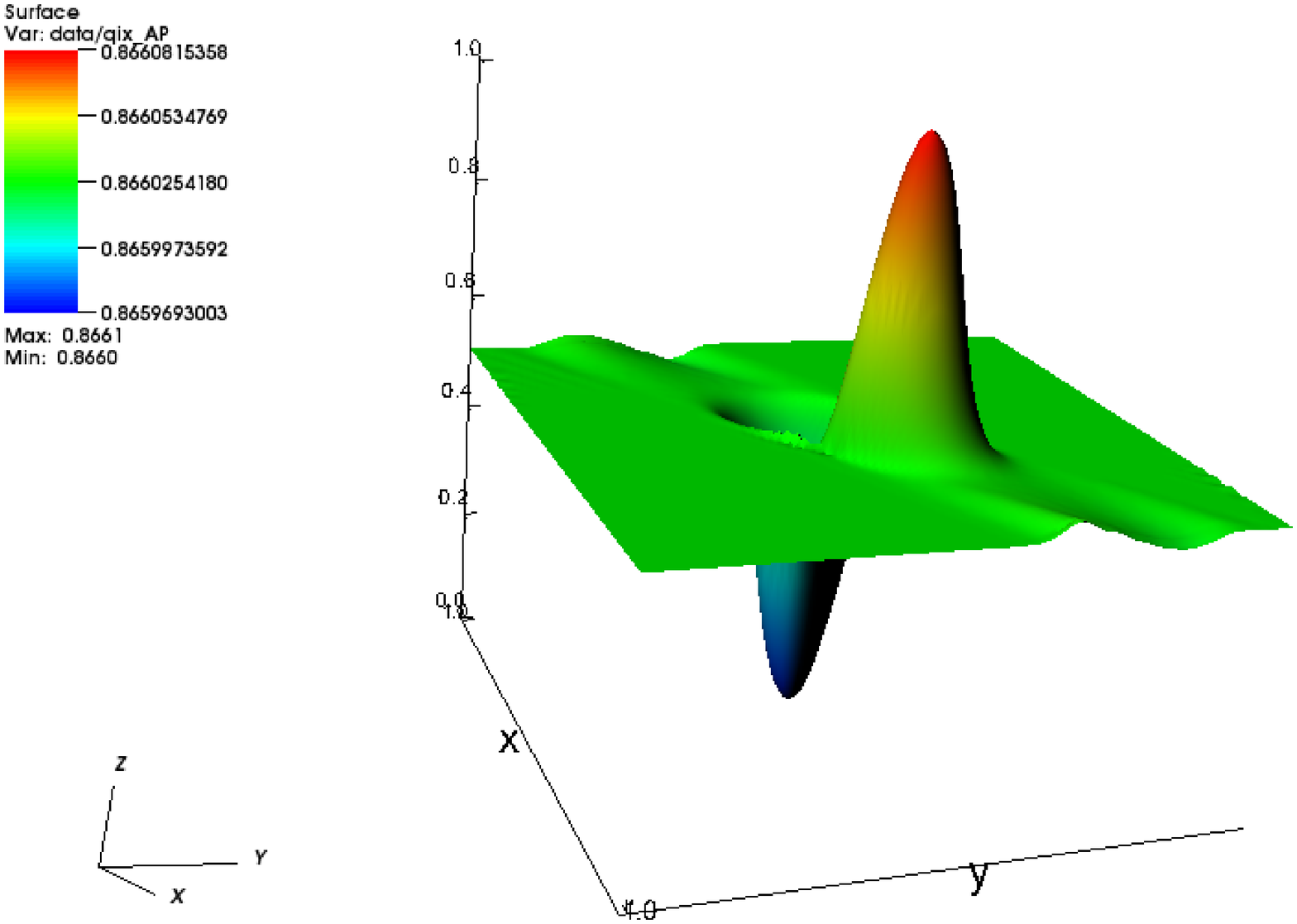} & \includegraphics[scale=0.25]{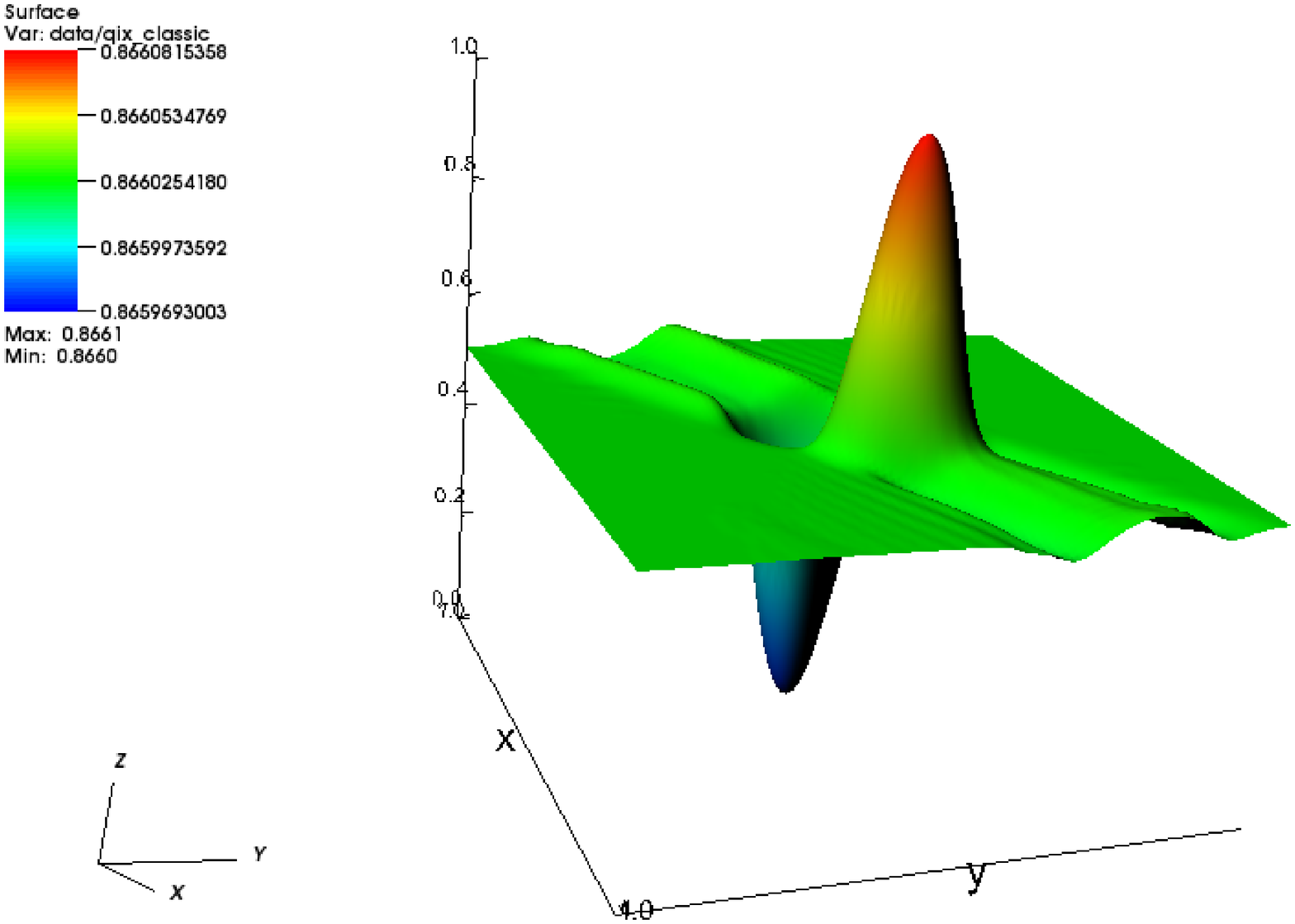} \\
$q_{i,x}^{\tau}$ (AP scheme) & $q_{i,x}^{\tau}$ (classical scheme) \\
\includegraphics[scale=0.25]{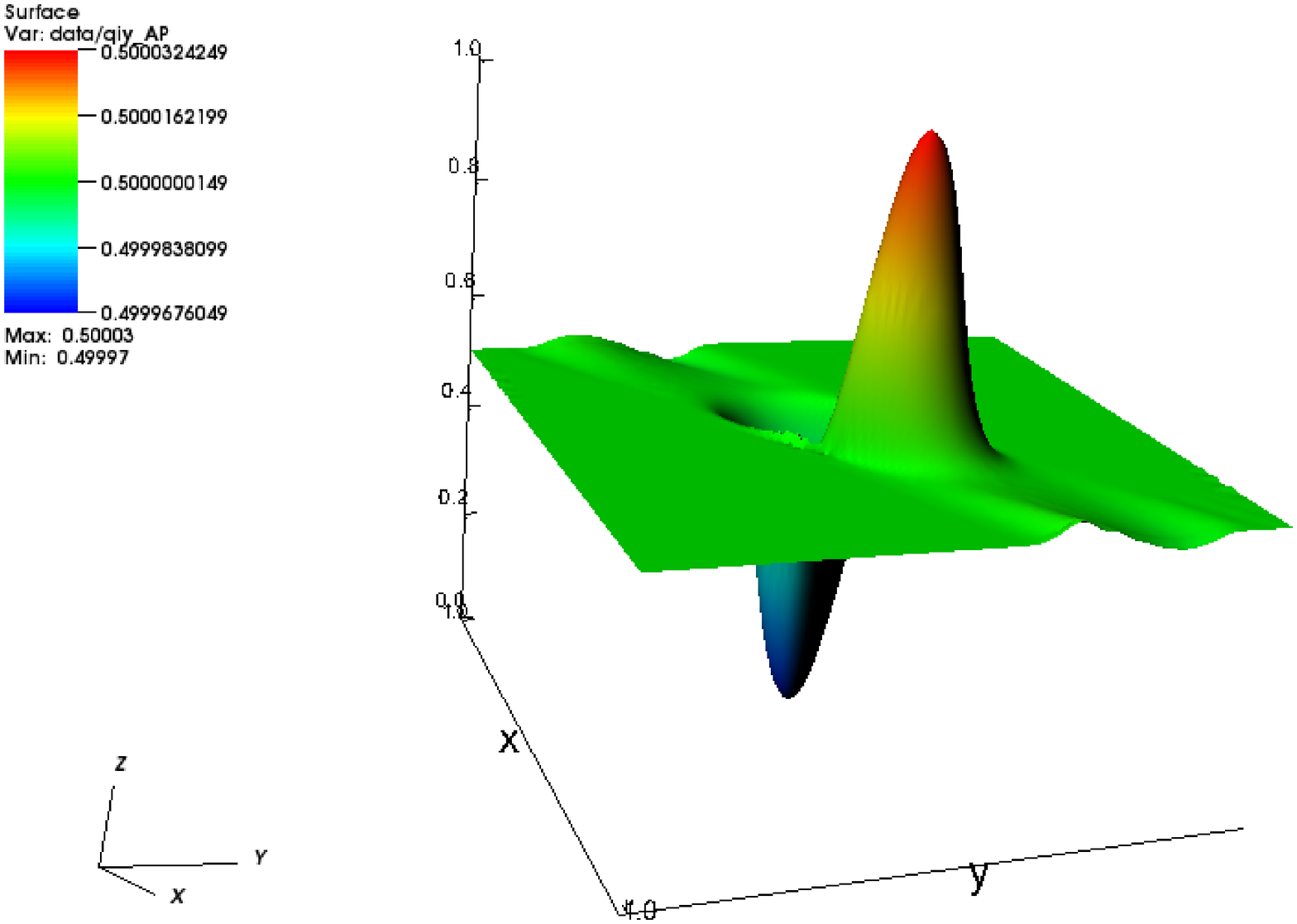} & \includegraphics[scale=0.25]{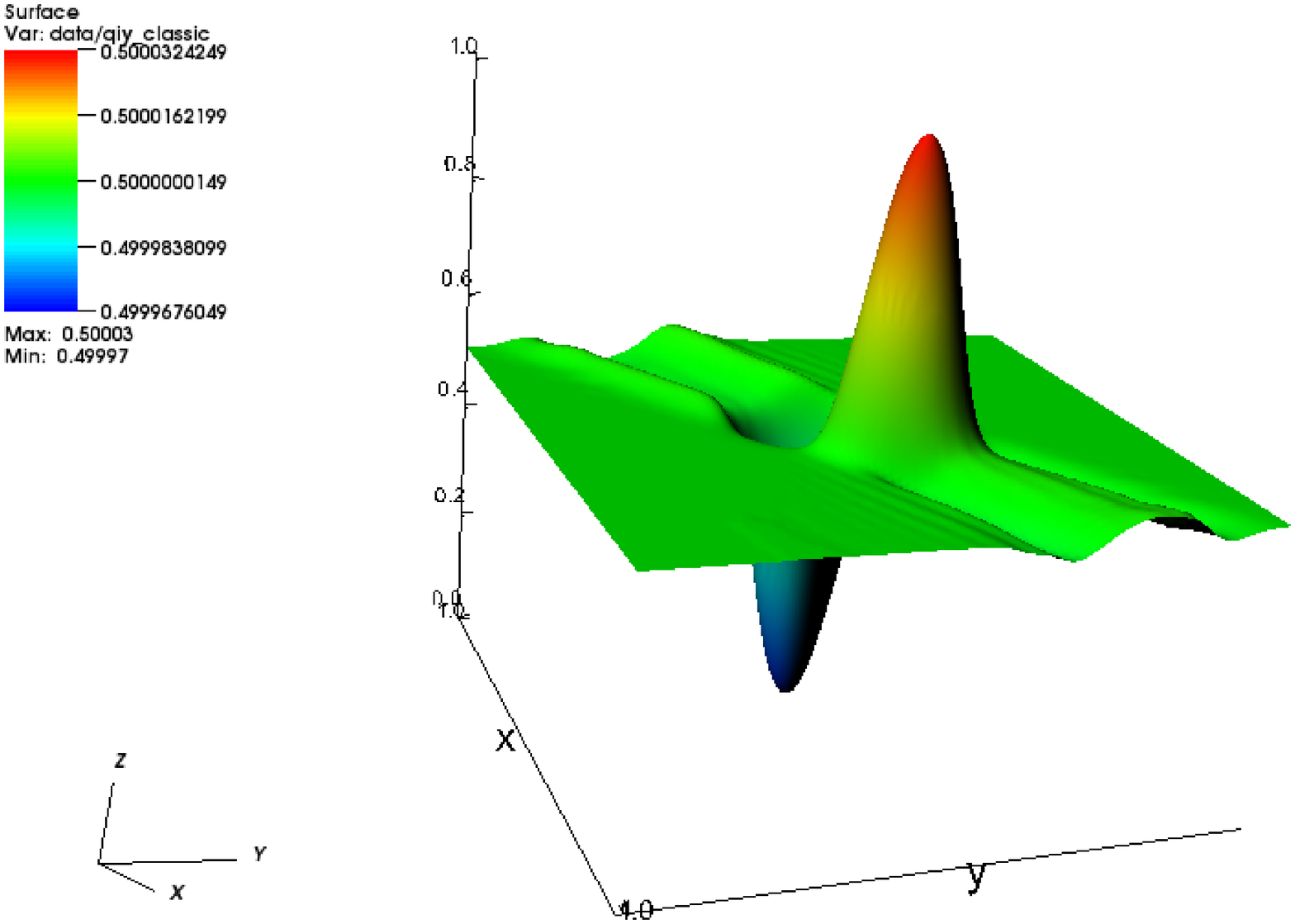} \\
$q_{i,y}^{\tau}$ (AP scheme) & $q_{i,y}^{\tau}$ (classical scheme)
\end{tabular}
\begin{figure}[ht]
\caption{Resolved case at time $t = 6 \times 10^{-6}$: $x$ and $y$ components of the ion momentum $\mathbf{q}_{i}^{\tau}$ as functions of $(x,y)$ computed with the AP scheme (left) and the classical scheme (right).} \label{qi_resolved}
\end{figure}
\end{center}
\begin{center}
\begin{tabular}{cc}
\includegraphics[scale=0.25]{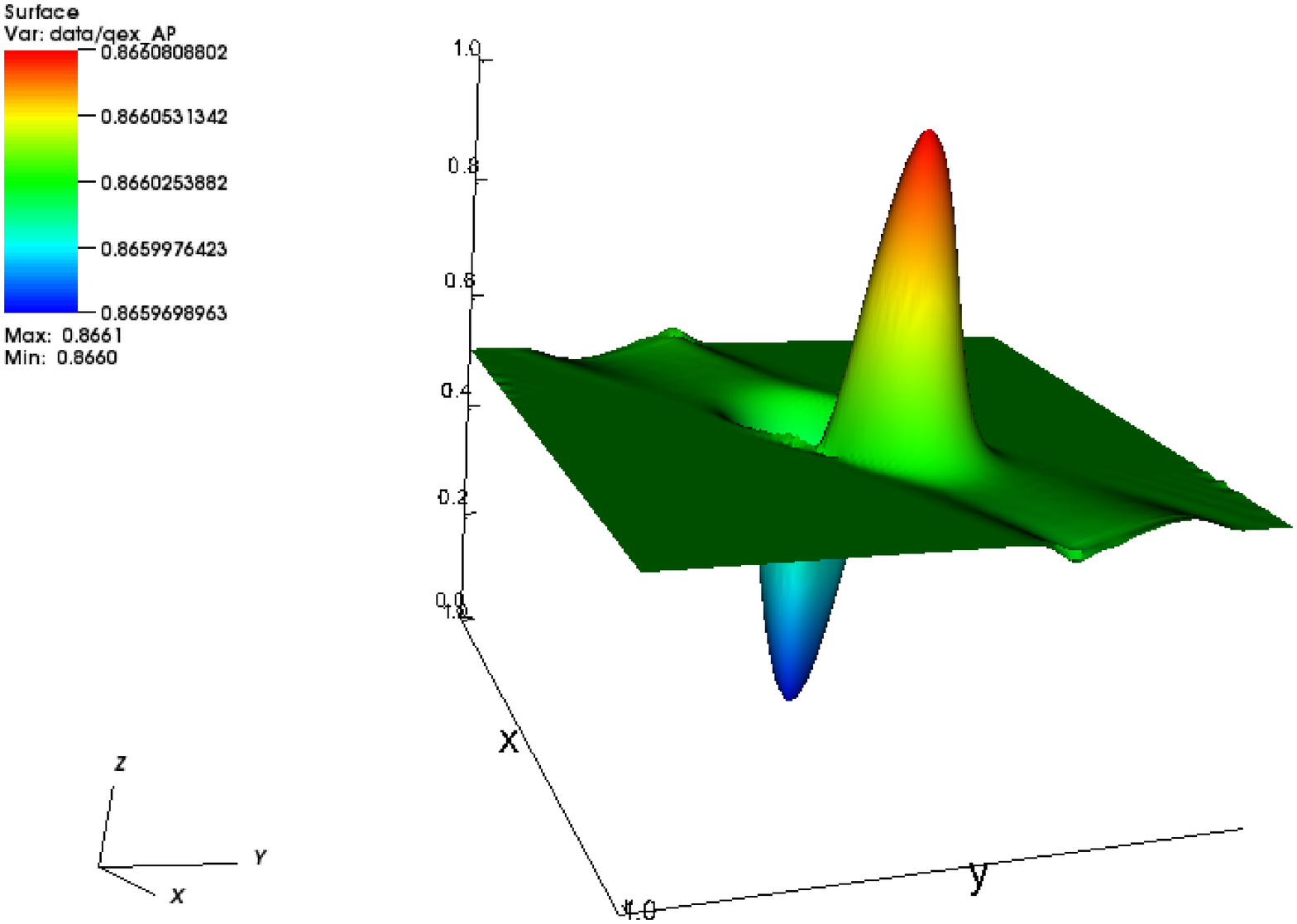} & \includegraphics[scale=0.25]{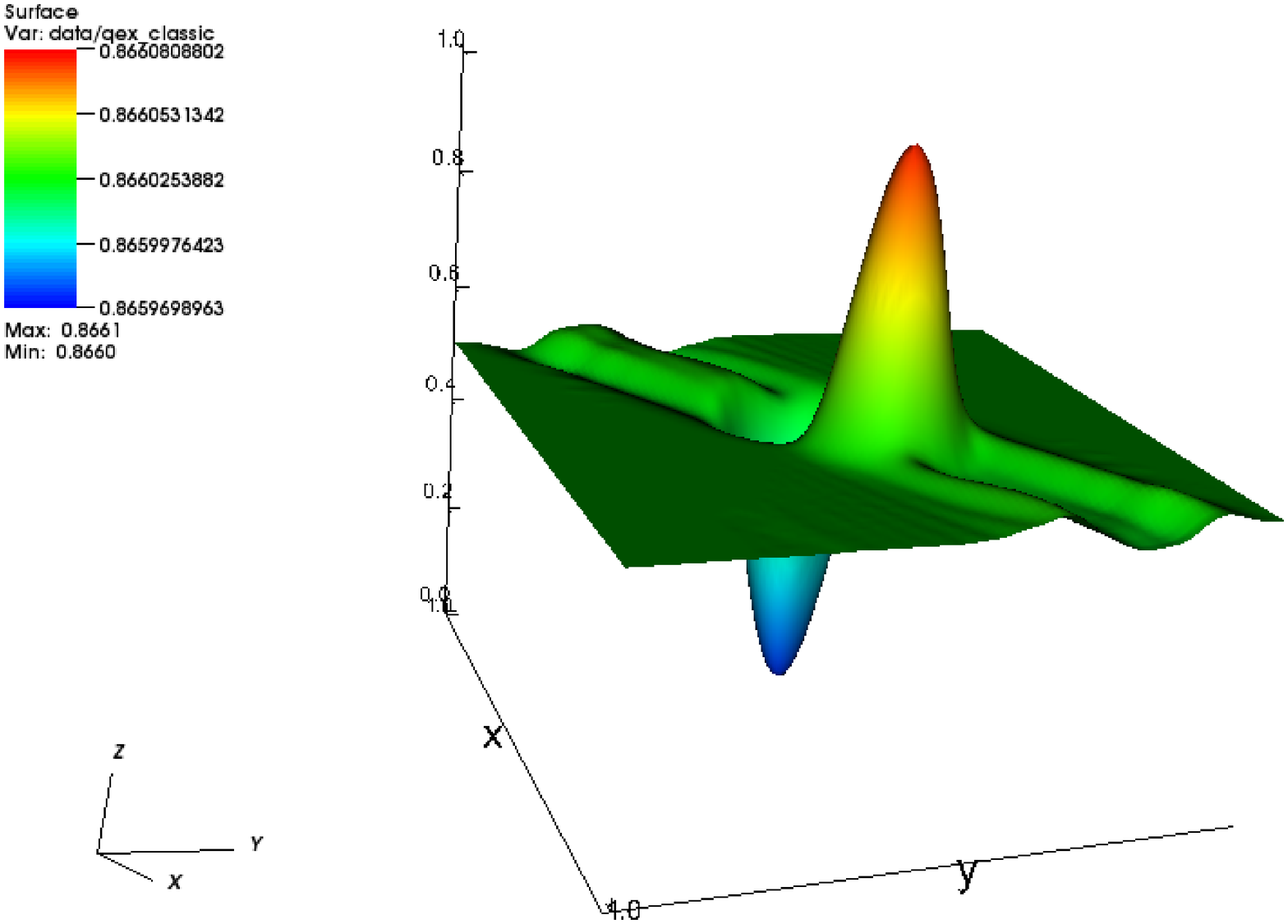} \\
$q_{e,x}^{\tau}$ (AP scheme) & $q_{e,x}^{\tau}$ (classical scheme) \\
\includegraphics[scale=0.25]{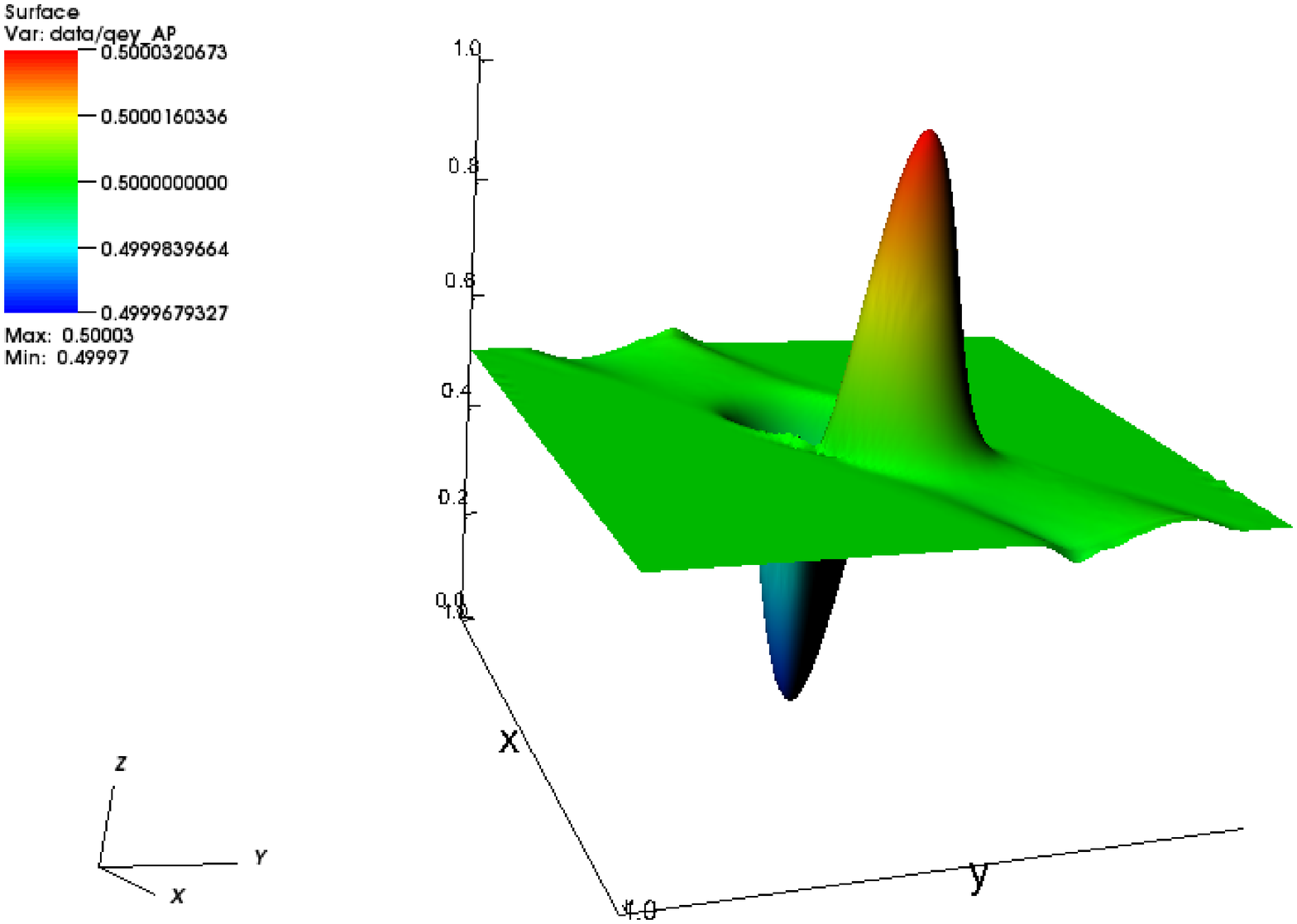} & \includegraphics[scale=0.25]{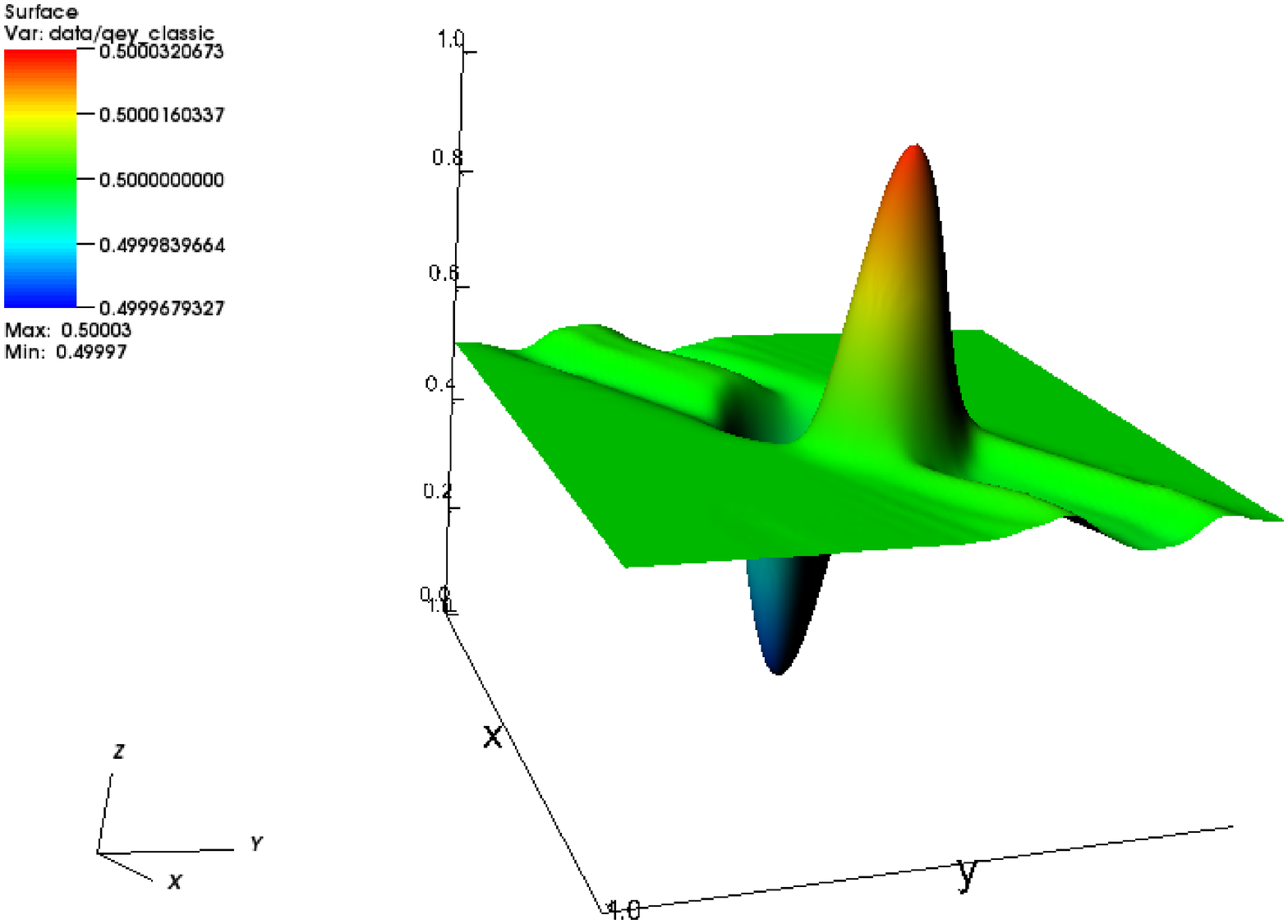} \\
$q_{e,y}^{\tau}$ (AP scheme) & $q_{e,y}^{\tau}$ (classical scheme)
\end{tabular}
\begin{figure}[ht]
\caption{Resolved case at time $t = 6 \times 10^{-6}$: $x$ and $y$ components of the electron momentum $\mathbf{q}_{e}^{\tau}$ as functions of $(x,y)$ computed with the AP scheme (left) and the classical scheme (right).} \label{qe_resolved}
\end{figure}
\end{center}

\indent We also assume that the physical domain $\Omega$ is $[1,2] \times [1,2]$ and is meshed by a $100 \times 100$ uniform mesh. We also precise the initial datas by taking $\tau = 10^{-8}$, $\epsilon = 1$, $T_{e} = 3$, $C = 10^{-2}$, $\alpha = \frac{2\pi}{3}$, $\eta = 80$, $(x_{0},y_{0}) = (\frac{3}{2},\frac{3}{2})$, $n_{0} = 1$ and $\phi_{0} = 0$. \\
\indent In Figures \ref{qi_resolved}-\ref{qe_resolved}, we present some results in the resolved situation for both classical and AP schemes and $\Delta t = 5 \times 10^{-9}$ is taken as time step. As we can see in these figures, the results which are produced by the AP scheme are very close to the classical method's ones. 


\begin{center}
\begin{tabular}{cc}
\includegraphics[scale=0.25]{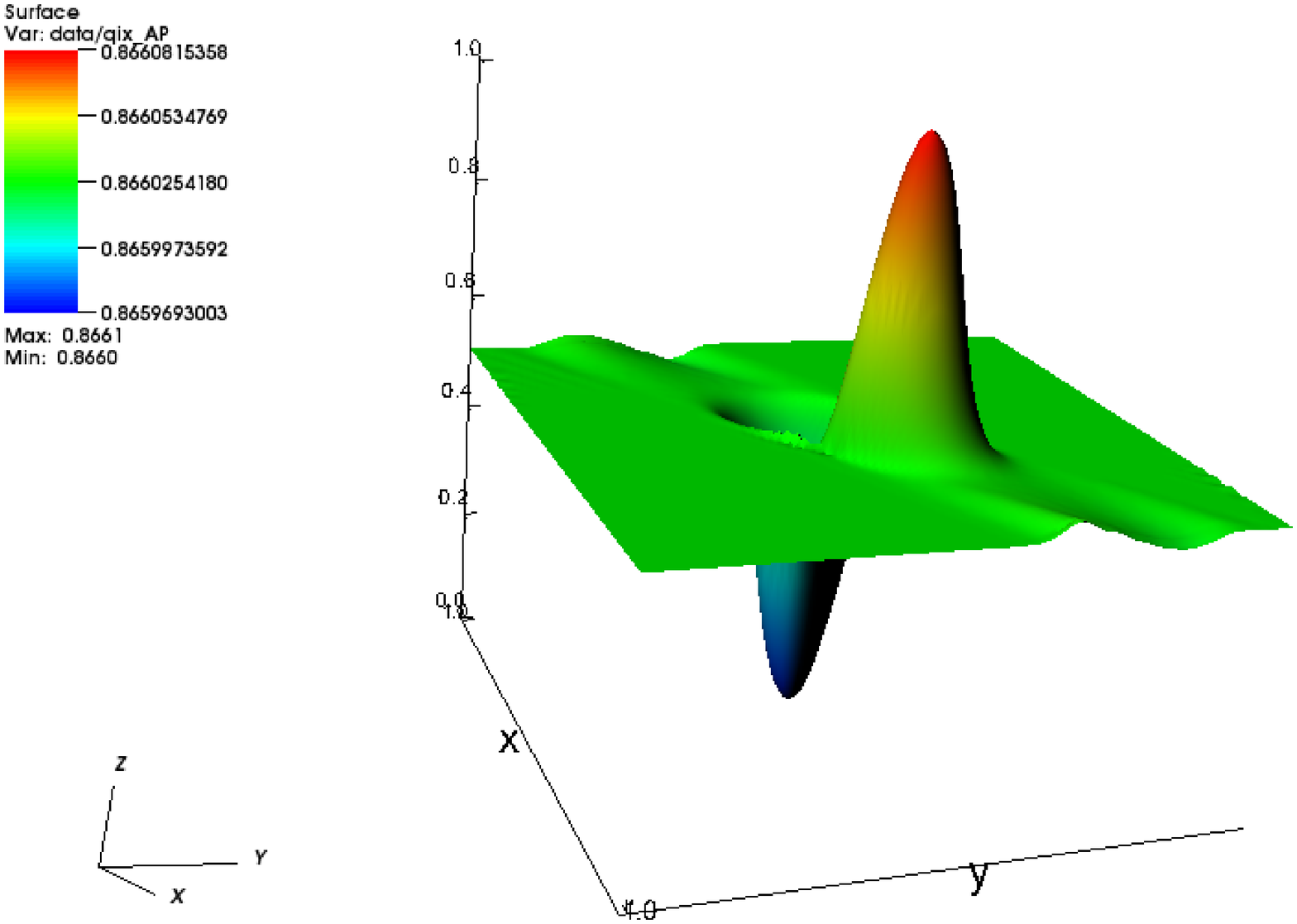} & \includegraphics[scale=0.25]{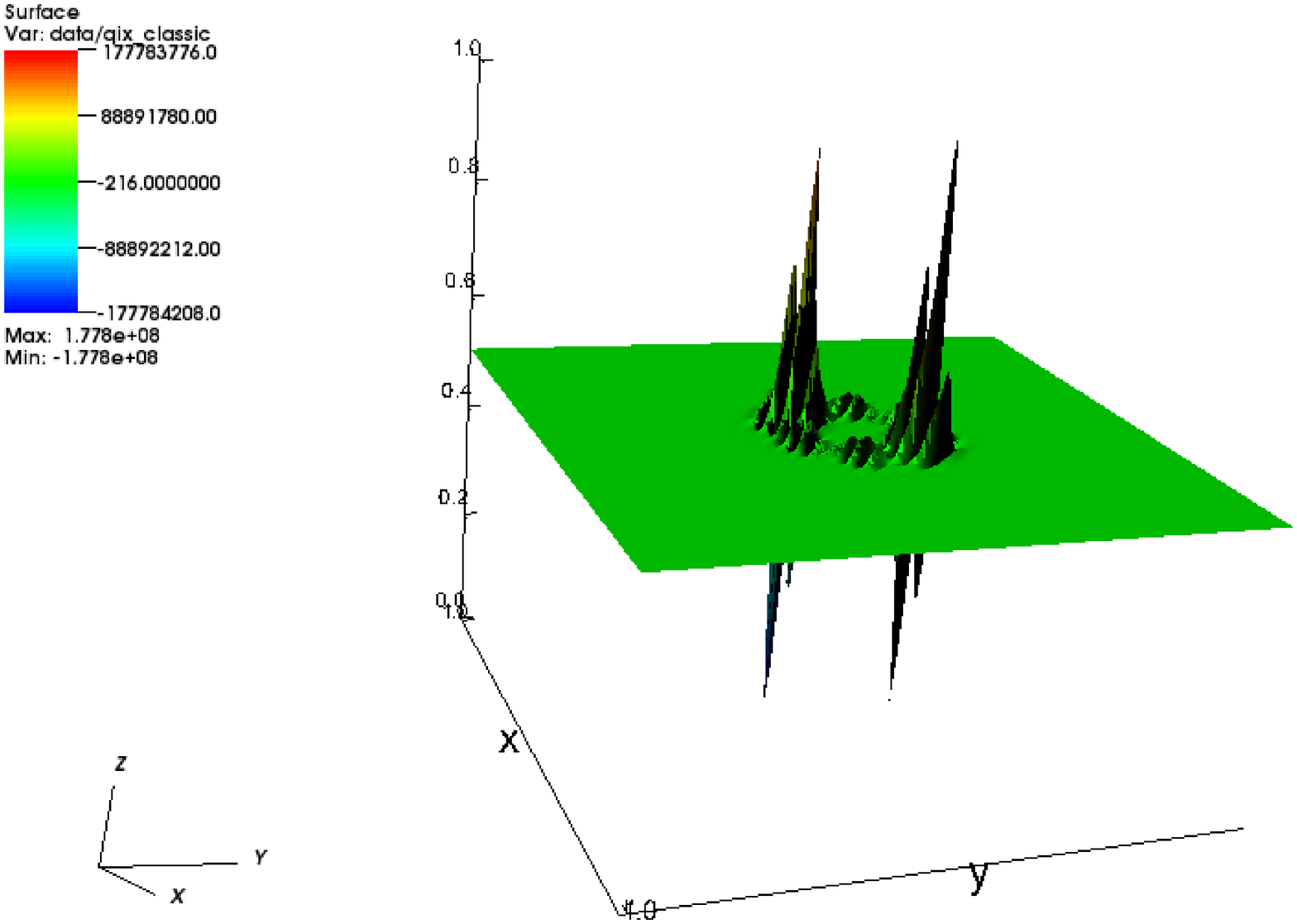} \\
$q_{i,x}^{\tau}$ (AP scheme) & $q_{i,x}^{\tau}$ (classical scheme) \\
\includegraphics[scale=0.25]{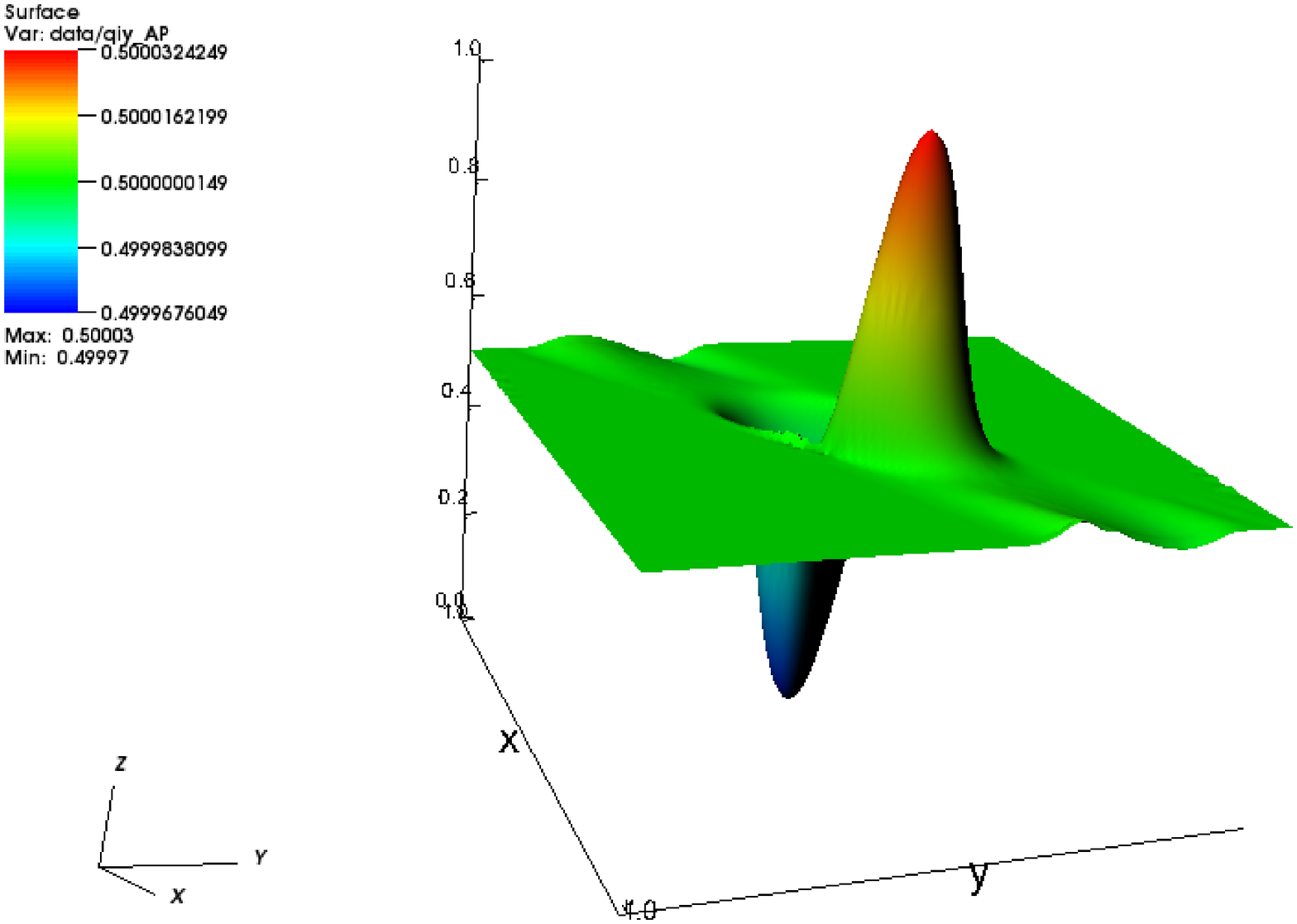} & \includegraphics[scale=0.25]{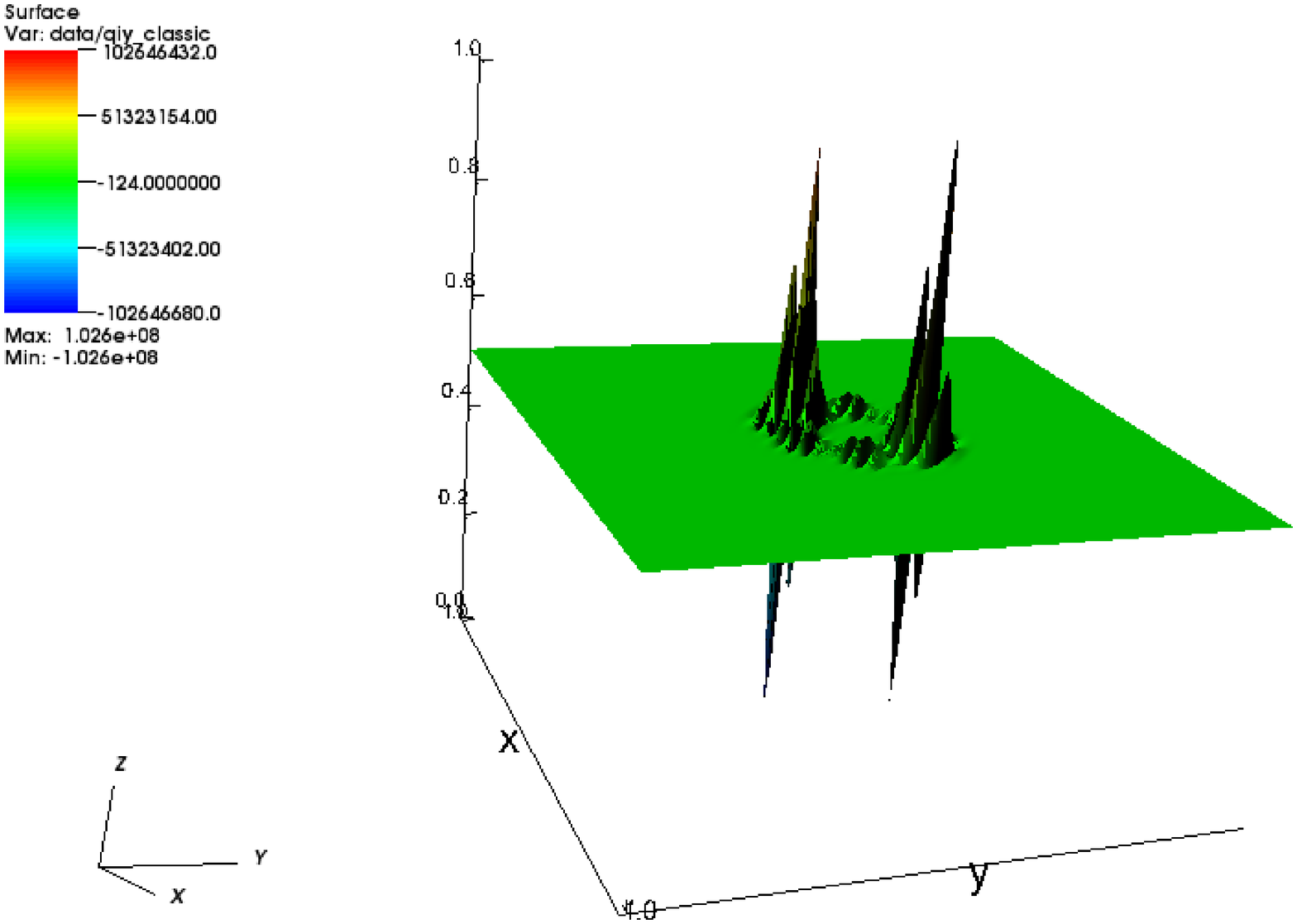} \\
$q_{i,y}^{\tau}$ (AP scheme) & $q_{i,y}^{\tau}$ (classical scheme)
\end{tabular}
\begin{figure}[ht]
\caption{Under-resolved case at time $t = 6 \times 10^{-6}$: $x$ and $y$ components of the ion momentum $\mathbf{q}_{i}^{\tau}$ as functions of $(x,y)$ computed with the AP scheme (left) and the classical scheme (right).} \label{qi_underresolved}
\end{figure}
\end{center}
\begin{center}
\begin{tabular}{cc}
\includegraphics[scale=0.25]{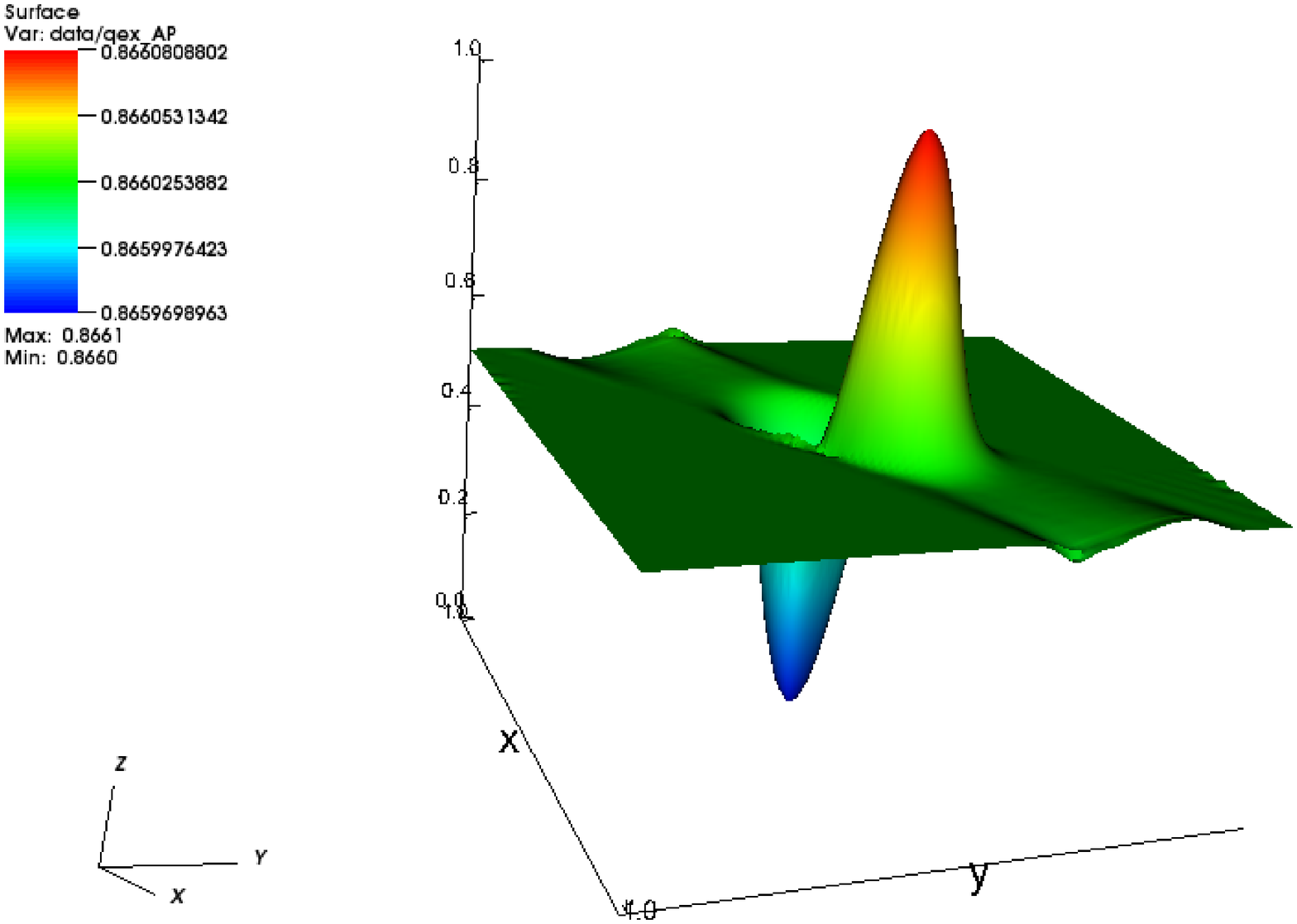} & \includegraphics[scale=0.25]{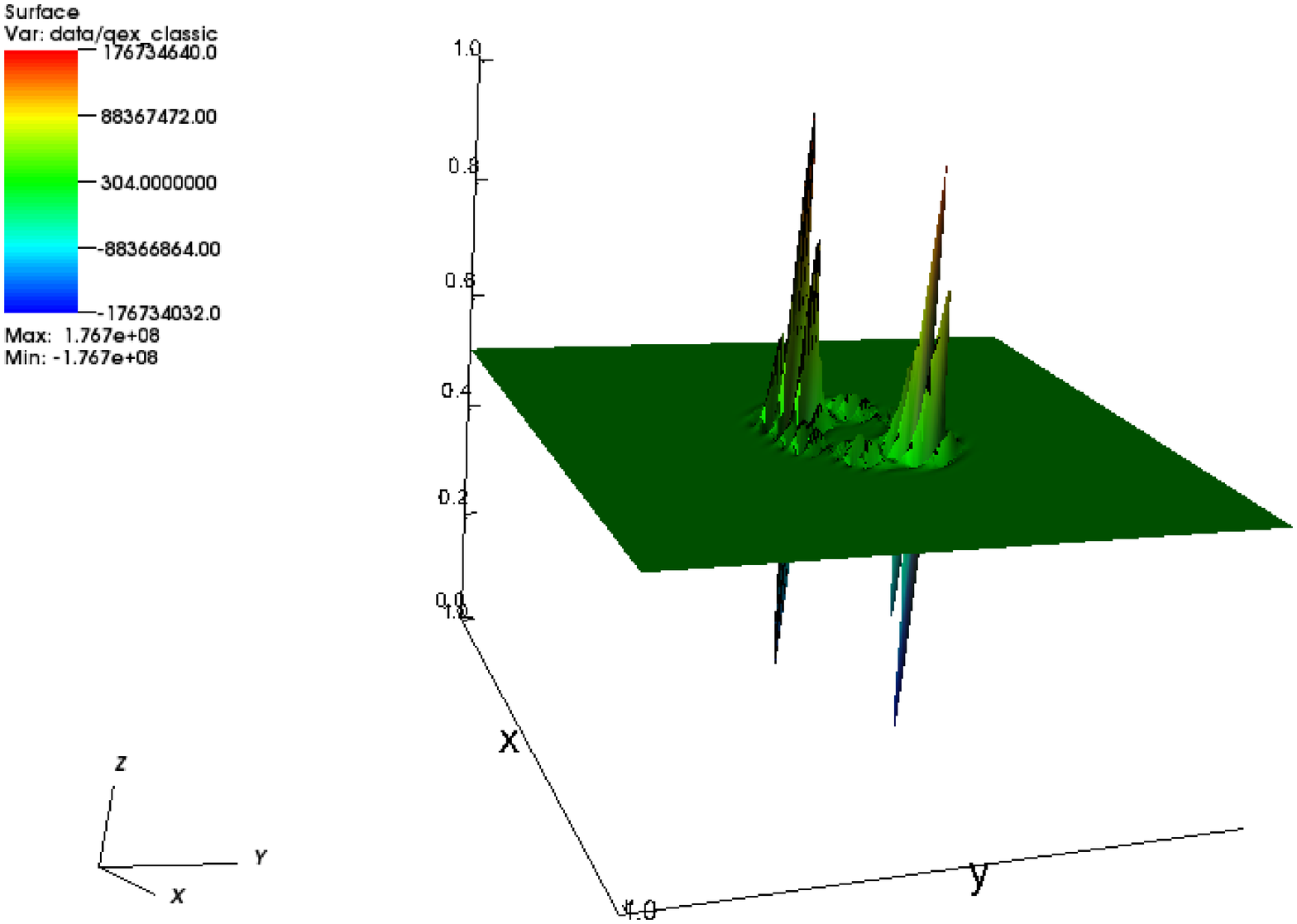} \\
$q_{e,x}^{\tau}$ (AP scheme) & $q_{e,x}^{\tau}$ (classical scheme) \\
\includegraphics[scale=0.25]{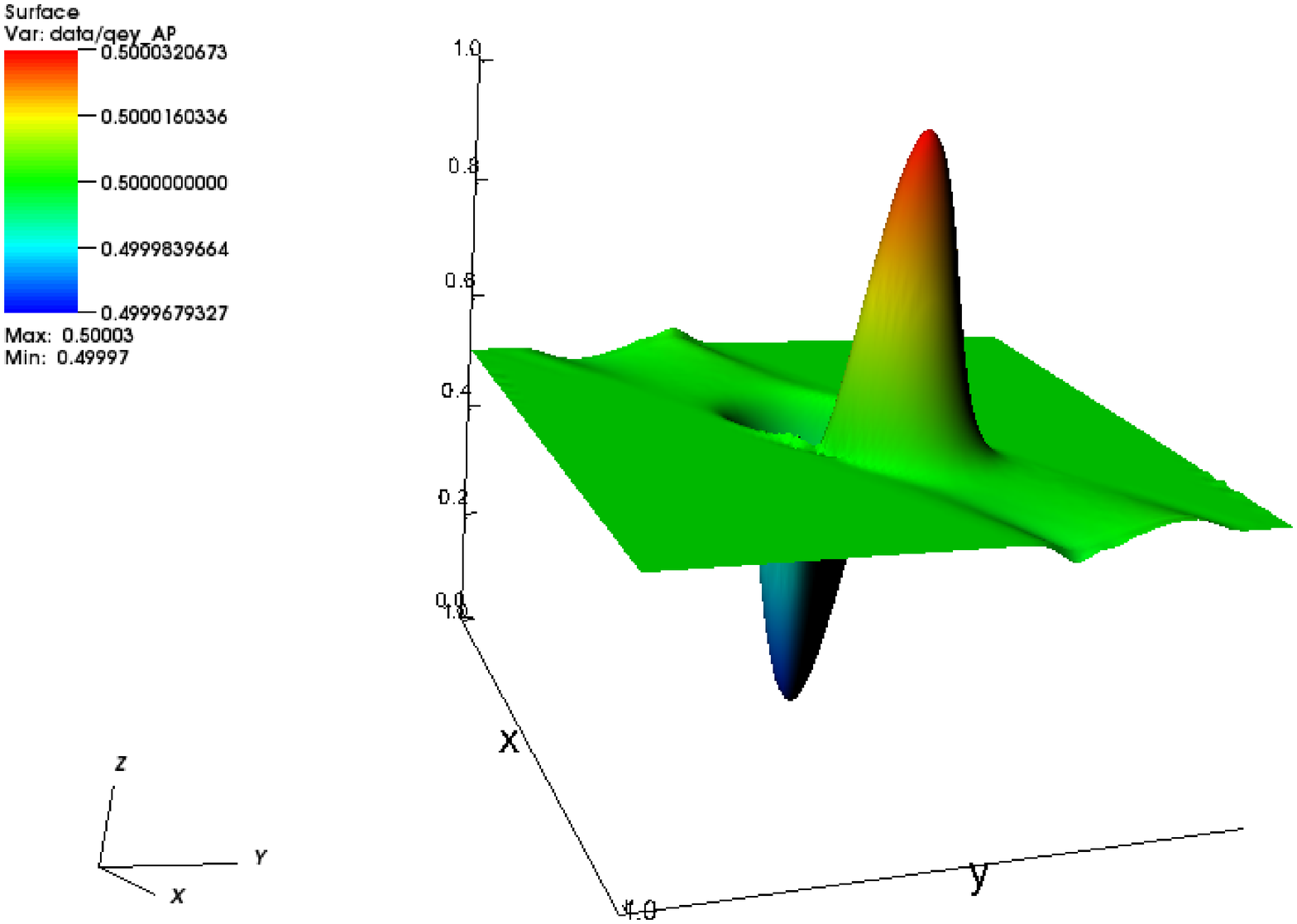} & \includegraphics[scale=0.25]{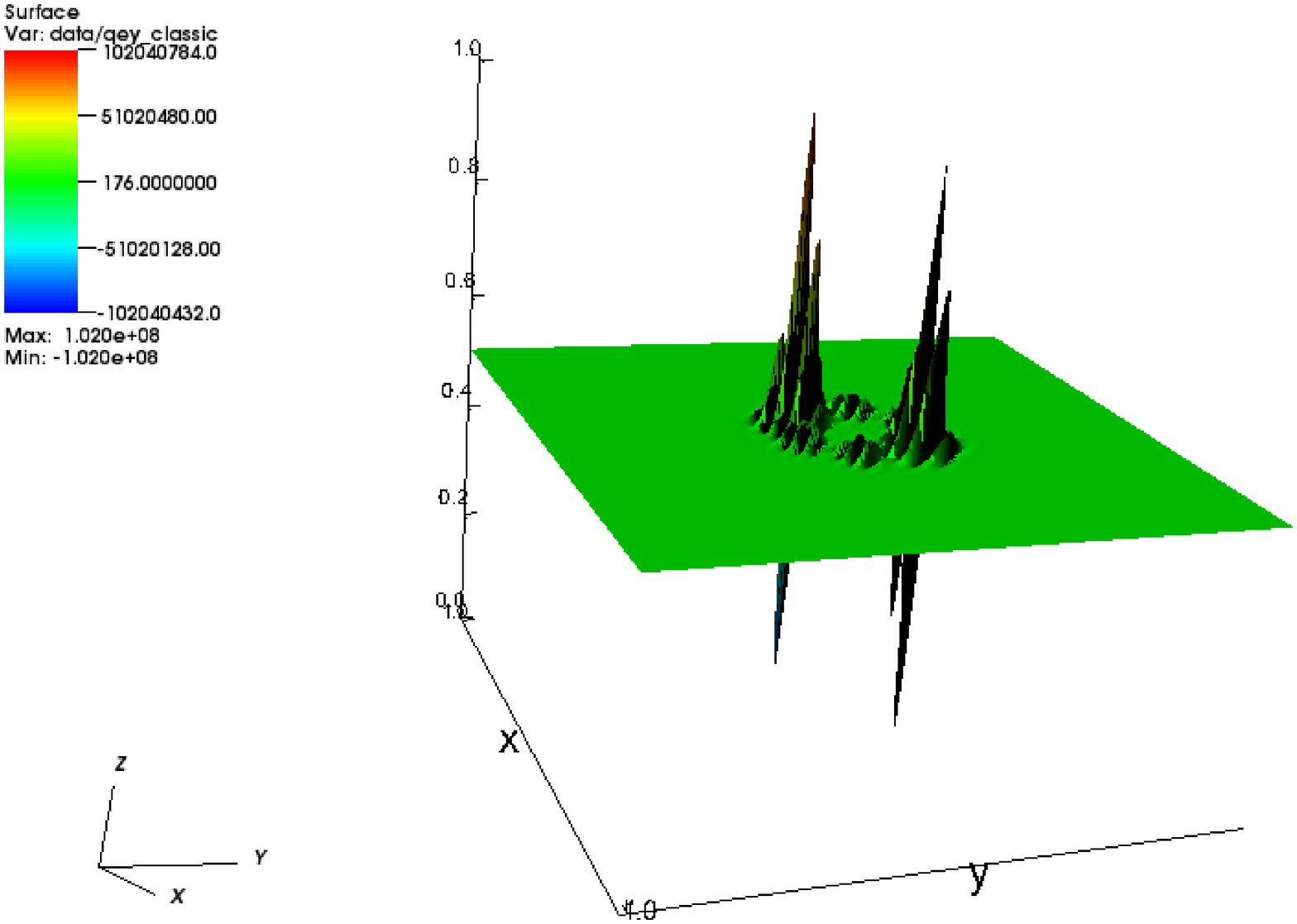} \\
$q_{e,y}^{\tau}$ (AP scheme) & $q_{e,y}^{\tau}$ (classical scheme)
\end{tabular}
\begin{figure}[ht]
\caption{Under-resolved case at time $t = 6 \times 10^{-6}$: $x$ and $y$ components of the electron momentum $\mathbf{q}_{e}^{\tau}$ as functions of $(x,y)$ computed with the AP scheme (left) and the classical scheme (right).} \label{qe_underresolved}
\end{figure}
\end{center}

\indent In Figures \ref{qi_underresolved}-\ref{qe_underresolved}, we present some simulations which are obtained in the under-resolved case, \textit{i.e.} where the time step is taken much larger than the time step which is required to insure the stability of the classical scheme. In the present case, we have chosen $\Delta t = 10^{-6}$, which is 200 times larger than the time step which has been used for the resolved case above. As we can see in these figures, the AP scheme remains stable and produces the same results as in the resolved case. However, the classical scheme blows up after a small number of time iterations, which is not surprising because the time step we have chosen is too large for satisfying the stability condition of this scheme. \\

\indent From this numerical experiment, we can conclude that the AP scheme we have developed for the perturbed two-fluid Euler-Lorentz model (\ref{ELPP_rescaled}) allows us to take a time step which does not satisfy the stability condition required for capturing the fast time variations of the solution. Furthermore, such a time step choice does not penalize the quality of the results which are obtained with the AP scheme.

\subsection{Impact of the perturbation parameter $C$}

Since the AP scheme we have built for the resolution of the perturbed Euler-Lorentz model (\ref{ELPP_rescaled}) has been validated in terms of quality of results, the impact of the value of $C$ needs to be investigated. Indeed, we recall that this parameter is linked with the constants $C_{i}$ and $C_{e}$ by the relations
\begin{equation}
C_{i} = \cfrac{T_{e}\,C}{1+T_{e}} \, , \qquad C_{e} = -\cfrac{T_{e}\,C}{\epsilon\,(1+T_{e})} \, ,
\end{equation}
and these constants are introduced to recover the uniqueness of the solution for the diffusion problem (\ref{eq_phi})-(\ref{BC_nphi_phi}) (see Section 4). Since $C_{i}$ and $C_{e}$ are introduced in (\ref{EL_rescaled}) through a perturbation of the mass conservation equations, these constants are assumed to be as close to 0 as possible. However, this is equivalent to assume that $C$ is close to 0, so the diffusion problem (\ref{ELPP_diffusion_phi})-(\ref{BC_nphi_phi}) is ill-conditioned since $\lambda_{2} = \cfrac{T_{e}\,C}{\Delta t^{2}\,(T_{e}-1)}$ (see Appendix \ref{reformulation_SD_appendix}) and degenerates into the non-unique solution problem (\ref{eq_phi})-(\ref{BC_nphi_phi}) when $C \to 0$. Then, it is necessary to investigate the consequences of the choice of $C$ on the stability of the AP scheme. \\

\begin{figure}[ht]
\begin{tabular}{cc}
\includegraphics[scale=0.24]{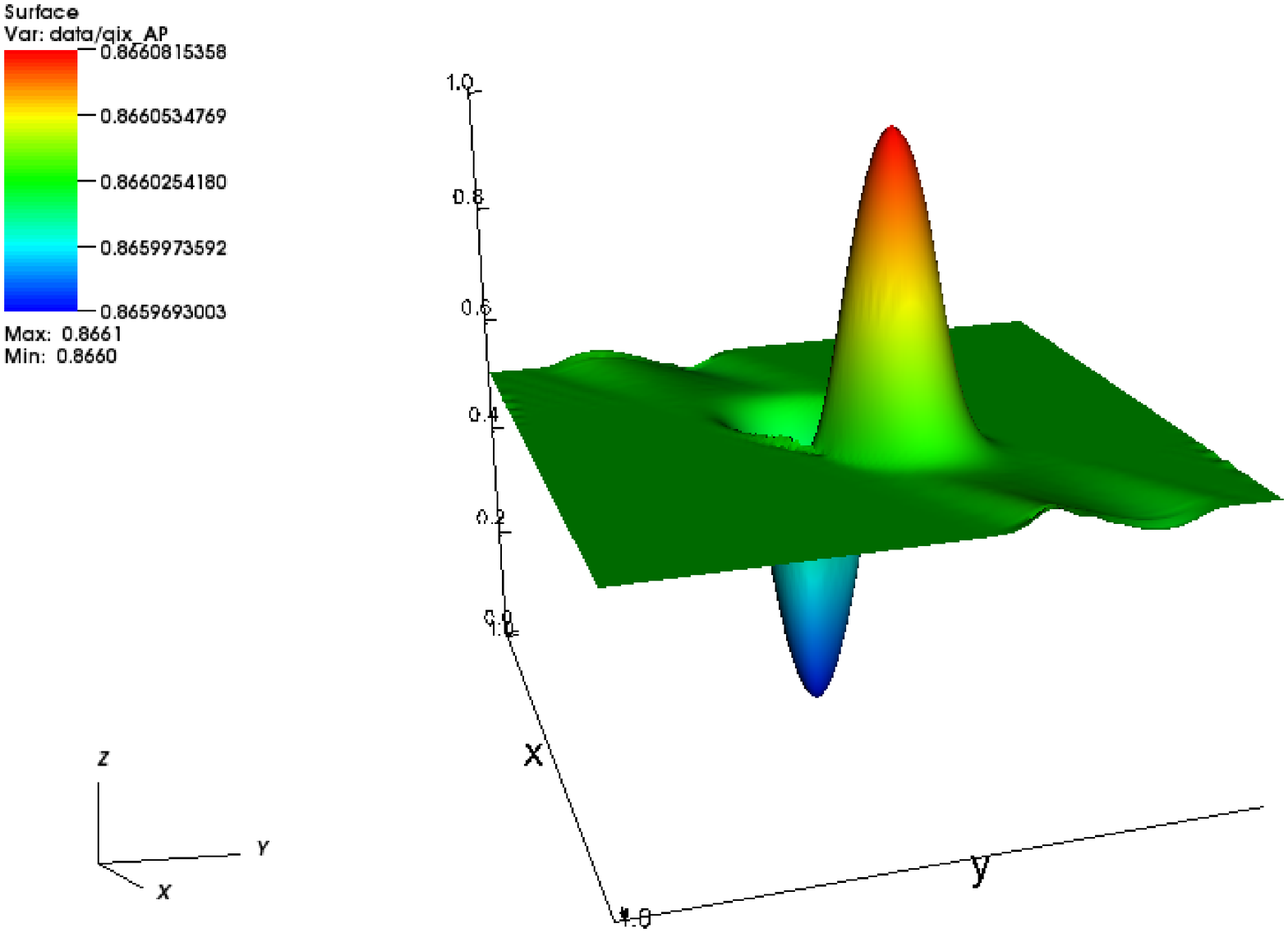} & \includegraphics[scale=0.24]{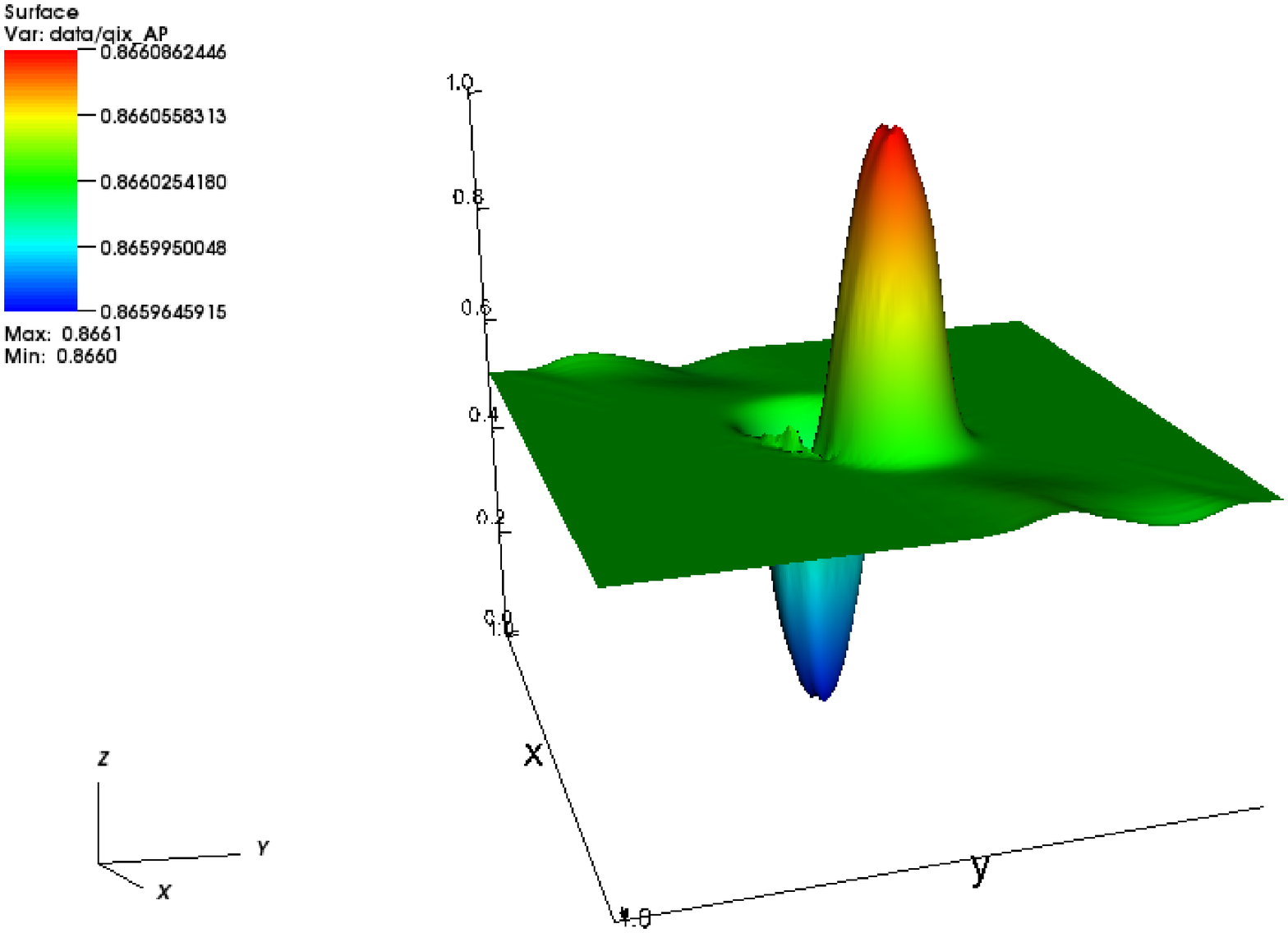} \\
$\Delta t = 10^{-6}$ & $\Delta t = 10^{-7}$
\end{tabular}
\begin{center}
\includegraphics[scale=0.24]{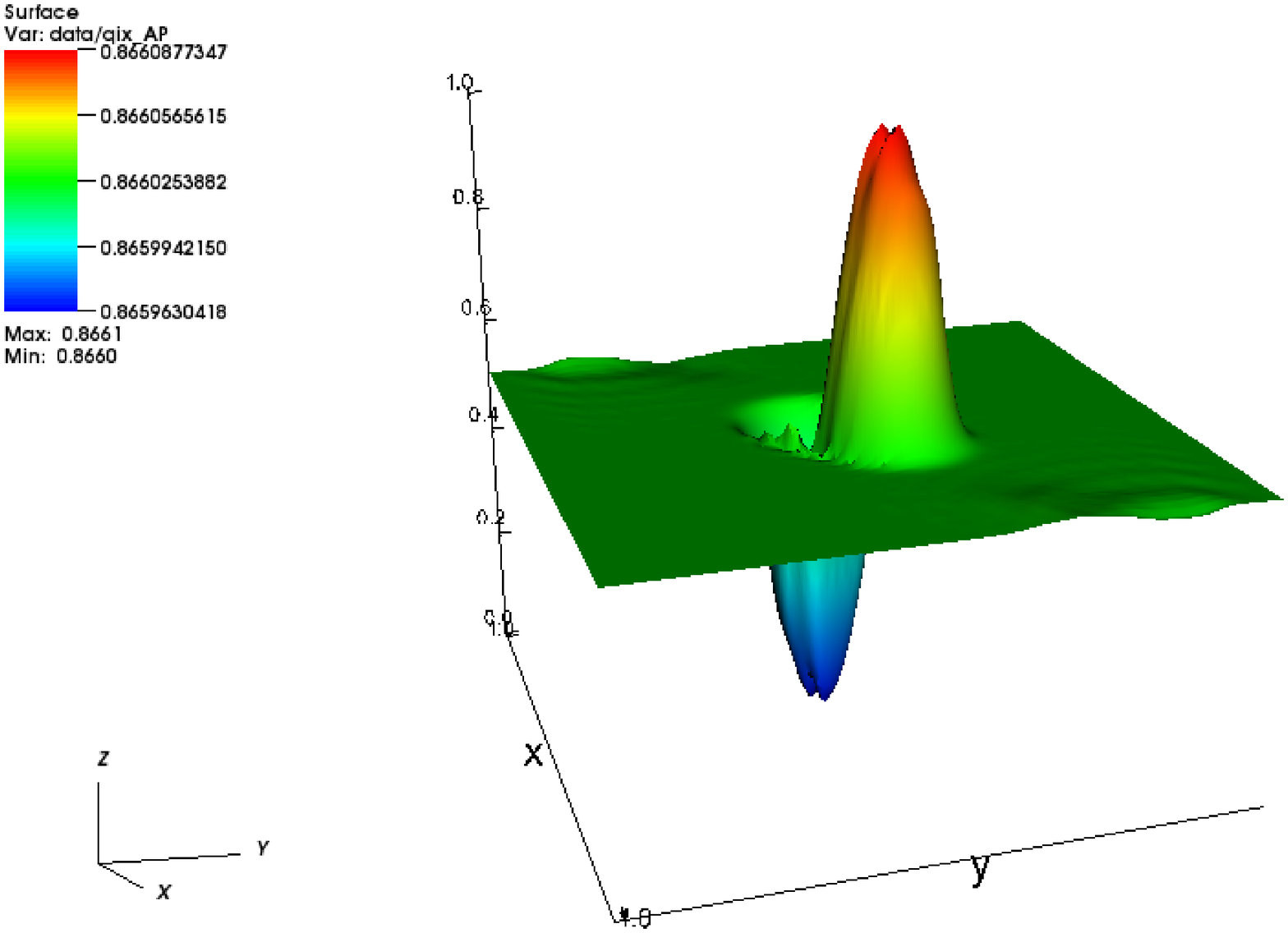} \\
$\Delta t = 10^{-8}$
\end{center}
\caption{$q_{i,x}^{\tau}$ at time $t = 6 \times 10^{-6}$ with $C = 10^{-2}$ and $\Delta t = 10^{-6}, 10^{-7}, 10^{-8}$.} \label{C_10m2}
\end{figure}

\indent In the last test sequence, we run the AP method with the initial datas which have been used in the previous paragraph, \textit{i.e.}
\begin{itemize}
\item We take a $100 \times 100$ uniform mesh over $\Omega = [1,2] \times [1,2]$,
\item The magnetic field is uniform and is defined as $\mathbf{B} = (\sin\alpha,-\cos\alpha,0)$ with $\alpha = \frac{2\pi}{3}$,
\item The initial electric potential is $\phi^{\tau,0} = \phi_{0}$ with $\phi_{0} = 0$,
\item The initial ion momentum and the initial electron momentum are defined by $\mathbf{q}_{i}^{\tau,0} = \mathbf{q}_{e}^{\tau,0} = \mathbf{B}$,
\item The initial density is $n^{\tau,0}$ defined by
\begin{equation}
n^{\tau,0}(x,y) = n_{0} + \tau\,\max \big( 0, 1-\eta\,(x-x_{0})^{2}-\eta\,(y-y_{0})^{2}\big) \, ,
\end{equation}
with $n_{0} = 1$ constant, $\eta = 80$ and $(x_{0},y_{0}) = (\frac{3}{2},\frac{3}{2})$,
\item We choose $T_{e} = 3$, $\epsilon = 1$ and $\tau = 10^{-8}$.
\end{itemize}

\begin{figure}[ht]
\begin{tabular}{cc}
\includegraphics[scale=0.24]{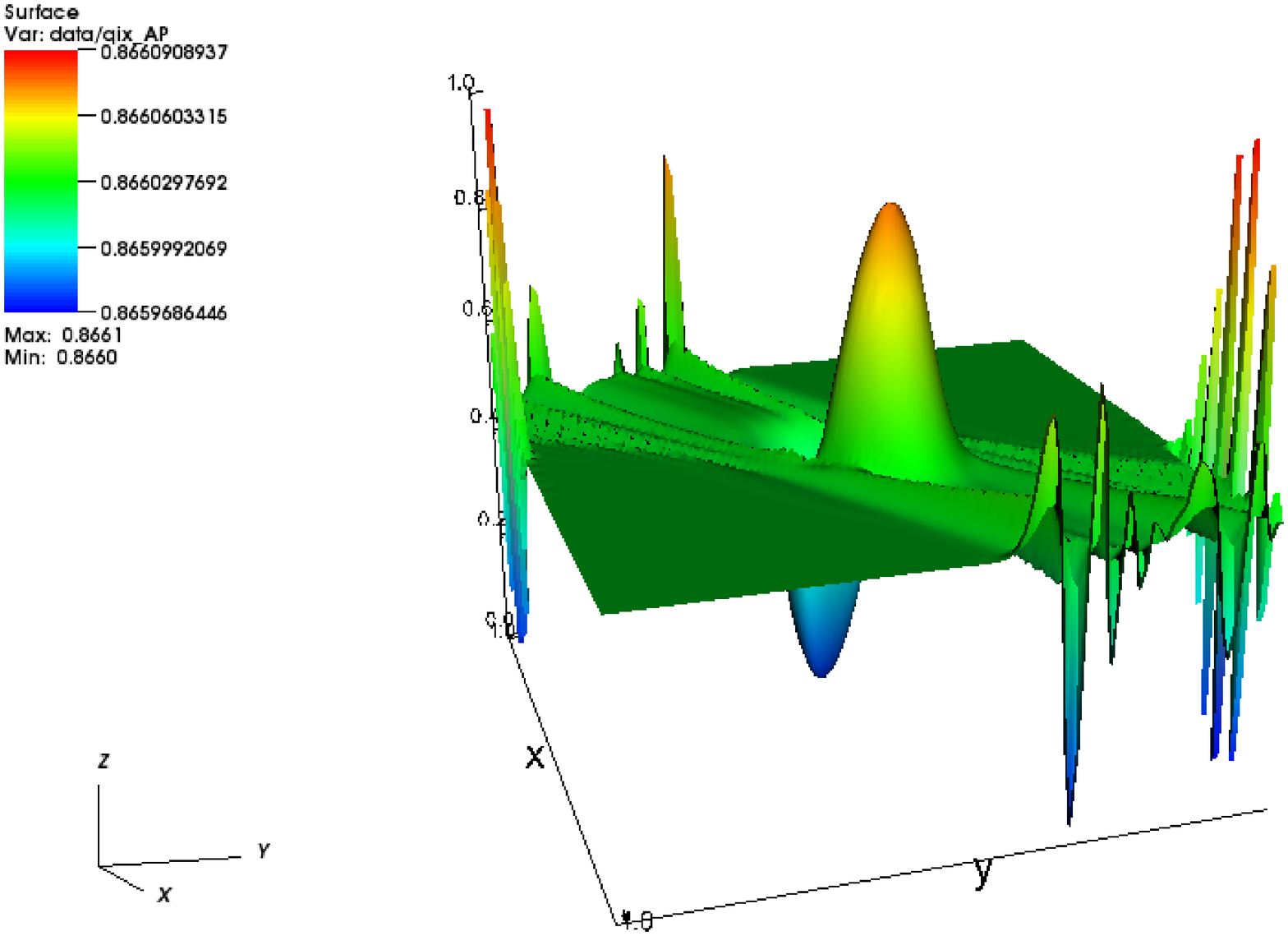} & \includegraphics[scale=0.24]{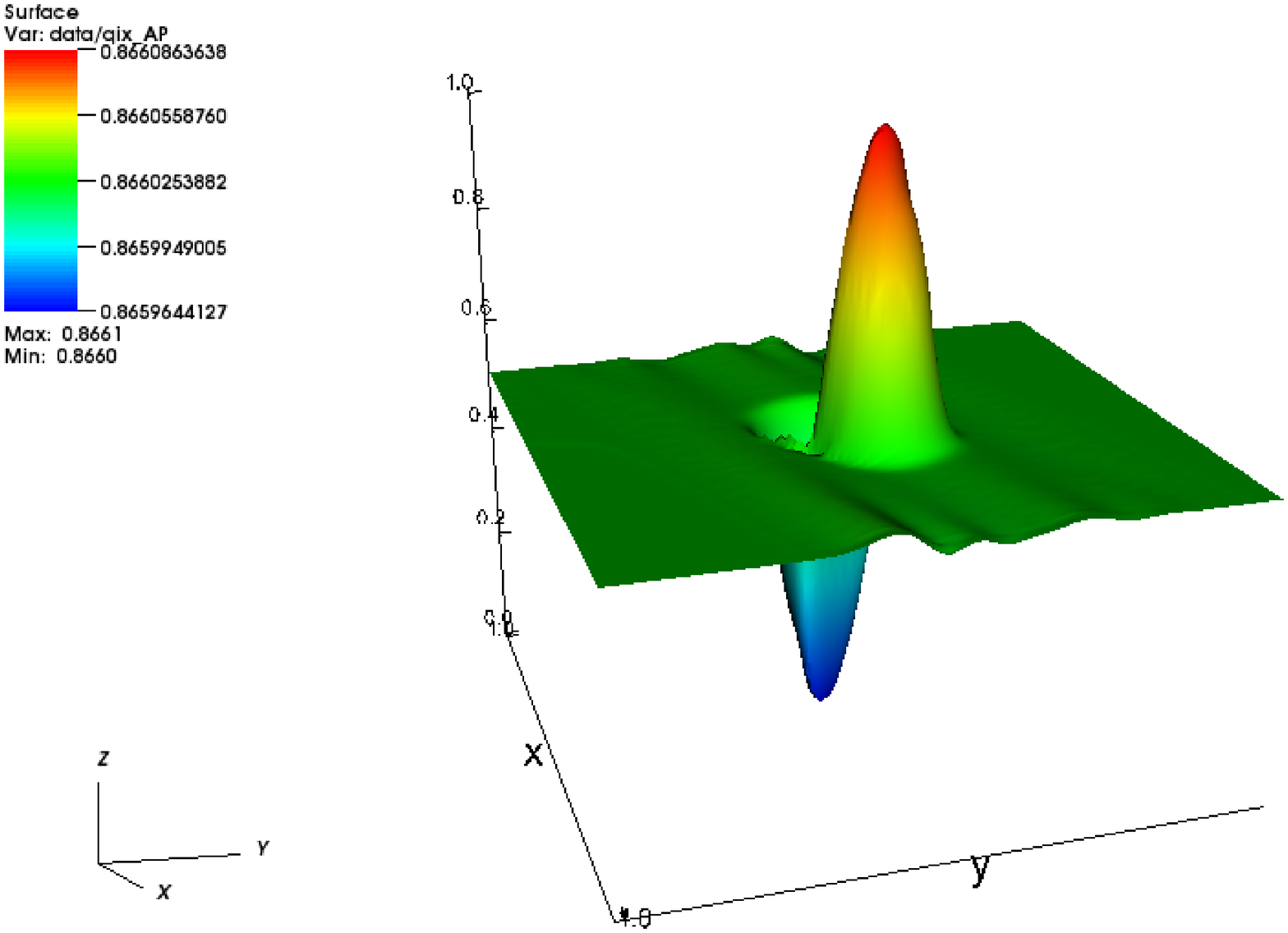} \\
$\Delta t = 10^{-6}$ & $\Delta t = 10^{-7}$
\end{tabular}
\begin{center}
\includegraphics[scale=0.24]{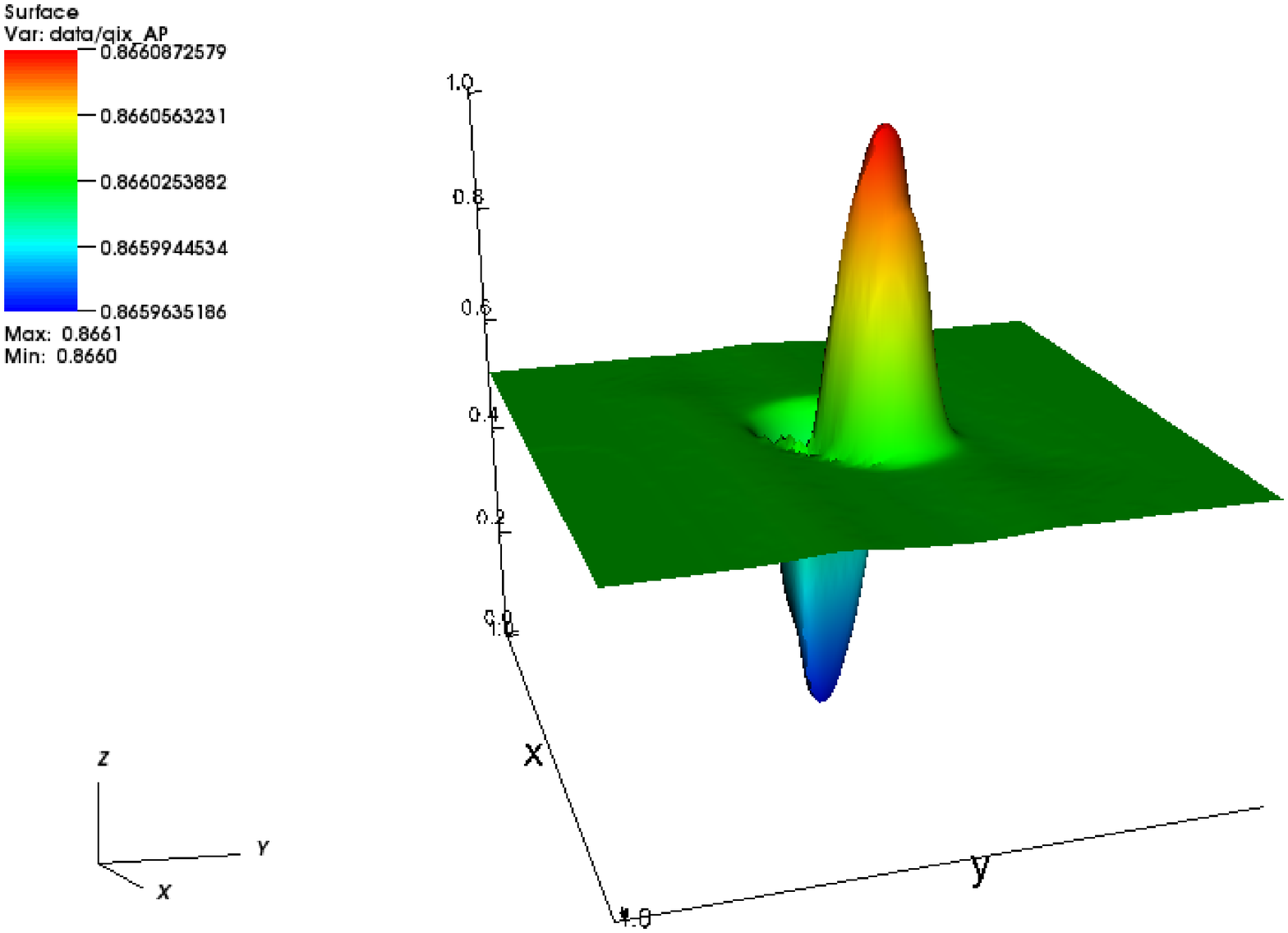} \\
$\Delta t = 10^{-8}$
\end{center}
\caption{$q_{i,x}^{\tau}$ at time $t = 4 \times 10^{-6}$ with $C = 10^{-3}$ and $\Delta t = 10^{-6}, 10^{-7}, 10^{-8}$.} \label{C_10m3}
\end{figure}

\begin{figure}[ht]
\begin{tabular}{cc}
\includegraphics[scale=0.24]{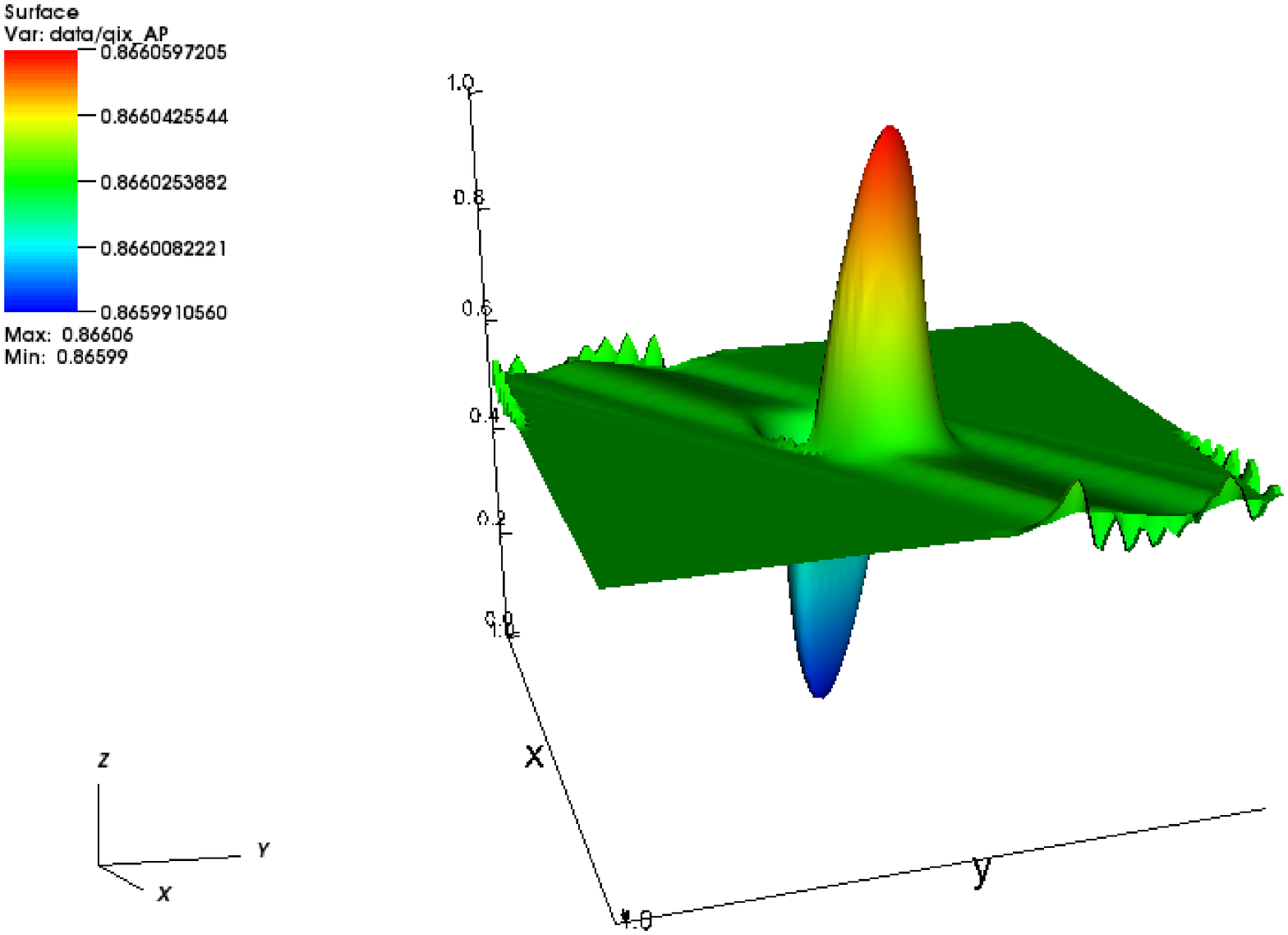} & \includegraphics[scale=0.24]{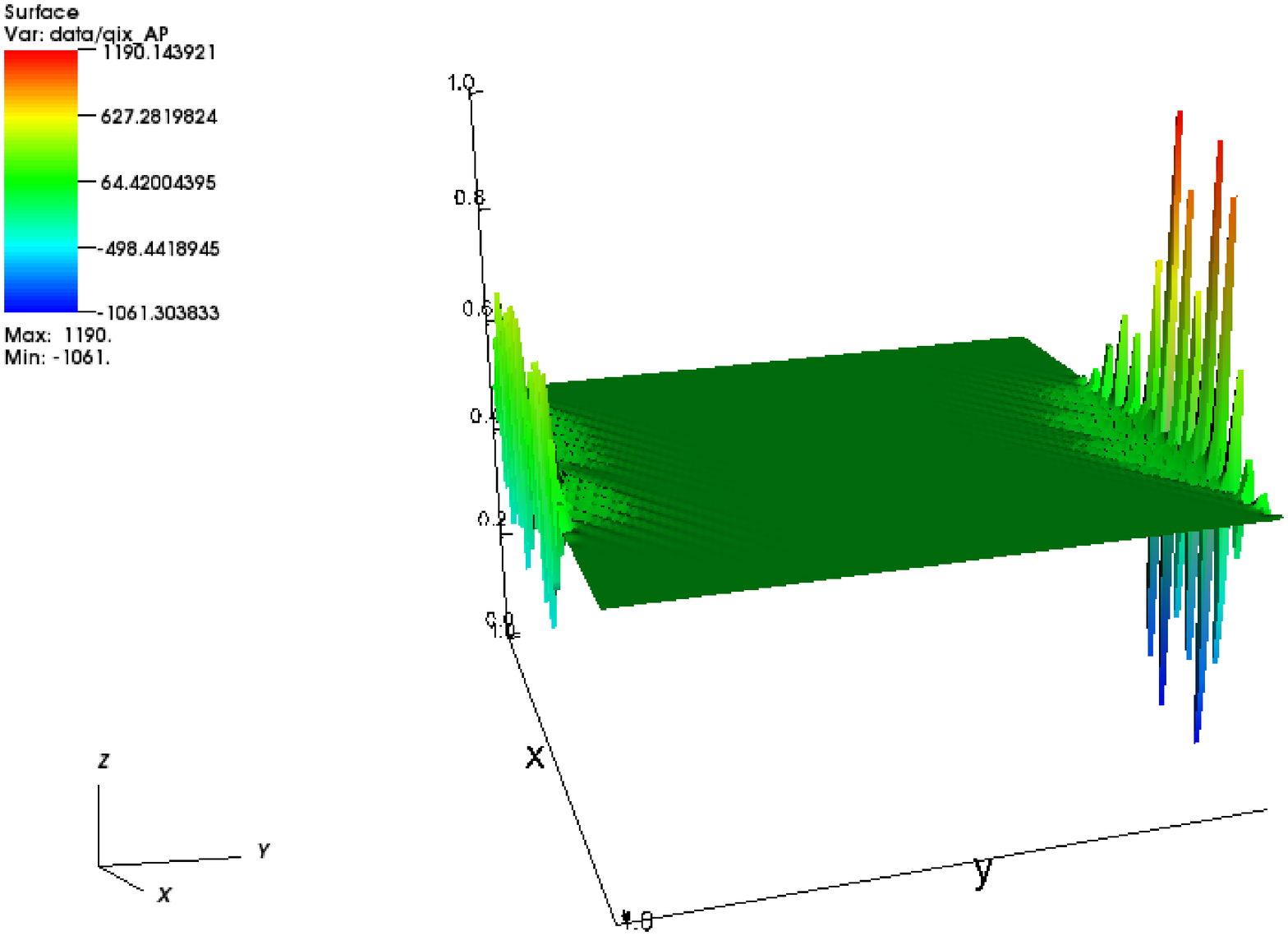} \\
$\Delta t = 10^{-6}$ & $\Delta t = 10^{-7}$
\end{tabular}
\begin{center}
\includegraphics[scale=0.24]{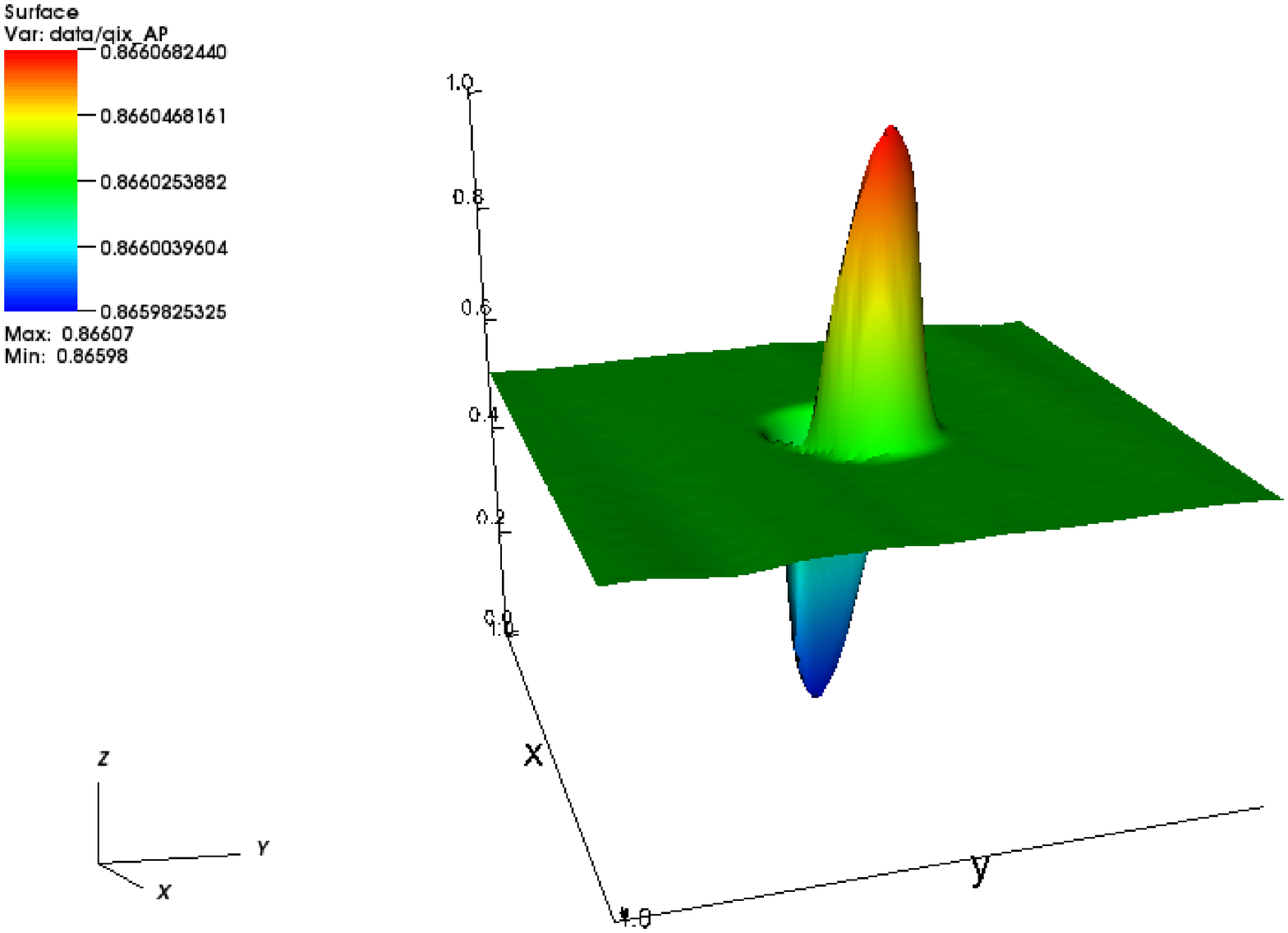} \\
$\Delta t = 10^{-8}$
\end{center}
\caption{$q_{i,x}^{\tau}$ at time $t = 2 \times 10^{-6}$ with $C = 10^{-4}$ and $\Delta t = 10^{-6}, 10^{-7}, 10^{-8}$.} \label{C_10m4}
\end{figure}

\indent In Figure \ref{C_10m2}, we plot the $x$-component of the ion momentum $\mathbf{q}_{i}^{\tau}$ which is obtained with the AP scheme (\ref{semi-discrete_ELPP}) at time $t = 6 \times 10^{-6}$ with $C = 10^{-2}$ and a time step $\Delta t$ which is equal to $10^{-6}$, $10^{-7}$ and $10^{-8}$. We can remark that the AP method is stable for $\Delta t \leq 10^{-6}$ since all the results are similar. \\
\indent Now, we do again this numerical experiment with $C = 10^{-3}$ instead of $C = 10^{-2}$. As we can remark in Figure \ref{C_10m3} where $q_{i,x}^{\tau}$ is plotted at time $t = 4 \times 10^{-6}$, the AP method is stable with $\Delta t = 10^{-7}$ and $\Delta t = 10^{-8}$. However, we observe some important boundary effects in the $\Delta t = 10^{-6}$ case. \\
\indent Finally, we perform this numerical experiment with $C = 10^{-4}$. In Figure \ref{C_10m4}, we plot $q_{i,x}^{\tau}$ at time $t = 2 \times 10^{-6}$ with $\Delta t = 10^{-6}, 10^{-7}, 10^{-8}$. No numerical artifacts are present in the $\Delta t = 10^{-8}$ case whereas we observe some boundary effects if $\Delta t = 10^{-6}$ and even the blowing up of the method when $\Delta t = 10^{-7}$ is considered.

\indent The results which are provided by the whole test sequence above indicate that we have a stability condition for the AP scheme which clearly depends on $C$ and which would be of the form
\begin{equation}
\Delta t = \mathcal{O}(C) \, .
\end{equation}
As a consequence, we can say that the AP scheme we have developed for the perturbed two-fluid Euler-Lorentz model (\ref{ELPP_rescaled}) is Asymptotic-Preserving when $C > 0$ is fixed and when $\tau \to 0$. However, it is not Asymptotic-Preserving when $C \to 0$ and $\tau > 0$ is fixed.

\section{Conclusions and perspectives}

\indent In this paper, we studied the isothermal two-fluid Euler-Lorentz system coupled with a quasi-neutrality constraint in a low Mach number regime and a strong magnetic field regime. After having presented the model and its limit regime, we proposed a reformulation of this model which is compatible with the construction of an Asymptotic-Preserving scheme. Then we presented the time semi-discrete AP scheme itself and its reformulation leading to the resolution of some anisotropic diffusion equations for $n^{\tau,m+1}$ and $\phi^{\tau,m+1}$. The equation for $\phi^{\tau,m+1}$ being ill-posed, we restored the uniqueness of the solution of this equation by introducing a small perturbation in the mass conservation equations. Finally, we performed some numerical tests of this scheme by comparing the results from the AP scheme to those from a fully explicit method and we tested the influence of the perturbation parameter $C$ on the behaviour of the AP scheme. \\
\indent At this point, several work pathes can be investigated. The first one is to go back to the time semi-discretization described in section 4.1 by invoking boundary conditions for the computation of the electric potential which are different from Neumann conditions. The second one is to generalize the present work to some two-fluid Euler-Lorentz models involving other pressure laws for $p_{i}$ and $p_{e}$.

\appendix

\section{Reformulation of the Euler-Lorentz model} \label{reformulation_continuous}
\setcounter{equation}{0}

In this paragraph, we detail the reformulation procedure of the Euler-Lorentz model
\begin{subnumcases}{\label{EL_rescaled_appendix}}
\displaystyle \D_{t}n^{\tau} + \nabla_{\mathbf{x}} \cdot \mathbf{q}_{\alpha}^{\tau} = 0 \, , & \label{EL_rescaled_appendix_n} \\
\begin{split}
\displaystyle \epsilon_{\alpha}\,\tau\,\Big[\D_{t}\mathbf{q}_{\alpha}^{\tau} + \nabla_{\mathbf{x}} \cdot \Big( \cfrac{\mathbf{q}_{\alpha}^{\tau} \otimes \mathbf{q}_{\alpha}^{\tau}}{n^{\tau}} \Big) \Big] + &T_{\alpha}\,\nabla_{\mathbf{x}}n^{\tau} \\
&= \mathfrak{q}_{\alpha}\, \big[- n^{\tau} \, \nabla_{\mathbf{x}}\phi^{\tau} + \mathbf{q}_{\alpha}^{\tau} \times \mathbf{B} \big] \, ,
\end{split}
& \label{EL_rescaled_appendix_q} \\
\alpha \in \{i,e\}\, , &
\end{subnumcases}
leading to the model (\ref{EL_reformulated_final}). First, we separate the parallel and perpendicular parts of (\ref{EL_rescaled_appendix_q}) for each value of $\alpha$: we obtain
\begin{subnumcases}{\label{EL_tau_reformulated_decompo_appendix}}
\displaystyle \D_{t}n^{\tau} + \nabla_{\mathbf{x}} \cdot (\mathbf{q}_{\alpha}^{\tau})_{||} + \nabla_{\mathbf{x}} \cdot (\mathbf{q}_{\alpha}^{\tau})_{\perp} = 0 \, , & \label{EL_tau_reformulated_decompo_appendix_n} \\
\begin{split}
\D_{t}\big((\mathbf{q}_{\alpha}^{\tau})_{||}\big) - \big(\D_{t}(&\mathbf{b}\otimes\mathbf{b})\big) \, \mathbf{q}_{\alpha}^{\tau} + (\mathbf{b}\otimes\mathbf{b}) \, \nabla_{\mathbf{x}} \cdot \Big( \cfrac{\mathbf{q}_{\alpha}^{\tau} \otimes \mathbf{q}_{\alpha}^{\tau}}{n^{\tau}} \Big) \\
& + \cfrac{1}{\epsilon_{\alpha}\,\tau}\, (\mathbf{b}\otimes\mathbf{b})\, \big(T_{\alpha}\,\nabla_{\mathbf{x}}n^{\tau} + \mathfrak{q}_{\alpha}\,n^{\tau} \, \nabla_{\mathbf{x}}\phi^{\tau} \big) = 0 \, ,
\end{split}
& \label{EL_tau_reformulated_decompo_appendix_qpara} \\
\begin{split}
(\mathbf{q}_{\alpha}^{\tau})_{\perp} = \cfrac{1}{\|\mathbf{B}\|} \, \mathbf{b} \times \big(\mathfrak{q}_{\alpha}\,& T_{\alpha}\,\nabla_{\mathbf{x}}n^{\tau} + n^{\tau}\,\nabla_{\mathbf{x}}\phi^{\tau}\big) \\
&+ \cfrac{\mathfrak{q}_{\alpha}\,\epsilon_{\alpha}\,\tau}{\|\mathbf{B}\|} \,\mathbf{b} \times \Big[\D_{t}\mathbf{q}_{\alpha}^{\tau} + \nabla_{\mathbf{x}} \cdot \Big(\cfrac{\mathbf{q}_{\alpha}^{\tau} \otimes \mathbf{q}_{\alpha}^{\tau}}{n^{\tau}} \Big) \Big] \, , 
\end{split}
& \label{EL_tau_reformulated_decompo_appendix_qperp} \\
\alpha \in \{i,e\} \, . &
\end{subnumcases}
In order to obtain (\ref{EL_reformulated_final_zero_force}) for any $\alpha \in \{i,e\}$, we compute the divergence in space of (\ref{EL_tau_reformulated_decompo_appendix_qpara}) on one hand and the derivative in time of (\ref{EL_tau_reformulated_decompo_appendix_n}) on the other hand. We obtain
\begin{equation}
\begin{split}
\nabla_{\mathbf{x}} \cdot \big(\D_{t}\big((\mathbf{q}_{\alpha}^{\tau})_{||}\big)\big) &- \nabla_{\mathbf{x}}\cdot \Big(\big(\D_{t}(\mathbf{b}\otimes\mathbf{b})\big) \, \mathbf{q}_{\alpha}^{\tau} - (\mathbf{b}\otimes\mathbf{b}) \, \nabla_{\mathbf{x}} \cdot \Big( \cfrac{\mathbf{q}_{\alpha}^{\tau} \otimes \mathbf{q}_{\alpha}^{\tau}}{n^{\tau}} \Big) \Big) \\
&+ \cfrac{1}{\epsilon_{\alpha}\,\tau}\, \nabla_{\mathbf{x}} \cdot \big( (\mathbf{b}\otimes\mathbf{b})\, \big(T_{\alpha}\,\nabla_{\mathbf{x}}n^{\tau} + \mathfrak{q}_{\alpha}\,n^{\tau} \, \nabla_{\mathbf{x}}\phi^{\tau} \big) \big) = 0 \, ,
\end{split}
\end{equation}
and
\begin{equation}
\D_{t}^{2}n^{\tau} + \D_{t}\nabla_{\mathbf{x}} \cdot (\mathbf{q}_{\alpha}^{\tau})_{||} + \D_{t}\nabla_{\mathbf{x}} \cdot (\mathbf{q}_{\alpha}^{\tau})_{\perp} = 0 \, ,
\end{equation}
for any $\alpha \in \{i,e\}$. By doing some linear combinations of these equations, we obtain a system of 2 equations for $n^{\tau}$ and $\phi^{\tau}$ of the form
\begin{equation} \label{diffusion_continuous_nphi_tau}
\hspace{-0.2cm} \left\{
\begin{array}{l}
\displaystyle \D_{t}^{2}n^{\tau} - \cfrac{1}{\epsilon_{\alpha}\,\tau}\, \nabla_{\mathbf{x}} \cdot \big( (\mathbf{b}\otimes\mathbf{b})\, \big(T_{\alpha}\,\nabla_{\mathbf{x}}n^{\tau} + \mathfrak{q}_{\alpha}\,n^{\tau} \, \nabla_{\mathbf{x}}\phi^{\tau} \big) \big) \\
\displaystyle \quad = \nabla_{\mathbf{x}}\cdot \Big(\big(\D_{t}(\mathbf{b}\otimes\mathbf{b})\big) \, \mathbf{q}_{\alpha}^{\tau} - (\mathbf{b}\otimes\mathbf{b}) \, \nabla_{\mathbf{x}} \cdot \Big( \cfrac{\mathbf{q}_{\alpha}^{\tau} \otimes \mathbf{q}_{\alpha}^{\tau}}{n^{\tau}} \Big) -\D_{t}((\mathbf{q}_{\alpha}^{\tau})_{\perp}) \Big) \, , \\
\alpha \in \{i,e\} \, ,
\end{array}
\right.
\end{equation}
which are exactly the equations (\ref{EL_reformulated_final_zero_force}). Then the Euler-Lorentz model (\ref{EL_rescaled_appendix}) is equivalent to the combination of (\ref{EL_tau_reformulated_decompo_appendix_qpara}), (\ref{EL_tau_reformulated_decompo_appendix_qperp}) and (\ref{diffusion_continuous_nphi_tau}).

\section{Reformulation of the semi-discrete problem} \label{reformulation_SD_appendix}
\setcounter{equation}{0}

This paragraph is devoted to the reformulation of the semi-discrete problems (\ref{semi-discrete_origin}) and (\ref{ELPP_rescaled}). Since the model (\ref{semi-discrete_origin}) is not more than (\ref{ELPP_rescaled}) with $C_{i} = C_{e} = 0$, we present the reformulation of the semi-discrete scheme for the perturbed Euler-Lorentz model. This scheme is recalled here:
\begin{subnumcases}{\label{semi-discrete_ELPP_appendix}}
\begin{split}
\cfrac{n^{\tau,m+1}-n^{\tau,m}}{\Delta t} + C_{\alpha}\, &\cfrac{\phi^{\tau,m+1}-\phi^{\tau,m}}{\Delta t} \\
& + \nabla_{\mathbf{x}} \cdot \big( (\mathbf{b}^{m+1} \otimes \mathbf{b}^{m+1}) \, \mathbf{q}_{\alpha}^{\tau,m+1} \big) \\
&+ \nabla_{\mathbf{x}} \cdot \big((\mathbb{I}-\mathbf{b}^{m+1} \otimes \mathbf{b}^{m+1})\,\mathbf{q}_{\alpha}^{\tau,m}\big) = 0 \, ,
\end{split}
& \label{semi-discrete_ELPP_n_appendix} \\
\begin{split}
\cfrac{\mathbf{q}_{\alpha}^{\tau,m+1}-\mathbf{q}_{\alpha}^{\tau,m}}{\Delta t} &+ \nabla_{\mathbf{x}} \cdot \Big(\cfrac{\mathbf{q}_{\alpha}^{\tau,m} \otimes \mathbf{q}_{\alpha}^{\tau,m}}{n^{\tau,m}} \Big) + \cfrac{T_{\alpha}}{\epsilon_{\alpha}\,\tau}\, \nabla_{\mathbf{x}}n^{\tau,m+1} \\
&= \cfrac{\mathfrak{q}_{\alpha}}{\epsilon_{\alpha}\,\tau}\, \big[ - n^{\tau,m+1}\,\nabla_{\mathbf{x}}\phi^{\tau,m+1} + \mathbf{q}_{\alpha}^{\tau,m+1} \times \mathbf{B}^{m+1} \big] \, ,
\end{split}
& \label{semi-discrete_ELPP_q_appendix} \\
\alpha \in \{i,e\} \, .
\end{subnumcases}

\indent First, we separate the parallel and the perpendicular parts of (\ref{semi-discrete_ELPP_q_appendix}) according to $\mathbf{b}^{m+1}$. More precisely, we obtain the equations (\ref{SD_qperp}) by performing the vector product of $\mathbf{b}^{m+1}$ by (\ref{semi-discrete_ELPP_q_appendix}). Concerning the equations (\ref{SD_qpara}), we obtain them by multiplying the tensor $(\mathbf{b}^{m+1} \otimes \mathbf{b}^{m+1})$ by (\ref{semi-discrete_ELPP_q_appendix}). \\
\indent In order to obtain the diffusion equations (\ref{ELPP_diffusion_n}) and (\ref{ELPP_diffusion_phi}), we put (\ref{SD_qpara}) with $\alpha = i$ (resp. $\alpha = e$) in (\ref{semi-discrete_ELPP_n_appendix}) with $\alpha = i$ (resp. $\alpha = e$). We obtain a system of two diffusion equations satisfied by $(n^{\tau,m+1},\phi^{\tau,m+1})$:
\begin{equation}
\left\{
\begin{array}{l}
\begin{split}
&\cfrac{n^{\tau,m+1}-n^{\tau,m}}{\Delta t} + C_{\alpha}\,\cfrac{\phi^{\tau,m+1}-\phi^{\tau,m}}{\Delta t} + \nabla_{\mathbf{x}} \cdot \mathbf{q}_{\alpha}^{\tau,m} \\
&\quad - \Delta t \, \nabla_{\mathbf{x}} \cdot \Big( (\mathbf{b}^{m+1} \otimes \mathbf{b}^{m+1}) \, \nabla_{\mathbf{x}} \cdot \Big( \cfrac{\mathbf{q}_{\alpha}^{\tau,m} \otimes \mathbf{q}_{\alpha}^{\tau,m}}{n^{\tau,m}} \Big) \Big) \\
&\quad - \cfrac{\Delta t}{\epsilon_{\alpha}\,\tau} \, \nabla_{\mathbf{x}} \cdot \Big((\mathbf{b}^{m+1} \otimes \mathbf{b}^{m+1}) \, \big( T_{\alpha}\,\nabla_{\mathbf{x}} n^{\tau,m+1} \\
&\qquad \qquad \qquad \qquad \qquad \qquad \qquad \quad + \mathfrak{q}_{\alpha}\,n^{\tau,m+1} \, \nabla_{\mathbf{x}} \phi^{\tau,m+1}\big) \Big) = 0 \, ,
\end{split}
\\
\alpha \in \{i,e\} \, .
\end{array}
\right.
\end{equation}
By doing some linear combinations, these diffusion equations write
\begin{equation}
\begin{split}
-&\nabla_{\mathbf{x}} \cdot \big( (\mathbf{b}^{m+1} \otimes \mathbf{b}^{m+1})\, \nabla_{\mathbf{x}} n^{\tau,m+1}\big) \\
& \qquad \qquad + \tau\, \cfrac{1+\epsilon}{\Delta t^{2} \, (1+T_{e})} \, n^{\tau,m+1} + \tau \, \cfrac{C_{i} + \epsilon\,C_{e}}{\Delta t^{2}\, (1+T_{e})} \, \phi^{\tau,m+1} \\
&= \cfrac{\tau}{1+T_{e}} \, \Bigg[ -\cfrac{1}{\Delta t} \, \nabla_{\mathbf{x}} \cdot (\mathbf{q}_{i}^{\tau,m}+\epsilon\,\mathbf{q}_{e}^{\tau,m}) + \cfrac{1+\epsilon}{\Delta t^{2}} \, n^{\tau,m} + \cfrac{C_{i}+\epsilon\,C_{e}}{\Delta t^{2}} \, \phi^{\tau,m} \\
& \qquad \qquad \qquad + \nabla_{\mathbf{x}} \cdot \Big( (\mathbf{b}^{m+1} \otimes \mathbf{b}^{m+1})\, \big[ \nabla_{\mathbf{x}} \cdot (\cfrac{\mathbf{q}_{i}^{\tau,m} \otimes \mathbf{q}_{i}^{\tau,m}}{n^{\tau,m}}) \\
& \qquad \qquad \qquad \qquad \qquad \qquad \qquad \qquad \qquad + \epsilon\,\nabla_{\mathbf{x}} \cdot (\cfrac{\mathbf{q}_{e}^{\tau,m} \otimes \mathbf{q}_{e}^{\tau,m}}{n^{\tau,m}}) \big] \Big) \Bigg] \, ,
\end{split}
\end{equation}
and
\begin{equation}
\begin{split}
&-\nabla_{\mathbf{x}} \cdot \big( n^{\tau,m+1}\, (\mathbf{b}^{m+1} \otimes \mathbf{b}^{m+1})\, \nabla_{\mathbf{x}}\phi^{\tau,m+1}\big) \\
& \qquad \qquad + \tau\, \cfrac{T_{e}\,C_{i} - \epsilon\,C_{e}}{\Delta t^{2}\, (T_{e}-1)} \, \phi^{\tau,m+1} + \tau\, \cfrac{T_{e}-\epsilon}{\Delta t^{2}\,(T_{e}-1)} \, n^{\tau,m+1} \\
&= \tau \times \cfrac{T_{e}}{T_{e}-1} \, \Bigg[ -\cfrac{1}{\Delta t}\, \nabla_{\mathbf{x}} \cdot \big( \mathbf{q}_{i}^{\tau,m} - \cfrac{\epsilon}{T_{e}}\, \mathbf{q}_{e}^{\tau,m}\big) + \cfrac{T_{e}-\epsilon}{\Delta t^{2}\, T_{e}} \, n^{\tau,m} \\
& \qquad + \cfrac{T_{\epsilon}\,C_{i}-\epsilon\,C_{e}}{\Delta t^{2}\,T_{e}} \, \phi^{\tau,m} + \nabla_{\mathbf{x}} \cdot \Big( (\mathbf{b}^{m+1} \otimes \mathbf{b}^{m+1})\, \big[ \nabla_{\mathbf{x}} \cdot ( \cfrac{\mathbf{q}_{i}^{\tau,m} \otimes \mathbf{q}_{i}^{\tau,m}}{n^{\tau,m}} ) \\
& \qquad \qquad \qquad \qquad \qquad \qquad \qquad \qquad \qquad - \cfrac{\epsilon}{T_{e}} \, \nabla_{\mathbf{x}} \cdot ( \cfrac{\mathbf{q}_{e}^{\tau,m} \otimes \mathbf{q}_{e}^{\tau,m}}{n^{\tau,m}} ) \big] \Big) \Bigg] \, .
\end{split}
\end{equation}
If we consider the constraints (\ref{constraints_CiCe}) for $C_{i}$ and $C_{e}$, we make the diffusion equations uncoupled. Firstly, we compute $n^{\tau,m+1}$ by solving
\begin{equation} \label{ELPP_diffusion_n_appendix}
-\nabla_{\mathbf{x}} \cdot \big((\mathbf{b}^{m+1} \otimes \mathbf{b}^{m+1}) \, \nabla_{\mathbf{x}}n^{\tau,m+1}\big) + \tau \, \lambda_{1} \, n^{\tau,m+1} = \tau \, R^{\tau,m+1} \, ,
\end{equation}
with
\begin{equation}
\begin{split}
\lambda_{1} &= \cfrac{1+\epsilon}{\Delta t^{2}\,(1+T_{e})} \, , \\
R^{\tau,m+1} &= \cfrac{1}{1+T_{e}} \, \Bigg[ -\cfrac{1}{\Delta t} \, \nabla_{\mathbf{x}} \cdot (\mathbf{q}_{i}^{\tau,m}+\epsilon\,\mathbf{q}_{e}^{\tau,m}) + \cfrac{1+\epsilon}{\Delta t^{2}} \, n^{\tau,m} \\
& \qquad \qquad + \nabla_{\mathbf{x}} \cdot \Big( (\mathbf{b}^{m+1} \otimes \mathbf{b}^{m+1})\, \big[ \nabla_{\mathbf{x}} \cdot (\cfrac{\mathbf{q}_{i}^{\tau,m} \otimes \mathbf{q}_{i}^{\tau,m}}{n^{\tau,m}}) \\
& \qquad \qquad \qquad \qquad \qquad \qquad \qquad + \epsilon\,\nabla_{\mathbf{x}} \cdot (\cfrac{\mathbf{q}_{e}^{\tau,m} \otimes \mathbf{q}_{e}^{\tau,m}}{n^{\tau,m}}) \big] \Big) \Bigg] \, ,
\end{split}
\end{equation}
then we use it to compute $\phi^{\tau,m+1}$ by solving
\begin{equation} \label{ELPP_diffusion_phi_appendix}
-\nabla_{\mathbf{x}} \cdot \big(n^{\tau,m+1}\,(\mathbf{b}^{m+1} \otimes \mathbf{b}^{m+1}) \, \nabla_{\mathbf{x}}\phi^{\tau,m+1}\big) + \tau \, \lambda_{2} \, \phi^{\tau,m+1} = \tau \, S^{\tau,m+1} \, ,
\end{equation}
with
\begin{equation}
\begin{split}
\lambda_{2} &= \cfrac{T_{e}\,C}{\Delta t^{2}\,(T_{e}-1)} \, , \\
S^{\tau,m+1} &= \cfrac{T_{e}}{T_{e}-1} \, \Bigg[ -\cfrac{1}{\Delta t}\, \nabla_{\mathbf{x}} \cdot \big( \mathbf{q}_{i}^{\tau,m} - \cfrac{\epsilon}{T_{e}}\, \mathbf{q}_{e}^{\tau,m}\big) \\
& \qquad \qquad + \cfrac{\epsilon-T_{e}}{\Delta t^{2}\, T_{e}} \, (n^{\tau,m+1}-n^{\tau,m}) + \cfrac{C}{\Delta t^{2}} \, \phi^{\tau,m} \\
& \qquad \qquad + \nabla_{\mathbf{x}} \cdot \Big( (\mathbf{b}^{m+1} \otimes \mathbf{b}^{m+1})\, \big[ \nabla_{\mathbf{x}} \cdot ( \cfrac{\mathbf{q}_{i}^{\tau,m} \otimes \mathbf{q}_{i}^{\tau,m}}{n^{\tau,m}} ) \\
& \qquad \qquad \qquad \qquad \qquad \qquad \qquad - \cfrac{\epsilon}{T_{e}} \, \nabla_{\mathbf{x}} \cdot ( \cfrac{\mathbf{q}_{e}^{\tau,m} \otimes \mathbf{q}_{e}^{\tau,m}}{n^{\tau,m}} ) \big] \Big) \Bigg] \, .
\end{split}
\end{equation}

\section{Reformulation of the fully-discrete problem} \label{reformulation_FD}
\setcounter{equation}{0}

In this paragraph, we develop the reformulation procedure for the finite volume scheme which is detailed in Section 5. This scheme writes
\begin{subnumcases}{\label{FD_ELPP_appendix}}
\begin{split}
&\cfrac{{n^{\tau,m+1}}_{|_{K}}-{n^{\tau,m}}_{|_{K}}}{\Delta t} + \Big(\nabla_{h} \cdot \big((\mathbf{b}^{m+1}\otimes\mathbf{b}^{m+1})\,\mathbf{q}_{\alpha}^{\tau,m+1}\big)\big)_{|_{K}} \\
& + \Big(\nabla_{h}^{FV} \cdot \big((\mathbb{I}-\mathbf{b}^{m+1} \otimes \mathbf{b}^{m+1})\,\mathbf{q}_{\alpha}^{\tau,m}\big)\Big)_{|_{K}} = -C_{\alpha}\,\cfrac{{\phi^{\tau,m+1}}_{|_{K}}-{\phi^{\tau,m}}_{|_{K}}}{\Delta t} \, ,
\end{split}
& \label{FD_ELPP_n_appendix} \\
\begin{split}
&\cfrac{{\mathbf{q}_{\alpha}^{\tau,m+1}}_{|_{K}}-{\mathbf{q}_{\alpha}^{\tau,m}}_{|_{K}}}{\Delta t} + \Big(\nabla_{h}^{FV} \cdot \big(\cfrac{\mathbf{q}_{\alpha}^{\tau,m} \otimes \mathbf{q}_{\alpha}^{\tau,m}}{n^{\tau,m}} \big) \Big)_{|_{K}} \\
& = -\cfrac{1}{\epsilon_{\alpha}\,\tau} \, \big[ T_{\alpha}\,\nabla_{h} n^{\tau,m+1} + \mathfrak{q}_{\alpha}\,(n^{\tau,m+1} \,\nabla_{h}\phi^{\tau,m+1} - \mathbf{q}_{i}^{\tau,m+1} \times \mathbf{B}^{m+1})\big]_{|_{K}} \, , 
\end{split}
& \label{FD_ELPP_q_appendix} \\
\alpha \in \{i,e\} \, .
\end{subnumcases}

As in the previous appendices, we separate the parallel part and the perpendicular part of (\ref{FD_ELPP_q_appendix}) according to ${\mathbf{b}^{m+1}}_{|_{K}}$. By multiplying the tensor $({\mathbf{b}^{m+1}}_{|_{K}} \otimes {\mathbf{b}^{m+1}}_{|_{K}})$ by (\ref{FD_ELPP_q_appendix}), we obtain (\ref{qpara_FD}) for each $\alpha$. Concerning $\big((\mathbf{q}_{\alpha}^{\tau,m+1})_{\perp}^{m+1}\big)_{|_{K}}$, we compute the vector product of ${\mathbf{b}^{m+1}}_{|_{K}}$ and (\ref{FD_ELPP_q_appendix}) and we obtain (\ref{qperp_FD}). \\
\indent In order to obtain the discrete diffusion equations (\ref{eq_nmp1_FD}) and (\ref{eq_phimp1_FD}), we follow the same procedure as in the semi-discrete case (see Appendix \ref{reformulation_SD_appendix}): we replace $(\mathbf{q}_{i}^{\tau,m+1})_{||}^{m+1}$ (resp. $(\mathbf{q}_{e}^{\tau,m+1})_{||}^{m+1}$) by its expression given by (\ref{qpara_FD}) with $\alpha = i$ (resp. $\alpha = e$) and we obtain two diffusion equation of the form
\begin{equation}
\left\{
\begin{array}{l}
\begin{split}
-\Big(&\nabla_{h}\cdot \big( (T_{\alpha}\,\nabla_{h}n^{\tau,m+1} + \mathfrak{q}_{\alpha}\,n^{\tau,m+1}\,\nabla_{h}\phi^{\tau,m+1})_{||}^{m+1}\big)\Big)_{|_{K}} \\
&\qquad \qquad \qquad \qquad + \cfrac{\epsilon_{\alpha}\,\tau}{\Delta t^{2}}\, (n^{\tau,m+1} + C_{\alpha} \, \phi^{\tau,m+1}) \\
&= \epsilon_{\alpha}\,\tau \, \Big[ \nabla_{h} \cdot \Big( \big( -\cfrac{1}{\Delta t}\, \mathbf{q}_{\alpha}^{\tau,m} + \nabla_{h}^{FV} \cdot (\cfrac{\mathbf{q}_{\alpha}^{\tau,m} \otimes \mathbf{q}_{\alpha}^{\tau,m}}{n^{\tau,m}}) \big)_{||}^{m+1}\Big) \\
&\qquad \qquad + \cfrac{1}{\Delta t^{2}}\, n^{\tau,m} + \cfrac{C_{\alpha}}{\Delta t^{2}}\, \phi^{\tau,m} - \cfrac{1}{\Delta t} \, \nabla_{h}^{FV} \cdot \big( (\mathbf{q}_{\alpha}^{\tau,m})_{\perp}^{m+1}\big) \Big]_{|_{K}} \, ,
\end{split}
\\
\alpha \in \{i,e\} \, .
\end{array}
\right.
\end{equation}

Finally, we consider the constraints (\ref{constraints_CiCe}) to make these 2 equations uncoupled and, up to some linear combinations, they can be rewritten under the form
\begin{equation} \label{eq_nmp1_FD_appendix}
-\Big(\nabla_{h}\cdot \big( (\nabla_{h}n^{\tau,m+1})_{||}^{m+1}\big)\Big)_{|_{K}} + \lambda_{1}\,\tau\,{n^{\tau,m+1}}_{|_{K}} = \tau\,{R^{\tau,m+1}}_{|_{K}} \, ,
\end{equation}
and
\begin{equation} \label{eq_phimp1_FD_appendix}
-\Big(\nabla_{h}\cdot \big( n^{\tau,m+1}\,(\nabla_{h}\phi^{\tau,m+1})_{||}^{m+1}\big)\Big)_{|_{K}} + \lambda_{2}\,\tau\,{\phi^{\tau,m+1}}_{|_{K}} = \tau\,{S^{\tau,m+1}}_{|_{K}} \, ,
\end{equation}
with $\lambda_{1}$, $\lambda_{2}$, ${R^{\tau,m+1}}_{|_{K}}$ and ${S^{\tau,m+1}}_{|_{K}}$ defined by
\begin{equation}
\begin{split}
\lambda_{1} &= \cfrac{1+\epsilon}{\Delta t^{2}\,(1+T_{e})} \, , \qquad \lambda_{2} = \cfrac{T_{e}\,C}{\Delta t^{2}\,(T_{e}-1)} \, , \\
{R^{\tau,m+1}}_{|_{K}} &= \cfrac{1}{1+T_{e}} \, \Bigg[ \cfrac{1+\epsilon}{\Delta t^{2}} \, n^{\tau,m} + \nabla_{h} \cdot \Big(\big( -\cfrac{1}{\Delta t} \, (\mathbf{q}_{i}^{\tau,m}+\epsilon\,\mathbf{q}_{e}^{\tau,m}) \\
&\qquad + \nabla_{h}^{FV} \cdot (\cfrac{\mathbf{q}_{i}^{\tau,m} \otimes \mathbf{q}_{i}^{\tau,m}}{n^{\tau,m}}) + \epsilon\, \nabla_{h}^{FV} \cdot (\cfrac{\mathbf{q}_{e}^{\tau,m} \otimes \mathbf{q}_{e}^{\tau,m}}{n^{\tau,m}}) \big)_{||}^{m+1} \Big) \\
&\qquad - \cfrac{1}{\Delta t} \, \Big(\nabla_{h}^{FV} \cdot \big( (\mathbf{q}_{i}^{\tau,m})_{\perp}^{m+1}\big) + \epsilon \, \nabla_{h}^{FV} \cdot \big( (\mathbf{q}_{e}^{\tau,m})_{\perp}^{m+1}\big) \Big) \Bigg]_{|_{K}} \, , \\
{S^{\tau,m+1}}_{|_{K}} &= \cfrac{T_{e}}{T_{e}-1} \, \Bigg[ \cfrac{\epsilon-T_{e}}{\Delta t^{2}\,T_{e}} \, (n^{\tau,m+1} - n^{\tau,m}) + \cfrac{C}{\Delta t^{2}} \, \phi^{\tau,m} \\
&\qquad + \nabla_{h} \cdot \Big( \big( -\cfrac{1}{\Delta t}\, (\mathbf{q}_{i}^{\tau,m}-\cfrac{\epsilon}{T_{e}}\,\mathbf{q}_{e}^{\tau,m}) \\
&\qquad + \nabla_{h}^{FV}\cdot (\cfrac{\mathbf{q}_{i}^{\tau,m} \otimes \mathbf{q}_{i}^{\tau,m}}{n^{\tau,m}}) - \cfrac{\epsilon}{T_{e}} \, \nabla_{h}^{FV} \cdot (\cfrac{\mathbf{q}_{e}^{\tau,m} \otimes \mathbf{q}_{e}^{\tau,m}}{n^{\tau,m}}) \big)_{||}^{m+1} \Big) \\
&\qquad - \cfrac{1}{\Delta t} \Big( \nabla_{h}^{FV} \cdot \big( (\mathbf{q}_{i}^{\tau,m})_{\perp}^{m+1}\big) - \cfrac{\epsilon}{T_{e}} \, \nabla_{h}^{FV} \cdot \big((\mathbf{q}_{e}^{\tau,m})_{\perp}^{m+1}\big) \Big) \Bigg]_{|_{K}} \, .
\end{split}
\end{equation}

\section*{Acknowledgments} This work has been supported by the french magnetic fusion programme
FR-FCM, by the CEA-Cadarache in the frame of the contract 'APPLA'
(\# V3629.001 av. 2), by the INRIA large-scale initiative 'FUSION', by the University Paul Sabatier
in the frame of the contract 'MOSITER' and by the Fondation 'Sciences et Technologies pour 
l'A\'eronautique et l'Espace', in the frame of the project 'Plasmax' 
(contract \# RTRA-STAE/2007/PF/002).


\begin{thebibliography}{1}


\bibitem{Beer-Hammett} \textsc{Beer, M., Hammett, G.}, \textit{Toroidal gyrofluid equations for simulations of tokamak turbulence}, Phys. Plasmas, \textbf{3}-11 (1996), 4046-4064.

\bibitem{Belaouar-Crouseilles-Degond-Sonnendrucker} \textsc{Belaouar, R., Crouseilles, N., Degond, P., Sonnendr\"ucker, E.}, \textit{An asymptotically stable semi-lagrangian scheme in the quasi-neutral limit}, J. Sci. Comput. \textbf{41} (2009), 341-365.

\bibitem{Brizard_PhD} \textsc{Brizard, A.} \textit{Nonlinear gyrokinetic tokamak physics}, PhD thesis of Princeton University (1990).

\bibitem{Hahm-Brizard} \textsc{A. Brizard, T.-S. Hahm}, \textit{Foundations of nonlinear gyrokinetic theory}, Rev. Mod. Phys. \textbf{79} (2007), 421-468.

\bibitem{BDD} \textsc{Brull, S., Degond, P., Deluzet, F.}, \textit{Degenerate anisotropic elliptic problems and magnetized plasmas simulations}, to appear in Commun. Comput. Phys.

\bibitem{Buet-Cordier-Lucquin-Mancini} \textsc{Buet, C., Cordier, S., Lucquin-Desreux, B., Mancini, S.}, \textit{Diffusion limit of the Lorentz model: Asymptotic-Preserving schemes}, Model. Math. Anal. Numer. \textbf{36}-4 (2002), 631-655.

\bibitem{Buet-Despres} \textsc{Buet, C., Despr\'es, B.}, \textit{Asymptotic-Preserving and positive schemes for radiation hydrodynamics}, J. Comput. Phys. \textbf{215} (2006), 717-740.

\bibitem{Carrillo-Goudon-Lafitte} \textsc{Carrillo, J.-A., Goudon, T., Lafitte, P.}, \textit{Simulation of fluid and particles flows: Asymptotic-Preserving schemes for bubbling and flowing regimes}, J. Comput. Phys. \textbf{223}-1 (2007), 208-234.

\bibitem{crisp1} \textsc{Crispel, P., Degond, P., Vignal, M.-H.}, \textit{Quasi-neutral fluid models for current-carrying plasmas}, J. Comput. Phys. \textbf{223}-1 (2007), 208-234.

\bibitem{crisp2} \textsc{Crispel, P., Degond, P., Vignal, M.-H.}, \textit{An asymptotic preserving scheme for the two-fluid Euler-Poisson model in the quasi-neutral limit}, J. Comput. Phys. \textbf{205}-2 (2005), 408-438.

\bibitem{crisp3} \textsc{Crispel, P., Degond, P., Vignal, M.-H.}, \textit{A plasma expansion model based on the full Euler-Poisson system},
Math. Models Methods Appl. Sci. \textbf{17}-7 (2007), 1129-1158.

\bibitem{Crouseilles-Lemou} \textsc{Crouseilles, N., Lemou, M.} \textit{An Asymptotic-Preserving scheme based on a micro-macro decomposition for collisional Vlasov equations: diffusion and high-field scaling limits}, submitted.

\bibitem{Narski} \textsc{Degond, P., Deluzet, F., Lozinski, A., Narski, J., Negulescu, C.}, \textit{Duality-based asymptotic-preserving method for highly anisotropic diffusion equations}, to appear in Comm. Math. Sci.

\bibitem{Navoret} \textsc{Degond, P., Deluzet, F., Navoret, L., Sun, A.-B., Vignal, M.-H.}, \textit{Asymptotic-Preserving Particle-In-Cell method for the Vlasov-Poisson system near quasi-neutrality}, J. Comput. Phys. \textbf{229}-16 (2010), 5630-5652.

\bibitem{Degond-Deluzet-Negulescu} \textsc{Degond, P., Deluzet, F., Negulescu, C.}, \textit{An Asymptotic-Preserving scheme for strongly anisotropic elliptic problems}, Multiscale Model. Simul. \textbf{8}-2 (2010), 645-666.

\bibitem{DDSV} \textsc{Degond, P., Deluzet, F., Sangam, A., Vignal, M.-H.}, \textit{An asymptotic preserving scheme for the Euler equations in a strong magnetic field}, J. Comput. Phys. \textbf{228} (2009), 3540-3558.

\bibitem{Savelief} \textsc{Degond, P., Liu, H., Savelief, D., Vignal, M.-H.}, \textit{Numerical approximation of the Euler-Poisson-Boltzmann model in the quasi-neutral limit}, to appear in J. Sci. Comput.

\bibitem{Degond-Liu-Vignal} \textsc{Degond, P., Liu, J.-G., Vignal, M.-H.}, \textit{Analysis of an Asymptotic-Preserving scheme for the Euler-Poisson system in the quasi-neutral limit}, J. Numer. Anal. \textbf{46}-3 (2008), 1298-1322.

\bibitem{Degond-Tang} \textsc{Degond, P., Tang, M.}, \textit{All speed scheme for the low Mach number limit of the isentropic Euler equation}, to appear in Commun. Comput. Phys.

\bibitem{Dhaeseleer} \textsc{D'haeseleer, W.-D., Hitchon, W.-N.-G., Callen, J.-D., Shohet, J.-L.}, \textit{Flux coordinates and magnetic field structure. A guide to a fundamental tool of plasma theory}, Springer Series in Computational Physics, Springer-Verlag, Berlin (1991).

\bibitem{Dorland-Hammett} \textsc{Dorland, W., Hammett, G.}, \textit{Gyrofluid turbulence models with kinetic effects}, Phys. Fluids B \textbf{5}-3 (1993), 812-835.

\bibitem{Dubin} \textsc{Dubin, D.-H., Krommes, J.-A., Oberman, C., Lee, W.-W.}, \emph{Nonlinear gyrokinetic equations}, Phys. Fluids \textbf{26}-12 (1983), 3524-3535.

\bibitem{Falchetto} \textsc{Falchetto, G.-L., Ottaviani, M.} \textit{Effect of Collisional Zonal-Flow Damping on Flux-Driven Turbulent Transport}, Phys. Rev. Lett. \textbf{92}-2 (2004), 025002.


\bibitem{Filbet-Jin_stiff} \textsc{Filbet, F., Jin, S.} \textit{A class of Asymptotic-Preserving schemes for kinetic equations and related problems with stiff sources}, J. Comput. Phys. \textbf{229}-20 (2010), 7625-7648.

\bibitem{Filbet-Jin_BGK} \textsc{Filbet, F., Jin, S.} \textit{An Asymptotic-Preserving Scheme for the ES-BGK model of the Boltzmann equation}, J. Sci. Comput. \textbf{46}-2 (2011), 204-224.

\bibitem{Frenod-Mouton} \textsc{Fr\'enod, E., Mouton, A.} \textit{Two-dimensional Finite Larmor Radius approximation in canonical gyrokinetic coordinates}, J. Pure Appl. Math.: Advances Appl. \textbf{4}-2 (2010), 135-166.

\bibitem{Garbet} \textsc{Garbet, X., Bourdelle, C., Hoang, G.-T., Maget, P., Benkadda, S., Beyer, P., Figarella, C., Voitsekovitch, I., Agullo, O., Bian, N.}, \textit{Global simulations of ion turbulence with magnetic shear reversal}, Phys. Plasmas \textbf{8}-6 (2001), 2793-2803.

\bibitem{Grandgirard_JCP} \textsc{Grandgirard, V., Brunetti, M., Bertrand, P., Besse, N., Garbet, X., Gendrih, P., Manfredi, G., Sarazin, Y., Sauter, O., Sonnendr\"ucker, E., Vaclacik, J., Villard, L.}, \textit{A drift-kinetic semi-lagrangian 4D code for ion turbulence simulation}, J. Comput. Phys. \textbf{217} (2006), 395-423.

\bibitem{Grandgirard} \textsc{Grandgirard, V., Sarazin, Y., Garbet, X., Dif-Pradalier, G., Gendrih, P., Crouseilles, N., Latu, G., Sonnendr\"ucker, E., Besse, N., Bertrand, P.}, \textit{Computing ITG turbulence with a full-f semi-lagrangian code}, Comm. Nonlinear Sci. Numer. Simul. \textbf{13}-1 (2008), 81-87.

\bibitem{Hammett-Beer-Dorland-Cowley-Smith} \textsc{Hammett, G.-W., Beer, M.-A., Dorland, W., Cowley, S.-C., Smith, S.-A.}, \textit{Developments in the gyrofluid approach to tokamak turbulence simulations}, Plasmas Phys. Control. Fusion \textbf{35}-8 (1993), 973-985.

\bibitem{Hasegawa-Mima} \textsc{Hasegawa, A., Mima, K.}, \textit{Stationary Spectrum of Strong Turbulence in Magnetized Nonuniform Plasma}, Phys. Rev. Lett. \textbf{39}-4 (1977), 205-208.

\bibitem{Hasegawa-Wakatani} \textsc{Hasegawa, A., Wakatani, M.}, \textit{Plasma Edge Turbulence}, Phys. Rev. Lett. \textbf{50}-9 (1983), 682-686.

\bibitem{Hazeltine} \textsc{Hazeltine, R.-D., Meiss, J.-D.} \textit{Plasma confinement}, Dover Publications (2003).

\bibitem{Heikkinen} \textsc{Heikkinen, J.-A., Janhunen, S.-J., Kiviniemi, T.-P., Ogando, F.}, \textit{Full-f gyrokinetic method for particle simulation of tokamak transport}, J. Comput. Phys. \textbf{227}-11 (2008), 5582-5609.

\bibitem{Jin} \textsc{Jin, S.}, \textit{Efficient Asymptotic-Preserving (AP) schemes for some multiscale kinetic equations}, J. Sci. Comput. \textbf{21}-2 (1999), 451-454.

\bibitem{Klar} \textsc{Klar, A.}, \textit{An Asymptotic-Preserving numerical scheme for kinetic equations in the low Mach number limit}, J. Numer. Anal. \textbf{36} (2009), 1507-1527.

\bibitem{Lee} \textsc{W.-W. Lee}, \emph{Gyrokinetic approach in particle simulation}, Phys. Fluids \textbf{26}-2 (1983), 555-562.

\bibitem{Lee_2} \textsc{W.-W. Lee}, \emph{Gyrokinetic particle simulation model}, J. Comput. Phys. \textbf{72}-1 (1987), 243-269.

\bibitem{Lemou-Mieussens} \textsc{Lemou, M., Mieussens, L.}, \textit{A new Asymptotic-Preserving scheme based on micro-macro decomposition for linear kinetic equations in the diffusion limit}, J. Sci. Comput. \textbf{31} (2008), 334-368.

\bibitem{LeVeque} \textsc{LeVeque, R.}, \textit{Finite volume methods for hyperbolic problems}, Cambridge texts in Applied mathematics (2002).

\bibitem{Littlejohn} \textsc{R.-G. Littlejohn}, \emph{A guiding center Hamiltonian : A new approach}, J. Math. Phys. \textbf{20}-12 (1979), 2445-2458.

\bibitem{McClarren-Lowrie} \textsc{McLarren, R.-G., Lowrie, B.}, \textit{The effects of slope limiting on Asymptotic-Preserving schemes numerical methods for hyperbolic conservation laws}, J. Comput. Phys. \textbf{227} (2008), 9711-9726.

\bibitem{Miyamoto} \textsc{Miyamoto, K.}, \textit{Controlled fusion and plasma physics}, Chapman \& Hall (2007).

\bibitem{Ottaviani} \textsc{Ottaviani, M.}, \textit{An alternative approach to field-aligned coordinates for plasma turbulence simulations}, arXiv:1002.0748.

\bibitem{Ottaviani-Manfredi} \textsc{Ottaviani, M., Manfredi, G.}, \textit{The gyro-radius scaling of ion thermal transport from global numerical simulations of ion temperature gradient driven turbulence}, Phys. Plasmas \textbf{6}-8 (1999), 3267-3275.

\bibitem{Rusanov} \textsc{Rusanov, V.-V.}, \textit{The calculation of the iteraction of non-stationary shock waves and obstacles}, J. Comp. Math. Phys. \textbf{1} (1961), 267-279.

\end{thebibliography}
\end{document}